\documentclass[pra,aps,superbib,citeautoscript,twocolumn,longbibliography,superscriptaddress]{revtex4-1}

\usepackage[colorlinks=true,urlcolor=blue,citecolor=blue,linkcolor=blue]{hyperref}

\usepackage{amsmath,bm}
\usepackage{amstext}
\usepackage{epsfig}
\usepackage{xcolor}
\usepackage{subfig}
\usepackage{graphicx,amsmath,amssymb,tabularx}
\usepackage{multirow}
\usepackage{array}
\usepackage{dsfont}
\usepackage{caption}
\usepackage{cleveref}
\usepackage{ulem}
\usepackage[toc]{appendix}
\usepackage{physics}
\usepackage{mathrsfs}  
\usepackage{amsbsy}

\definecolor{dgreen}{rgb}{0,.5,0}
\definecolor{dred}{rgb}{.7,.0,.0}

\DeclareMathOperator*{\argmin}{arg\,min}
\DeclareMathAlphabet\mathbfcal{OMS}{cmsy}{b}{n}

\newcommand{\bxi}{\bm{\xi}}

\newcommand{\br}{{\bf r}}
\newcommand{\dr}{d{\bf r}}

\newcommand{\ie}{{\it i.e.}}

\newcommand{\be}{\begin{eqnarray}}
\newcommand{\ee}{\end{eqnarray}}
\DeclareMathAlphabet\mathbfcal{OMS}{cmsy}{b}{n}

\newcommand{\contract}[2]{\left( #1 \middle\vert #2 \right)}

\begin{document}

\title{Ensemble density functional theory of excited states: Exact $N$-centered formalism and practical opportunities}

\author{Lucien Dupuy}
\email{lucien.dupuy@unistra.fr}
\affiliation{Laboratoire de Chimie Quantique,
Institut de Chimie, CNRS / Universit\'{e} de Strasbourg,
4 rue Blaise Pascal, 67000 Strasbourg, France}
\affiliation
{University of Strasbourg Institute for Advanced Study,
5, all\'{e}e du G\'{e}n\'{e}ral Rouvillois, F-67083 Strasbourg, France}
\author{Toni Chiti}
\affiliation{Laboratoire de Chimie Quantique,
Institut de Chimie, CNRS / Universit\'{e} de Strasbourg,
4 rue Blaise Pascal, 67000 Strasbourg, France}
\author{J\'{e}r\'{e}my Morere}
\affiliation{\textit{Université de Lorraine,} CNRS, LPCT, \textit{F-54000 Nancy, France}}
\author{Emmanuel Fromager}
\affiliation{Laboratoire de Chimie Quantique,
Institut de Chimie, CNRS / Universit\'{e} de Strasbourg,
4 rue Blaise Pascal, 67000 Strasbourg, France}
\affiliation
{University of Strasbourg Institute for Advanced Study,
5, all\'{e}e du G\'{e}n\'{e}ral Rouvillois, F-67083 Strasbourg, France}

\begin{abstract}
Ground-state electronic structure calculations using Kohn--Sham density functional theory (KS-DFT) offer an unprecedented balance between efficiency and accuracy, now paradigmatic to the fields of quantum chemistry and condensed matter physics. KS-DFT can be extended to model electronic excitations through a density mapping onto a non-interacting ensemble state in which, unlike in thermal theories, the weights assigned to the excited states vary independently. Thanks to its numerous appeals, like the adequate treatment of multiple excitations for which the widely-used time-dependent extension of DFT struggles, ensemble DFT (eDFT) has lately become a vibrant area of research. Recently, an enlarged type of ensemble, referred to as $N$-centered (Nc) ensemble, has been introduced to describe within the same unified formalism both neutral and charged electronic excitations. This perspective paper provides a detailed exposition of exact Nc-eDFT, with a comprehensive review of its formal developments. To cut practical computational tools out of the exact theory, three original strategies are presented, complementing existing approaches. The first one, related to the design of ensemble density-functional approximations, consists in recycling regular ground-state functionals by dressing them with a weight-dependent scaling function deduced from exact properties of eDFT. We then explore quasi-degenerate formulations of ensemble density-functional perturbation theory, suggesting alternative definitions for the ensemble Hartree, exchange, and correlation energies, individually, and paving the way toward robust orbital-dependent eDFAs. Finally, we revisit and generalize the concept of quantum bath for an ensemble of non-interacting states, laying the foundations of an in-principle exact (in the sense of lattice eDFT) quantum embedding theory of excited states.
\end{abstract}

\maketitle




\section{Introduction}


For the computation of ground-state properties in molecules and materials, Kohn--Sham density functional theory~\cite{hktheo,KS} (KS-DFT) is firmly established as the workhorse of electronic structure theory~\cite{Teale2022_DFT_exchange}. Indeed, thanks to sustained development of ever more accurate density functional approximations~\cite{perdew1996generalized,Burke2012perspectiveDFA,Goerigk2017dftzoo} (DFAs), it reached an unprecedented balance between numerical efficiency and precision, allowing simulation of systems with up to thousands of atoms~\cite{Tanaka2024largescaleDFT}.
Description of electronic excitations through its time-dependent extension~\cite{runge1984density,casida1995timedependent,Casida_tddft_review_2012} (TDDFT) is comparatively less satisfactory because of several practical difficulties. One is the absence from the computed spectrum of multiple excitations (two holes/two particles in the case of double excitations, for example) under the linear-response regime with the widely-used adiabatic approximation~\cite{Casida_tddft_review_2012,elliott2011perspectives}, as a frequency-dependent (non-adiabatic) exchange-correlation (xc) kernel is required to model them~\cite{Lacombe2023_Non-adiabatic}. This is an area of active development, with important progress made for weakly-correlated systems where it was shown that dressed TDDFT~\cite{Maitra_DTDDFT04} provides the frequency dependence in an inexpensive and accurate manner~\cite{Dar2023_TDDFT_DE,Dar2025_DTDDFT_CI,baranova2025_excited-state_densities}, provided the double excitation is coupling to one or more single excitation(s) while being well-separated from the others. Another important limitation of linear response TDDFT is the breakdown of its single-reference picture in the vicinity of conical intersections~\cite{Casida_tddft_review_2012,Matsika2021elstructCI}. Achieving a balanced description of long- and short-range interactions, as required to describe charge-transfer states~\cite{maitra2017charge}, is also challenging without making the functional significantly more expensive to compute.

Setting aside neutral excitations and turning to the prediction of charged excitations within DFT, the most established approach is the DFT for fractional electron numbers of Perdew, Parr, Levy, and Balduz~\cite{perdew1982density} (PPLB). Its practical use is also impaired by an important limitation however, that is the need to model density-functional derivative discontinuities when crossing an integer electron number~\cite{perdew1982density,mori2008localization,cohen2008fractional,cohen2008insights,stein2010fundamental,zheng2011improving,cohen2011challenges,kraisler2013piecewise,kraisler2014fundamental,perdew2017understanding}. Standard (semi-)local DFAs lack this feature, leading for example to a systematic underestimation of the fundamental gap in solids~\cite{perdew2017understanding}, and unsatisfactory modeling of orbital relaxation effects when computing Fukui functions~\cite{Hellgren12_Effect}. Derivative discontinuities also play a role in the description of molecular dissociation and charge transfer processes~\cite{hodgson2017interatomic,Kraisler21_From}, but the aforementioned formal disparity in treatment of charged and neutral excitations makes it difficult to turn the analysis of such features in model systems into practical breakthroughs.

Ensemble DFT (eDFT), which is in the present context an in-principle exact alternative to TDDFT, describes variationally static ensembles of ground and low-lying excited states~\cite{Cernatic2022,gould2026ensemblization}. Introduced by Theophilou~\cite{JPC79_Theophilou_equi-ensembles,theophilou_book} and then further developed by Gross, Oliveira, and Kohn~\cite{gross1988rayleigh,gross1988density} (TGOK) for the prediction of neutral excitation processes, TGOK-DFT does not suffer from TDDFT's limitations regarding the description of double excitations~\cite{sagredo2018can,marut2020weight,Gould2021_Double} and degeneracies~\cite{Ullrich2001EDFTdegen}, while having essentially the same computational cost as a regular ground-state KS-DFT calculation. These are the main reasons why it is currently enjoying a renaissance with numerous analyses~\cite{deur2018exploring,deur2019ground,Fromager_2020,scott2024exact,Giarrusso2023_Exact,Giarrusso_excDFT25,Loos2025excstUEG,gould2017hartree,gould2019asymptotic,PRL20_Gould_Hartree_def_from_ACDF_th,PRL19_Gould_DD_correlation,gould2025stationaryconditionsexcitedstates}, developments of DFAs~\cite{yang2014exact,gould2018charge,loos2020weightdependent,Yang2021_Second,Gould2020_Approximately,PRL20_Gould_Hartree_def_from_ACDF_th,Gould2023_Electronic,gould2024local,Gould2025_PRL_tate-Specific,gould2026ensemblization}, and formal extensions~\cite{filatov2015spin,Filatov21SSR,Gould2021_Ensemble_ugly,Fromager2025indvElevel,dupuy2025_exact_static,daas2025ensembletimedependentdensityfunctional}. In this work, we will focus on one of these extensions, named $N$-centered eDFT~\cite{senjean2018unified,Cernatic2024_Neutral} (Nc-eDFT), which provides a unified density-functional picture of both neutral {\it and} charged electronic excitations. Previous investigations established it as a promising framework to tackle open quantum systems~\cite{Senjean_2020}, describe exchange-correlation (xc) derivative discontinuity contributions to gaps through (in-principle simpler to model) analytical ensemble weight derivatives~\cite{senjean2018unified,PRA21_Hodgson_exact_Nc-eDFT_1D,Cernatic2022}, and seamlessly extend key properties of either charged or neutral excitations to both, such as the exactification of Koopman's theorem~\cite{PRA21_Hodgson_exact_Nc-eDFT_1D,Cernatic2022,Cernatic2024_Neutral} through allowed constant shifts in the xc potential.

The purpose of the present Perspective paper is twofold: First, as a complement to a recent review of eDFT by Gould and coworkers~\cite{gould2026ensemblization}, we would like to present and discuss the latest theoretical developments of Nc-eDFT (which incorporates TGOK-DFT, by construction), thus giving a clear view of where the exact formalism currently stands, what can be computed with it, and what formal developments we envision next. Second, we will outline and experiment with several strategies to cut practical tools out of the exact formalism.
Regarding the latter point, note that several directions have already been taken (and continue to be) within eDFT, either through orbital-based eDFAs~\cite{yang2014exact,pribramjones2014excitations,yang2017direct,gould2018charge,sagredo2018can,deur2018exploring,senjean2018unified,deur2019ground,Senjean_2020,Yang2021_Second}, or full-fledged density functional developments~\cite{Gould2020_Approximately,PRL20_Gould_Hartree_def_from_ACDF_th,Gould2023_Electronic,Gould2025_PRL_tate-Specific} (we refer the reader to Ref.~\onlinecite{gould2026ensemblization} for further details). The alternative strategies presented in this work have (to the best of our knowledge) not been  explored so far and thus bring additional perspective to the field. As a first strategy, we propose to recycle regular (KS-DFT) functionals by dressing them with an ensemble weight-dependent scaling function, with an illustration on how exact properties of eDFT allow to build such a function. The second strategy deals with the quasi-degenerate formulation of ensemble density-functional perturbation theory, either from the perspective of Rayleigh--Schr\"{o}dinger perturbation theory~\cite{Lindgren1986} or its unitary Van Vleck analog~\cite{VanVleck29_On,SOKOLOV2024121}. Our third strategy is to leverage the recent successes of the local potential functional embedding theory of pure ground states~\cite{makhlouf2025_local_potential,makhlouf2026generalizedlocalpotentialfunctional} (LPFET), which is an in-principle exact lattice-DFT-based reformulation of density matrix embedding theory~\cite{knizia2012density}, by extending, in a general and comprehensive way, the concept of quantum bath to many-electron ensembles.

The paper is structured as follows. Sec.~\ref{sec:review_section_NcEDFT} is a review of Nc-eDFT where key features of the exact theory are highlighted and discussed. In the next Sec.~\ref{sec:densities_and_lrfunc_from_eDFT}, we report recent exact results derived in the context of TGOK-DFT: The derivation of an ensemble density-functional stationarity condition for ground and excited energy levels, and the (related) extraction of excited-state properties such as the density or the static density-density linear response function. Implications for the rationalization and improvement of $\Delta$-SCF methods are presented in Sec.~\ref{sec:Delta_scf_from_eDFT}. As a broader and practical perspective, the three strategies outlined previously are introduced and discussed in detail in Sec.~\ref{sec:routes_practice}.
Conclusions and additional perspectives are finally given in Sec.~\ref{sec:conclusions}.


\section{$N$-centered eDFT: Key features of the exact theory}\label{sec:review_section_NcEDFT}


\subsection{$N$-centered ensemble formalism}

Perhaps the most appealing feature of the Nc ensemble framework, introduced in Ref.~\cite{senjean2018unified} and then generalized in Refs.~\onlinecite{Cernatic2024_Neutral,cernatic2024extended_doubles} to any type of electronic excitation, is to trivially extend all our knowledge of TGOK ensembles’ formal properties to the study of charged excitations. Embracing this notion in full brings insights on the widest known family of ensembles satisfying a variational principle. An Nc ensemble is defined by its density matrix operator:
\begin{equation} \label{eq:densop}
    \hat{\Gamma}^{\bxi} = \left(1-\sum_{\lambda>0}\frac{N_\lambda}{N} \xi_\lambda \right) \ket{\Psi_0} \bra{\Psi_0} + \sum_{\lambda>0} \xi_\lambda \ket{\Psi_\lambda} \bra{\Psi_\lambda},
\end{equation}
where $\ket{\Psi_0}$ is the reference $N$-electron ground state of the system under study. All $\ket{\Psi_{\lambda>0}}$ can either be (neutral) $N$-electron excited states 
or charged excited states, \ie, ground or excited states with an (integer) electron number $N_\lambda = N \pm p\,,\; p \in \mathbb{N}^*$. In fact, any mixture of both charged and neutral excitations under appropriate weight constraints, being all weights within a given $\mathcal{N}$-electron sector of the Fock space are monotonically decreasing with the energy~\cite{gross1988rayleigh},
\begin{equation} \label{eq:weight_ordering}
    \xi_{\lambda-1} \overset{N_{\lambda-1}=N_{\lambda}=\mathcal{N}}{\geq} \xi_{\lambda} \geq 0,
\end{equation}
 satisfy a variational principle. This is the focus of Sec.~\ref{sec:var_principle} together with the introduction of the KS-DFT formulation of the theory. As evident from Eq.~\eqref{eq:densop}, the weight $\xi_0$ associated to $\ket{\Psi_0}$ is directly given by the set of weights $\{\xi_{\lambda>0}\}:=\bxi$ assigned to the excited states,
 \begin{equation} \label{eq:xi_0}
     \xi_{0} = \left(1-\sum_{\lambda>0}\frac{N_\lambda}{N} \xi_\lambda \right) \,.
 \end{equation}
 It is important to note that, in contrast to regular PPLB~\cite{perdew1982density} and TGOK~\cite{JPC79_Theophilou_equi-ensembles,gross1988density} ensembles, an Nc ensemble is not necessarily normalized:
 \begin{equation}
     \sum_{\nu \geq 0} \xi_{\nu} = 1 + \sum_{\lambda > 0} \frac{N-N_{\lambda}}{N} \xi_{\lambda} \, .
 \end{equation}
It has to be this way to ensure that the net number of electrons described by the ensemble is systematically equal to the (so-called central) integer number $N_0=N$ of electrons in the reference ground state $\Psi_0$, \ie,  
\be\label{eq:net_number_electrons_preserved}
\Tr\left[\hat{\Gamma}^{\bxi}\hat{N}\right]=\sum_{\nu\geq 0}\xi_\nu N_\nu=N,
\ee
where $\Tr$ denotes the trace and $\hat{N}$ is the electron counting operator. The above property holds for any set of weights, even if charged excitations are included into the ensemble, hence the name ``$N$-centered'' given to the formalism. If we introduce the second-quantized electron density operator at position $\br$,  
\be\label{eq:dens_op_SQ}
\displaystyle \hat{n}({\bf r})=\sum_\sigma \hat{\Psi}^\dagger_\sigma({\br})\hat{\Psi}_\sigma({\br}),
\ee
$\sigma=\uparrow,\downarrow$ denoting a one-electron spin state, the counting operator reads $\hat{N}=\int d\br\, \hat{n}({\bf r})$ and Eq.~(\ref{eq:net_number_electrons_preserved}) translates equivalently as the fact that the Nc ensemble density,
\begin{equation} \label{eq:nc_dens}
\begin{split}
    n^{\bxi}({\br})&= {\rm Tr} \left[ \hat{\Gamma}^{\bxi} \hat{n}({\bf r}) \right] 
    \\
    &= \left(1-\sum_{\lambda>0}\frac{N_\lambda}{N} \xi_\lambda \right) n_{\Psi_0}({\br}) + \sum_{\lambda>0} \xi_\lambda n_{\Psi_\lambda}({\br}),
\end{split}
\end{equation}
where the electronic density of any pure-state (normalized) wave function $\Psi$ is evaluated as follows,
\be\label{eq:dens_Psi_from_dens_op}
n_{\Psi}({\br})=\langle\Psi \vert\hat{n}({\bf r})\vert\Psi\rangle,
\ee
integrates to $N$, by construction. The motivation for making such a choice, which is unusual and therefore surprizing at first sight, was originally motivated by a wish to describe charged electronic excitations with the mathematical language of eDFT for neutral excitations~\cite{senjean2018unified}. The implications of this change of paradigm will be highlighted in the following.  

\subsection{Variational principle for $N$-centered ensembles 
and subsequent KS-DFT} \label{sec:var_principle}

Let us consider the general electronic Hamiltonian expression,
\be\label{eq:physical_Hamil}
\hat{H}=\hat{T}+\hat{W}_{\rm ee}+\hat{V}_{\rm ext},
\ee
where the kinetic energy $\hat{T}$, the electronic repulsion $\hat{W}_{\rm ee}$, and the external local potential $\hat{V}_{\rm ext}$ (which is usually the attractive nuclear potential in quantum chemistry) operators read in second quantization as follows,  
\be\label{eq:kinetic_ener_op_SQ} 
\hat{T}=-\dfrac{1}{2}\sum_{\sigma}\int d{\bf r}\,\hat{\Psi}^\dagger_\sigma({\br})\nabla_{\br}^2\hat{\Psi}_\sigma({\br}) \,,
\ee
\be
\hat{W}_{\rm ee}=\dfrac{1}{2}\sum_{\sigma,\sigma'}\int d{\bf r}\int d{\bf r}'\dfrac{\hat{\Psi}^\dagger_\sigma({\br})\hat{\Psi}^\dagger_{\sigma'}({\br}')\hat{\Psi}_{\sigma'}({\br}')\hat{\Psi}_\sigma({\br})}{\vert {\bf r}-{\bf r}'\vert} \,,
\ee
and 
\be\label{eq:ext_pot_operator_sq}
\hat{V}_{\rm ext}=\int d{\bf r}\,v_{\rm ext}({\bf r})\hat{n}({\bf r}), 
\ee
respectively. In a given $\mathcal{N}$-electron sector of the Fock space, the Hamiltonian can be written equivalently in the usual first-quantized form 
\be\label{eq:first_quantized_physical_Hamil}
\hat{H}\equiv -\dfrac{1}{2}\sum^{\mathcal{N}}_{i=1}\nabla_{\br_i}^2+\sum_{1\leq i<j}^{\mathcal{N}}\dfrac{1}{\vert {\bf r}_i-{\bf r}_j\vert}+\sum^{\mathcal{N}}_{i=1}v_{\rm ext}({\bf r}_i).
\ee
The Nc ensemble energy is then given by
\begin{equation} \label{eq:Nc_ens_E}
\begin{split}
    E^{\bxi} &= {\rm Tr} \left[  \hat{\Gamma}^{\bxi} \hat{H} \right] 
    \\
    &= \left(1-\sum_{\lambda>0}\frac{N_\lambda}{N} \xi_\lambda \right) E_0 + \sum_{\lambda>0} \xi_\lambda E_\lambda,
\end{split}
\end{equation}
where $E_0$ is the reference $N$-electron ground-state energy of the physical Hamiltonian $\hat{H}$, while $E_{\lambda>0}$ is a charged or neutral excited-state (with respect to the reference $N$-electron ground state) energy, \ie, the energy of an $N$-electron excited state or, if the number $N_\lambda$ of electrons in the excited state differs from $N$, a ground- or excited-state energy. Provided all weights $\{\xi_\lambda\}_{N_\lambda=\mathcal{N}}$ of any $\mathcal{N}$-electron sector of the Fock space are non-negative and monotonically decreasing with the energy (so that the $\mathcal{N}$-electron ground-state energy can be identified from the largest weight in that sector, the first excited-state energy from the second largest one, and so on), the complete ensemble energy can be determined variationally (while keeping the weights {\it fixed}), \ie
\be\label{eq:general_Nc_VP_no_convexity_assumption}
E^{\bxi}=\min_{\hat{\gamma}^{\bxi}}\Tr\left[\hat{\gamma}^{\bxi} \hat{H}\right], 
\ee
where $\hat{\gamma}^{\bxi}$ is a trial Nc ensemble density matrix operator. Note that, while in the conventional PPLB DFT of fractional electron numbers~\cite{perdew1982density}, the energy is assumed to be convex with respect to the number of electrons, thus allowing for a controlled variation of the latter through adjustments of the chemical potential~\cite{Cernatic2022}, it is completely unnecessary to make such an assumption in the present context of Nc-eDFT, by construction of the theory. As a result, even when the convexity hypothesis breaks down~\cite{Gonzalez26_Counterexamples,Di-Marino24_Ground}, the exact Nc ensemble energy should be recovered from the minimization, so that the exact ionization potentials and electron affinities can be extracted (see Sec.~\ref{sec:extraction_ind_ener_levels_NceDFT}).\\  
   
The density-functionalization of Eq.~(\ref{eq:general_Nc_VP_no_convexity_assumption}) is achieved from Levy's constrained search formalism\cite{levy1979universal}:
 \begin{equation}\label{eq:VP_dens_func_NceDFT}
    \begin{split}
             E^{\bxi}  &= \underset{n}{\rm min} \left\{ \underset{\hat{\gamma}^{\bxi}\rightarrow n}{\rm min}  {\rm Tr} \left[ \hat{\gamma}^{\bxi} \hat{H} \right] \right\} 
             \\
             & = \underset{n}{\rm min} \left\{ F^{\bxi} [n] + \int d {\bf r} \, v_{\rm ext}({\bf r}) \, n({\bf r}) \right\} ,
    \end{split}
 \end{equation}
where the density constraint $\hat{\gamma}^{\bxi}\rightarrow n$ reads more explicitly
\begin{equation} 
    n_{\hat{\gamma}^{\bxi}}({\bf r}) := {\rm Tr} \left[ \hat{\gamma}^{\bxi} \hat{n}({\bf r}) \right] = n({\bf r}),
\end{equation}
and
\be\label{eq:LL_nce_func_separate_def}
F^{\bxi} [n]= \underset{\hat{\gamma}^{\bxi}\rightarrow n}{\rm min}  {\rm Tr} \left[ \hat{\gamma}^{\bxi} (\hat{T}+\hat{W}_{\rm ee}) \right]
\ee
is the analog for Nc ensembles of Levy--Lieb's functional~\cite{levy1979universal,LFTransform-Lieb}. 
In this context, the Hartree-xc (Hxc) density functional is defined from the non-interacting (kinetic energy only) version of $F^{\bxi} [n]$,
\be\label{eq:Nce_TS_func_def}
T_{\rm s}^{\bxi} [n]=\underset{\hat{\gamma}^{\bxi}\rightarrow n}{\rm min}  {\rm Tr} \left[ \hat{\gamma}^{\bxi} \hat{T} \right],
\ee
as follows, by analogy with regular KS theory~\cite{KS,Cernatic2024_Neutral},
\begin{equation}\label{eq:Nc-Hxc_func_def}
    E_{\rm Hxc}^{\bxi} [n]= F^{\bxi} [n] - T_{\rm s}^{\bxi} [n].
\end{equation}
As readily seen, once a set of ensemble weight values $\bxi$ has been chosen, the Nc ensemble Hxc energy can be evaluated as a density functional for any (Nc ensemble-representable) $N$-electron density $n$. In other words, in Nc-eDFT, $n$ and $\bxi$ are {\it independent} variables, like in TGOK-DFT~\cite{gross1988density} but unlike in PPLB-DFT~\cite{perdew1982density}, where the weights are deduced from the fractional number of electrons in the system and, therefore, from the density itself~\cite{Cernatic2022}. This subtle but major difference between Nc and PPLB ensembles has many fundamental and practical implications regarding the extraction of observables from the ensemble, as shown later in the following sections.  

Inserting Eqs.~(\ref{eq:Nce_TS_func_def}) and (\ref{eq:Nc-Hxc_func_def}) into the density-functional variational principle of Eq.~(\ref{eq:VP_dens_func_NceDFT}) leads to the following KS expression of the exact Nc ensemble energy:
\begin{equation} \label{eq:var_principle}
    \begin{split}
            E^{\bxi} & = \underset{n}{\rm min} \bigg\{ \underset{\hat{\gamma}^{\bxi}\rightarrow n}{\rm min} \Big\{ {\rm Tr} \left[ \hat{\gamma}^{\bxi} \hat{T} \right] + E_{\rm Hxc}^{\bxi} [n_{\hat{\gamma}^{\bxi}}] 
            \\
            &  \qquad \qquad + \int d {\bf r} \, v_{\rm ext}({\bf r}) \, n_{\hat{\gamma}^{\bxi}}({\bf r})   \Big\} \bigg\} 
             \\
             & = \underset{\hat{\gamma}^{\bxi}}{\rm min} \left\{  {\rm Tr} \left[ \hat{\gamma}^{\bxi} (\hat{T}+\hat{V}_{\rm ext}) \right] + E_{\rm Hxc}^{\bxi} [n_{\hat{\gamma}^{\bxi}}] \right\}
             .
    \end{split}
\end{equation}
By construction, the minimizing KS Nc ensemble density matrix operator
\be
\hat{\gamma}^{\bxi}_{\rm s}=\sum_{\nu\geq 0}\xi_\nu \ket{\Phi^{\bxi}_\nu}\bra{\Phi^{\bxi}_\nu}
\ee
reproduces the exact Nc ensemble density of the physical system $n^{\bxi}$ (see Eq.~(\ref{eq:nc_dens})), \ie,
\begin{equation} \label{eq:nc_KS_dens}
\begin{split}
   \Tr\left[\hat{\gamma}^{\bxi}_{\rm s}\hat{n}(\br)\right]
   &=
   \left(1-\sum_{\lambda>0}\frac{N_\lambda}{N} \xi_\lambda \right) n_{\Phi^{\bxi}_0} + \sum_{\lambda>0} \xi_\lambda n_{\Phi^{\bxi}_\lambda}
   \\
   &= n^{\bxi}({\bf r}),
\end{split}
\end{equation}
where the fictitious non-interacting wavefunctions are all solutions to the following self-consistent KS-like equation,
\begin{equation} \label{eq:KS_Schrodinger}
    \left[ \hat{T} + \int d {\bf r} \, v^{\bxi}_{\rm s}({\bf r}) \, \hat{n}({\bf r}) \right] \ket{\Phi^{\bxi}_\nu} = \mathcal{E}^{\bxi}_\nu \ket{\Phi^{\bxi}_\nu},\; \forall \nu\geq 0. 
\end{equation}
Like in regular ground-state KS-DFT, the Nc ensemble analog of the KS potential is the sum of the external potential and the Hxc potential:
\begin{equation} \label{eq:ks_pot}
v_{\rm s}^{\bxi}({\bf r})= v_{\rm ext}({\bf r}) + v^{\bxi}_{\rm Hxc}({\bf r}), 
\end{equation}
where
\be\label{eq:Nce_Hxc_pot_def}
v^{\bxi}_{\rm Hxc}({\bf r})=\left.v^{\bxi}_{\rm Hxc}[n]({\bf r})\right|_{n=n^{\bxi}}\equiv\frac{\delta E^{\bxi}_{\rm Hxc}[n]}{\delta n({\bf r})} \Big|_{n=n^{\bxi}}.
\ee
Note that, unlike the external potential and the eigenfunctions of the true physical system, the ensemble Hxc potential is {\it weight-dependent}, so that the KS wavefunctions are weight-dependent too. The in-principle exact extraction of individual excited-state densities (see Sec.~\ref{sec:extraction_ind_densities} for further details) relies on this key feature, as it appears clearly in the zero-weight limit ($\bxi\rightarrow 0$) of the theory~\cite{Cernatic2022}.\\  

The drastic simplification of the original many-electron problem lies in the fact that solving Eq.~(\ref{eq:KS_Schrodinger}) is equivalent to solving the following one-electron ensemble KS equation self-consistently,
\begin{equation}
    \left[-\frac{\nabla^2_{{\bf r}}}{2} +  v^{\bxi}_{\rm s}({\bf r}) \right] \varphi^{\bxi}_{k}({\bf r}) = \varepsilon^{\bxi}_{k} \, \varphi^{\bxi}_{k}({\bf r}).
\end{equation}
If we denote $\theta_{\nu,k}$ the integer occupation of the KS orbital $\varphi^{\bxi}_{k}({\bf r})$ in the KS state $\Phi^{\bxi}_\nu$, then the ensemble density simply reads 
\begin{equation}\label{eq:ens_dens_frac_occ_EKS_orbs}
\begin{split}
        n^{\bxi}({\bf r}) & = \sum_{\nu\geq 0} \xi_\nu \sum_k \theta_{\nu,k} \, \big|\varphi^{\bxi}_{k}({\bf r})\big|^2 
        \\
        & = \sum_{k} \left( \sum_{\nu\geq 0} \xi_\nu \, \theta_{\nu,k} \right) \big|\varphi^{\bxi}_{k}({\bf r})\big|^2,
\end{split}
\end{equation}
where the second line shows how weight-dependent fractional KS occupation numbers emerge at the ensemble KS (eKS) level of theory. The total individual-state KS energies are analogously determined from the eKS orbital energies as follows, 
\begin{equation}
    \mathcal{E}^{\bxi}_\nu = \sum_k \theta_{\nu,k} \, \varepsilon^{\bxi}_{k},\; \nu\geq 0.
\end{equation}

Even though, for clarity, we postpone to Sec.~\ref{sec:mrpt_eDFT} the discussion on how to separate the total ensemble Hxc energy into Hartree, exchange, and correlation contributions, which is a nontrivial problem in eDFT~\cite{gould2026ensemblization}, it is still instructive to define the Hx part of the energy, from which the definition of the ensemble correlation energy follows. For that purpose, a clarification is needed. As shown by Gould and Pittalis~\cite{gould2017hartree}, the KS ensemble should in fact be constructed in the non-interacting {\it limit} of eDFT, \ie, for electrons interacting infinitesimally, so that complications arising from the degeneracies in the strictly non-interacting case can be avoided. Consequently, the Nc ensemble density-functional KS system is properly defined as follows, 
\be\label{eq:KS_ens_non-int-limit}
\begin{split}
\hat{\gamma}_{\rm s}^{\bxi}[n]&\underset{\alpha\rightarrow 0^+}{=}\underset{\hat{\gamma}^{\bxi}\rightarrow n}{\argmin}  {\rm Tr} \left[ \hat{\gamma}^{\bxi} (\hat{T}+\alpha\hat{W}_{\rm ee}) \right]
\\
&\quad\equiv \sum_{\nu\geq 0} \xi_\nu\ket{\Phi^{\bxi}_\nu[n]}\bra{\Phi^{\bxi}_\nu[n]}
.
\end{split}
\ee
It can then be used to evaluate the total Hx ensemble energy functional, whose expression below is usually referred to as ensemble exact exchange (EEXX) functional~\cite{gould2026ensemblization,Cernatic2022}:
\be\label{eq:_standard_def_ens_Hx}
E_{\rm Hx}^{\bxi} [n]=\Tr\left[\hat{\gamma}_{\rm s}^{\bxi}[n]\hat{W}_{\rm ee}\right].
\ee
Note that the above definition guarantees that the ensemble correlation energy is strictly negative, according to the variational principle of Eq.~(\ref{eq:general_Nc_VP_no_convexity_assumption}):
\be
\begin{split}
E_{\rm c}^{\bxi} [n]
&=F^{\bxi}[n]-T_{\rm s}^{\bxi}[n]-E_{\rm Hx}^{\bxi}[n]
\\
&=\Tr\left[\left(\hat{\Gamma}^{\bxi}[n]-\hat{\gamma}_{\rm s}^{\bxi}[n]\right)\left(\hat{T}+\hat{W}_{\rm ee}\right)\right]
\\
&=\Tr\left[\left(\hat{\Gamma}^{\bxi}[n]-\hat{\gamma}_{\rm s}^{\bxi}[n]\right)\left(\hat{T}+\hat{W}_{\rm ee}+\hat{V}^{\bxi}[n]\right)\right]
\\
&<0,
\end{split}
\ee
where $\hat{V}^{\bxi}[n]=\int d\br\,v^{\bxi}[n](\br)\hat{n}(\br)$ is the local potential operator that generates, through the corresponding solutions to the Schr\"{o}dinger equation, the interacting ensemble density matrix operator with density $n$,
\be\label{eq:int_ens_dens_func_dens_matrix_op_exp}
\hat{\Gamma}^{\bxi}[n]
=\sum_{\nu\geq 0}\xi_\nu\ket{\Psi_\nu^{\bxi}[n]}\bra{\Psi_\nu^{\bxi}[n]},
\ee
\ie, the minimizer of Eq.~(\ref{eq:LL_nce_func_separate_def}).

\subsection{Extraction of ground and excited energy levels}\label{sec:extraction_ind_ener_levels_NceDFT}

While the Nc ensemble construction introduced in Eq.~(\ref{eq:densop}) has nothing physical, its {\it linear} variation with the ensemble weights makes it appealing as it allows for the extraction of physical properties. For example, any (ground or excited) energy level included into the ensemble can be computed, in principle exactly, following a {\it single} eKS-DFT calculation. This was only recently realized\cite{deur2019ground}, as the first envisioned route to individual energy levels involved combining results from ensembles with different sets of weights\cite{gross1988density}, each being determined by a separate self-consistent eKS calculation. The two central ideas leading to a single-calculation extraction of energies are the above-mentioned linear variation in weights of the ensemble energy (see Eq.~\eqref{eq:Nc_ens_E}), from which the exact expression below follows,
\begin{equation}\label{eq:general_extraction_ind_ener_from_Nce_ener}
\begin{split}
    \qquad  E_\nu 
    &=\left(E_\nu-\frac{N_\nu}{N}E_0\right)+\frac{N_\nu}{N}E_0 
    \\
    &=\sum_{\lambda>0}\delta_{\lambda\nu}\frac{\partial E^{\bxi}}{\partial \xi_{\lambda}}+\frac{N_\nu}{N}\left(E^{\bxi}-\sum_{\lambda>0}\xi_{\lambda}\frac{\partial E^{\bxi}}{\partial \xi_{\lambda}}\right)
    \\
    &= \frac{N_\nu}{N} E^{\bxi} + \sum_{\lambda>0} \left(\delta_{\lambda\nu}-\frac{N_\nu}{N} \xi_{\lambda}\right) \frac{\partial E^{\bxi}}{\partial \xi_{\lambda}}, \, \forall \nu\geq 0,
\end{split}
\end{equation}
and the expression of the to-be-differentiated ensemble energy in terms of the KS energies (see Eqs.~\eqref{eq:var_principle} and \eqref{eq:KS_Schrodinger}),
\begin{equation}\label{eq:Nce_ener_from_KS_ener}
    E^{\bxi} = \sum_{\nu \geq 0} \xi_{\nu} \, \mathcal{E}^{\bxi}_{\nu} + E_{\rm Hxc}^{\bxi}[n^{\bxi}] - \big( v_{\rm Hxc}^{\bxi} \big| n^{\bxi} \big),
\end{equation}
with 
\begin{equation}
    \big( v_{\rm Hxc}^{\bxi} \big| n^{\bxi} \big) = \int d {\bf r} \, v_{\rm Hxc}^{\bxi}({\bf r}) \, n^{\bxi}({\bf r}) \,.
\end{equation}

By applying the Hellmann--Feynmann theorem to the KS energies (see Eqs.~\eqref{eq:KS_Schrodinger} and \eqref{eq:ks_pot}) we get
\be
\begin{split}
\sum_{\nu \geq 0} \xi_{\nu}\dfrac{\partial \mathcal{E}^{\bxi}_{\nu}}{\partial \xi_{\lambda}}
&=\sum_{\nu \geq 0} \xi_{\nu}\left.\dfrac{\partial\left(v_{\rm Hxc}^{\bxi} \middle\vert n\right)}{\partial \xi_{\lambda}}\right|_{n=n_{\Phi_\nu^{\bxi}}}
\\
&=\left.\dfrac{\partial\left(v_{\rm Hxc}^{\bxi} \middle\vert n\right)}{\partial \xi_{\lambda}}\right|_{n=n^{\bxi}}
,
\end{split}
\ee
and since, according to Eq.~\eqref{eq:Nce_Hxc_pot_def}, $\left.\partial E_{\rm Hxc}^{\bm \eta}[n^{\bxi}]/\partial \xi_{\lambda}\right|_{{\bm \eta}=\bxi}=(v_{\rm Hxc}^{\bxi}\vert \partial n^{\bxi}/\partial \xi_{\lambda})$, we have
\begin{equation}
\frac{\partial E^{\bxi}}{\partial \xi_{\lambda}} 
    = \mathcal{E}^{\bxi}_{\lambda} - \frac{N_{\lambda}}{N} \mathcal{E}^{\bxi}_{0} + \frac{\partial E^{\bxi}_{\rm Hxc}[n]}{\partial \xi_{\lambda}} \Big|_{n=n^{\bxi}},
\end{equation}
so that we finally recover from Eq.~(\ref{eq:general_extraction_ind_ener_from_Nce_ener}) the key relation
\begin{equation} \label{eq:ind_e_nc}
\begin{split}
        E_\nu & \underset{\nu\geq 0}{=} \mathcal{E}^{\bxi}_{\nu} + \frac{N_\nu}{N} \left( E_{\rm Hxc}^{\bxi}[n^{\bxi}] - \big( v_{\rm Hxc}^{\bxi} \big| n^{\bxi} \big) \right)
        \\
        & \quad + \sum_{\lambda>0} \left(\delta_{\lambda\nu}-\frac{N_\nu}{N} \xi_{\lambda}\right) \frac{\partial E_{\rm Hxc}^{\bxi}[n]}{\partial \xi_{\lambda}}\Big|_{n=n^{\bxi}}.
\end{split}
\end{equation}
This expression of individual energy levels in terms of their KS counterparts, which was originally derived in Ref.~\citenum{Cernatic2024_Neutral}, readily confirms that once an eKS calculation has been performed (for one, in principle arbitrary, choice of weight values $\bxi$), thus yielding the ensemble density and the KS energies, all energy levels included into the ensemble can be computed, in principle exactly, from the knowledge of the Nc ensemble Hxc functional and its (density-functional and weight) derivatives. In the exact theory, the computed energy levels should obviously not depend on the choice of weights. While the use of weights as empirical parameters in practical (approximate) eDFT calculations is not desirable, because it is expected to be functional- and system-dependent~\cite{senjean2015linear}, using the invariance of computed physical quantities under weight variations as a constraint for the development of eDFAs seems like a more promising path to follow, as illustrated later in Sec.~\ref{sec:scaling_gs_DFAs}.\\

From Eq.~\eqref{eq:ind_e_nc} one can relate straightforwardly any energy gap between two physical states (belonging to the ensemble) to that of their KS analogs as follows,
\begin{equation} \label{eq:e_gap_nc}
\begin{split}
        &E_\kappa - E_\nu  = \mathcal{E}^{\bxi}_{\kappa} - \mathcal{E}^{\bxi}_{\nu} 
        \\
        &+\frac{N_\kappa-N_\nu}{N} \left( E_{\rm Hxc}^{\bxi}[n^{\bxi}] - \big( v_{\rm Hxc}^{\bxi} \big| n^{\bxi} \big) \right)
        \\
        &+ \sum_{\lambda>0} \left(\delta_{\lambda\kappa}-\delta_{\lambda\nu}-\frac{N_\kappa-N_\nu}{N} \xi_{\lambda}\right) \frac{\partial E_{\rm Hxc}^{\bxi}[n]}{\partial \xi_{\lambda}}\Big|_{n=n^{\bxi}},
\end{split}
\end{equation}
where the second term on the right-hand side of Eq.~\eqref{eq:e_gap_nc}, which does not exist in regular TGOK-DFT of neutral excitations~\cite{gross1988density} ($N_\kappa=N_\nu=N$ in this case), is a reminiscence of Levy--Zahariev's (LZ) shift in potential\cite{kraisler2013piecewise,levy2014ground} that naturally extends to Nc ensembles, as readily seen from Eq.~\eqref{eq:ind_e_nc}, as follows,
\begin{equation}
\begin{split}
        &v_{\rm Hxc}^{\bxi}[n]({\bf r})  
          \\
        &\rightarrow  v_{\rm Hxc}^{\bxi}[n]({\bf r})
        + \frac{\displaystyle E_{\rm Hxc}^{\bxi}[n]-( v_{\rm Hxc}^{\bxi}[n] \vert n)}{\displaystyle \int d{\bf r}\,n({\bf r})} 
         =:\bar{v}_{\rm Hxc}^{\bxi}[n]({\bf r}).
\end{split}
\end{equation}
Note that, like in the regular $N$-electron ground-state case~\cite{levy2014ground}, the LZ-shifted Hxc potential can be interpreted as the local (Nc ensemble here) Hxc energy per electron:  
\begin{equation} \label{eq:LZ_intrel_EV}
    E_{\rm Hxc}^{\bxi}[n] = \int d{\bf r} \, \bar{v}_{\rm Hxc}^{\bxi}[n]({\bf r}) \, n({\bf r}).
\end{equation}

At this point one should realize that, unlike in PPLB DFT, shifting the Nc ensemble Hxc potential by {\it any} constant $c$ has no impact on the Nc KS ensemble and, therefore, on the ensemble density (which always integrates to $N$, by construction), because the solutions to the eKS Eq.~(\ref{eq:KS_Schrodinger}) are constructed for each integer number of electrons, separately. Interestingly, once the LZ shift has been applied to the Hxc potential, thus shifting the KS energies as follows,
\begin{equation}
    \begin{split}
    \mathcal{E}^{\bxi}_{\nu} \,\rightarrow \, \bar{\mathcal{E}}^{\bxi}_{\nu} & = \mathcal{E}^{\bxi}_{\nu} 
    \\
    + & \frac{N_{\nu}}{N} \left(E_{\rm Hxc}^{\bxi}[n^{\bxi}] - \big( v_{\rm Hxc}^{\bxi} \big| n^{\bxi} \big)\right) \,,
    \end{split}
\end{equation}
the latter become invariant under any constant shift $c$ in the original Hxc potential, since
\begin{equation}
    \begin{split}
            &\mathcal{E}^{\bxi}_{\nu}- \frac{N_\nu}{N}  \int d {\bf r} \, v_{\rm Hxc}^{\bxi}({\bf r}) \, n^{\bxi}({\bf r}) 
            \\
            &=\big(\mathcal{E}^{\bxi}_{\nu} +N_{\nu} \, c \big) - \frac{N_\nu}{N}  \int d {\bf r} \, \big(v_{\rm Hxc}^{\bxi}({\bf r})+c \big) \, n^{\bxi}({\bf r}) \,.
    \end{split}
\end{equation}
As both the ensemble energy and the energy levels can be expressed in terms of the LZ-shifted KS energies respectively as follows, according to Eqs.~\eqref{eq:net_number_electrons_preserved}, \eqref{eq:Nce_ener_from_KS_ener} and \eqref{eq:ind_e_nc}, 
\begin{equation}
    E^{\bxi} = \sum_{\nu \geq 0} \bar{\mathcal{E}}^{\bxi}_{\nu},
\end{equation}
and
\begin{equation} \label{eq:ind_eb_nc}
        \forall \nu, \quad  E_\nu  = \bar{\mathcal{E}}^{\bxi}_{\nu}  + \sum_{\lambda>0} \left(\delta_{\lambda\nu}-\frac{N_\nu}{N} \xi_{\lambda}\right) \frac{\partial E_{\rm Hxc}^{\bxi}[n]}{\partial \xi_{\lambda}}\Big|_{n=n^{\bxi}},
\end{equation}
they are also invariant under constant shifts in the Nc ensemble Hxc potential, and so is {\it any} energy gap
\begin{equation} \label{eq:eb_gap_nc}
\begin{split}
        & E_\kappa - E_\nu  = \bar{\mathcal{E}}^{\bxi}_{\kappa} - \bar{\mathcal{E}}^{\bxi}_{\nu}
        \\
        &+ \sum_{\lambda>0} \left(\delta_{\lambda\kappa}-\delta_{\lambda\nu}-\frac{N_\kappa-N_\nu}{N} \xi_{\lambda}\right) \frac{\partial E_{\rm Hxc}^{\bxi}[n]}{\partial \xi_{\lambda}}\Big|_{n=n^{\bxi}}
        .
        \end{split}
\end{equation}
The exact Nc-eDFT expressions for energy levels in Eqs.~\eqref{eq:ind_e_nc} and~\eqref{eq:ind_eb_nc} apply to ground and excited states from any $N\pm p$-electron sector of the Fock space that are included into the ensemble. To be more precise, if a given $\mathcal{N}$-electron excited state is targeted, all the $\mathcal{N}$-electron states below in energy (down to the ground state) should be part of the ensemble, such that the variational principle of Eq.~\eqref{eq:general_Nc_VP_no_convexity_assumption} applies and eDFT can be formulated. Thus, the energy gap expressions of Eqs.~\eqref{eq:e_gap_nc} and~\eqref{eq:eb_gap_nc} provide a unified framework to compute and analyze both charged {\it and} neutral excitation energies. Interestingly, contributions of neutral double excitations, which are missed by linear response TDDFT under the adiabatic approximation~\cite{maitra2004double,cave2004dressed,Huix-Rotllant2011_Assessment,elliott2011perspectives,maitra2022double,Casida_tddft_review_2012,Lacombe2023_Non-adiabatic}, can be included explicitly in the KS formulation of Nc-eDFT~\cite{cernatic2024extended_doubles}. However, in this matter, it is important to keep in mind that the excitation processes occurring in the fictitious non-interacting eKS system might be (very) different from those occurring in the true interacting physical system, even though both share the exact same ensemble density, as show in Ref.~\citenum{cernatic2024extended_doubles} (see also Sec.~\ref{sec:Koopmans_exactification}).

\subsection{Systematic exactification of Koopmans' theorem for ground and excited states} \label{sec:Koopmans_exactification}

In order to reconnect with more traditional density-functional approaches to electronic excitations, such as PPLB DFT, we can alternatively decide not to rely on the gauge-invariant (``gauge'' referring here to constant shifts in the Hxc potential) LZ-shifted KS energies and, instead, exploit fully the flexibility of the Nc ensemble formalism (to simplify energy gap expressions, for example) through appropriate choices of gauge.  
As Eq.~\eqref{eq:e_gap_nc} applies seamlessly to any couple of states (ground or excited,  with the same number of electrons or not), it allows to address how shifts in the Hxc potential can match KS energy differences with that of the true system, thus exactifying Koopmans' theorem\cite{perdew1983physical} in all generality\cite{Cernatic2024_Neutral,cernatic2024extended_doubles}. If, for example, we target a specific charged transition $\nu\rightarrow\kappa$ ($N_\kappa\neq N_\nu$) and choose the Hxc potential, through the appropriate constant shift $c^{\bxi [\nu\rightarrow\kappa]}$, such that
\begin{equation} \label{eq:shift_IP_Koopmans}
\begin{split}
&\int d {\bf r} \, \left(v_{\rm Hxc}^{\bxi}({\bf r})+c^{\bxi [\nu\rightarrow\kappa]}\right) \, n^{\bxi}({\bf r})
\\
&=\int d {\bf r} \, v_{\rm Hxc}^{\bxi[\nu\rightarrow\kappa]}({\bf r}) \, n^{\bxi}({\bf r}) 
\\
&= \mathscr{D}^{\bxi [\nu\rightarrow\kappa]}_{\rm Hxc}[n^{\bxi}],
\end{split}
\end{equation}
or, equivalently,
\be\label{eq:expression_constant_shift_Koopmans}
c^{\bxi [\nu\rightarrow\kappa]}:=\dfrac{\mathscr{D}^{\bxi [\nu\rightarrow\kappa]}_{\rm Hxc}[n^{\bxi}]-\big( v_{\rm Hxc}^{\bxi} \big| n^{\bxi} \big)}{N},
\ee
where
\begin{equation} \label{eq:d_IP_Koopmans}
    \begin{split}
            &\mathscr{D}^{\bxi [\nu\rightarrow\kappa]}_{\rm Hxc}[n^{\bxi}] =   E_{\rm Hxc}^{\bxi}[n^{\bxi}] 
            \\
            & + \sum_{\lambda>0} \left(\dfrac{N}{N_\kappa-N_\nu}\left(\delta_{\lambda\kappa}-\delta_{\lambda\nu}\right)-{\xi_{\lambda}} \right) \frac{\partial E_{\rm Hxc}^{\bxi}[n]}{\partial \xi_{\lambda}}\Big|_{n=n^{\bxi}},
    \end{split}
\end{equation}
then Koopmans' theorem does apply, \ie,
\begin{equation} \label{eq:IP_Koopmans_nc}
        \begin{split}
                 E_\kappa - E_\nu  & = \mathcal{E}^{\bxi[\nu\rightarrow\kappa]}_{\kappa} - \mathcal{E}^{\bxi[\nu\rightarrow\kappa]}_{\nu}
                 \\
                 & = \sum_j \big( \theta_{\kappa,j} - \theta_{\nu,j} \big) \, \varepsilon^{\bxi[\nu\rightarrow\kappa]}_{j},
        \end{split}
\end{equation}
where $\mathcal{E}^{\bxi[\nu\rightarrow\kappa]}_{\nu,\kappa}$ and $\varepsilon^{\bxi[\nu\rightarrow\kappa]}_{j}$ are, respectively, the KS total and orbital energies computed with the adjusted Hxc potential $v_{\rm Hxc}^{\bxi[\nu\rightarrow\kappa]}$. In this context, Koopmans' theorem is an arbitrary choice more than a theorem, the actual theorem being the exact relation of Eq.~(\ref{eq:eb_gap_nc}). 
If the transition involves, in the fictitious KS ensemble, the ground or singly-excited (from the $N$-electron highest occupied molecular orbital (HOMO) $\varphi_N^{\bxi}$ to a virtual one $\varphi^{\bxi}_{i(\nu)>N}$, where $\nu>0$) $N$-electron state $\nu\geq 0$ on the one hand and, on the other hand, the singly-ionized (from $\varphi^{\bxi}_{i(\nu)\geq N}$) ground state $\kappa$, then we immediately recover from Eq.~(\ref{eq:IP_Koopmans_nc}) an exact and familiar ionization potential (IP) theorem that holds for both ground and excited $N$-electron states:       
\begin{equation}\label{eq:exact_IP_nu_from_NcEDFT}
     I_{\nu}^N\equiv E_0^{N-1}-E_{\nu}^N=- \varepsilon_{i(\nu)}^{\bxi[\nu\rightarrow\kappa]}.
\end{equation}
Let us emphasize that Eq.~(\ref{eq:IP_Koopmans_nc}) is more general that the above relation, as it can treat states with multiple-excitation character without any additional difficulty, at least in principle~\cite{cernatic2024extended_doubles}.\\

Turning to neutral transitions between $N$-electron states, the trick of Eq.~(\ref{eq:expression_constant_shift_Koopmans}) does not apply anymore, as the second term on the right-hand side of Eq.~(\ref{eq:e_gap_nc}) cancels out in this case ($N_\kappa=N_\nu=N$). It is nonetheless possible to achieve an exactification of Koopmans' theorem also in this case, as was shown recently~\cite{Cernatic2024_Neutral} using a very elegant idea pioneered by Levy\cite{levy1995excitation}. The neutral excitation of interest is to be decomposed as two {\it separate ionization} processes:
\begin{equation} \label{eq:decomp_io}
    E_{\nu}^N - E_{0}^N = I_{0}^N - I_{\nu}^N,
\end{equation}
with the ground- and excited-state IPs defined as
\begin{equation}
    \left\{ I_{\nu}^N = E_{0}^{N-1} - E_{\nu}^N \right\}_{\nu\geq 0}.
\end{equation}
Using specific shifts defined according to Eq.~(\ref{eq:expression_constant_shift_Koopmans}), we can enforce the match of KS and physical IPs separately, \ie, 
\begin{equation}\label{eq:IP_separate_exactification}
    \begin{split}
        I_{0}^N & = \mathcal{E}^{\bxi[0\rightarrow\kappa]}_{\kappa} - \mathcal{E}^{\bxi[0\rightarrow\kappa]}_{0},
        \\
        I_{\nu}^N & = \mathcal{E}^{\bxi[\nu\rightarrow\kappa]}_{\kappa} - \mathcal{E}^{\bxi[\nu\rightarrow\kappa]}_{\nu},\; \nu>0,
    \end{split}
\end{equation}
where $\kappa$ is the intermediate $(N-1)$-electron ground state, so that  
\begin{equation}
    \begin{split}
        &E_{\nu}^N - E_{0}^N 
        \\& = \left(\mathcal{E}^{\bxi[0\rightarrow\kappa]}_{\kappa} - \mathcal{E}^{\bxi[0\rightarrow\kappa]}_{0}\right) - \left(\mathcal{E}^{\bxi[\nu\rightarrow\kappa]}_{\kappa} - \mathcal{E}^{\bxi[\nu\rightarrow\kappa]}_{\nu}\right)
\\
&=\mathcal{E}^{\bxi[\nu\rightarrow\kappa]}_{\nu}-\mathcal{E}^{\bxi[0\rightarrow\kappa]}_{0}+\left(\mathcal{E}^{\bxi[0\rightarrow\kappa]}_{\kappa}-\mathcal{E}^{\bxi[\nu\rightarrow\kappa]}_{\kappa}\right),
      \end{split}
\end{equation}
or, equivalently, if we want to evaluate all KS energies from the ground-state ionization adjustment $[0\rightarrow\kappa]$, like in regular KS-DFT, 
\begin{equation}\label{eq:general_neutral_XE_shift_of_one_orb_ener}
    \begin{split}
        &E_{\nu}^N - E_{0}^N 
          \\
&=\left(\mathcal{E}^{\bxi[0\rightarrow\kappa]}_{\nu}+c^{\bxi [\nu\rightarrow\kappa]}-c^{\bxi [0\rightarrow\kappa]}\right)-\mathcal{E}^{\bxi[0\rightarrow\kappa]}_{0} 
        \\
        & = \sum_j \big( \theta_{\nu,j}- \theta_{0,j}\big) \, \varepsilon^{\bxi[0\rightarrow\kappa]}_{j} 
        +c^{\bxi [\nu\rightarrow\kappa]}-c^{\bxi [0\rightarrow\kappa]},
    \end{split}
\end{equation}
thus giving the desired Koopmans' theorem and generalizing previous works~\cite{levy1995excitation,gould2022single} to {\it any} type of neutral excitation, without ever exploiting or referring to the asymptotic behavior of the ensemble density. We refer the reader to Refs.~\citenum{Cernatic2022} and~\citenum{PRA21_Hodgson_exact_Nc-eDFT_1D} for a comprehensive comparison of standard proofs with that of Nc-eDFT. In the particular case where, in the eKS picture, the neutral excitation process consists of a single excitation from orbital $i$ to orbital $i(\nu)$ [we recall that the physical process might be more complex or very different, as shown in Ref.~\citenum{cernatic2024extended_doubles}], we recover the expressions of Refs.~\citenum{levy1995excitation} and~\citenum{gould2022single}:
\begin{equation}\label{eq:exact_single_neutral_from_KS}
\begin{split}
        E_{\nu}^N - E_{0}^N  
        &=\varepsilon^{\bxi[0\rightarrow\kappa]}_{i(\nu)}-\varepsilon^{\bxi[0\rightarrow\kappa]}_{i}+c^{\bxi [\nu\rightarrow\kappa]}-c^{\bxi [0\rightarrow\kappa]}
        \\
        &= \varepsilon^{\bxi[\nu\rightarrow\kappa]}_{i(\nu)} -\varepsilon^{\bxi[0\rightarrow\kappa]}_{i},
\end{split}
\end{equation}
where, as readily seen, the two orbital energies must be evaluated with different shifts, otherwise the exact excitation energy cannot be reproduced, as one would expect.\\ 

As a final comment, let us emphasize that the introduction in Eq.~(\ref{eq:decomp_io}) of the intermediate singly-ionized ground state is an arbitrary choice that preserves the exactness of the theory even if the true system undergoes multiple electronic excitations. In other words, Eq.~(\ref{eq:general_neutral_XE_shift_of_one_orb_ener}) remains formally exact also in this case. This has been illustrated in Ref.~\citenum{cernatic2024extended_doubles} in the case of double excitations. Obviously, if the KS excited state $\nu$ is described by the double $(i,j)\rightarrow \big( i(\nu)>N , \, j(\nu)>N\big)$ excitation, we may alternatively replace the state $\kappa$ in Eq.~(\ref{eq:IP_separate_exactification}) by the doubly-ionized ground state (that we denote $\kappa_{\rm d}$), for example, thus leading to
\begin{equation}
    \begin{split}
        &E_{\nu}^N - E_{0}^N  
        \\
        & = 
        \varepsilon^{\bxi[0\rightarrow\kappa_{\rm d}]}_{i(\nu)} + \varepsilon^{\bxi[0\rightarrow\kappa_{\rm d}]}_{j(\nu)}
        -\varepsilon^{\bxi[0\rightarrow\kappa_{\rm d}]}_{i} - \varepsilon^{\bxi[0\rightarrow\kappa_{\rm d}]}_{j}
        \\
        &\quad +2 \left(c^{\bxi [\nu\rightarrow\kappa_{\rm d}]}-c^{\bxi [0\rightarrow\kappa_{\rm d}]}\right)
        \\
        & = 
        \varepsilon^{\bxi[\nu\rightarrow\kappa_{\rm d}]}_{i(\nu)} + \varepsilon^{\bxi[\nu\rightarrow\kappa_{\rm d}]}_{j(\nu)}
        -\varepsilon^{\bxi[0\rightarrow\kappa_{\rm d}]}_{i} - \varepsilon^{\bxi[0\rightarrow\kappa_{\rm d}]}_{j},
        \end{split}
\end{equation}
which is the exact analog of Eq.~(\ref{eq:exact_single_neutral_from_KS}) for double excitations.\\

For clarity, let us mention one more time that eDFT does not enforce a state-by-state match between interacting and fictitious non-interacting KS states in any fashion. Thus, the character of a KS excitation is {\it not} always informative of that of the physical excitation, especially so in the strongly-correlated regime where the one-particle picture of electronic excitations breaks down~\cite{cernatic2024extended_doubles}. In spite of this loss of information, inherent to any DFT, the ability of eDFT to treat systems where multiple excitations play a role is genuine~\cite{gould2026ensemblization}, simply because, unlike in linear response TDDFT~\cite{Casida_tddft_review_2012}, these excitations can be included explicitly into the theory (at the KS level of description).

\subsection{Equivalence between Hxc density-functional derivative discontinuities and ensemble weight derivatives} \label{sec:dd_vs_wd}

Within conventional (PPLB) DFT, the correct prediction of fundamental gaps crucially depends on modeling derivative discontinuities of the Hxc density functional, \ie,  discontinuities that occur in the Hxc potential when crossing an integer electron number~\cite{perdew1983physical,perdew2017understanding}. This relates to the fact that, unlike in Nc-eDFT, the Hxc functional is exclusively a functional of the density (there is no weight dependence in that case~\cite{Cernatic2022}), which can integrate to a fractional electron number. The resulting Janak's theorem~\cite{janak1978proof,perdew1983physical} fixes uniquely (not up to a constant shift anymore) the Hxc potential between two successive integer numbers of electrons, thus leading to the above-mentioned discontinuities.\\ 

Even though we have been dealing previously with charged excitations, we have until now never mentioned explicitly the concept of derivative discontinuity in the context of Nc-eDFT. To be more precise, it was not necessary to pay attention to these subtleties to proceed with the derivation of charged excitation energies, which is quite convenient in practice. Nevertheless, such discontinuities occur also within Nc-eDFT~\cite{PRA21_Hodgson_exact_Nc-eDFT_1D,cernatic2024extended_doubles,Cernatic2024_Neutral}, in direct connection with the exactification of Koopmans' theorem (see Sec.~\ref{sec:Koopmans_exactification}). Indeed, as the latter uniquely fixes the Nc ensemble Hxc potential for a given charged transition $[\kappa\rightarrow\nu]$ (or, equivalently, $[\nu\rightarrow\kappa]$), according to Eqs.~\eqref{eq:shift_IP_Koopmans}-\eqref{eq:d_IP_Koopmans}, considering another charged transition $[\kappa'\rightarrow\nu]$ will automatically lead to another adjustement of the Hxc potential. Still, in this construction, both differ only by a constant shift from the Nc ensemble Hxc potential $v^{\bxi}_{\rm Hxc}$, which is unique up to a constant and ensures that the exact ensemble density $n^{\bxi}$ is reproduced by the KS ensemble. The difference can be expressed explicitly as follows,    
\begin{equation} \label{eq:vHXC_dxiE}
    \begin{split}
        &\int d {\bf r} \,\Big( v_{\rm Hxc}^{\bxi[\kappa\rightarrow\nu]}({\bf r})  - v_{\rm Hxc}^{\bxi[\kappa'\rightarrow\nu]}({\bf r}) \Big) \, \frac{n^{\bxi}({\bf r})}{N} 
        \\
        &=c^{\bxi[\kappa\rightarrow\nu]}-c^{\bxi[\kappa'\rightarrow\nu]}
        \\
        & = \sum_{\lambda>0} \left( \frac{\delta_{\lambda\kappa}-\delta_{\lambda\nu}}{N_{\kappa}-N_{\nu}} - \frac{\delta_{\lambda\kappa'}-\delta_{\lambda\nu}}{N_{\kappa'}-N_{\nu}} \right) \frac{\partial E_{\rm Hxc}^{\bxi}[n]}{\partial \xi_{\lambda}}\Big|_{n=n^{\bxi}},
    \end{split}
\end{equation}
thus suggesting a connection between derivative discontinuities and ensemble weight derivatives. Strictly speaking, what the above relation shows is not a discontinuity (with respect to some continuous variable that should be identified) but instead an adjustment of shift within a given Nc KS ensemble, when switching from one transition to another.\\  


A clearer connection with the concept of derivative discontinuities comes from the observation that, while the right-hand side of Eq.~(\ref{eq:vHXC_dxiE}) is evaluated for any variationally-valid set of weights including states $\Psi_\nu,\Psi_{\kappa}$, and $\Psi_{\kappa'}$, the transition-specific shifts on the left-hand side can be addressed in separate smaller ensembles. For clarity, we first consider the fundamental gap problem and follow the reasoning of Refs.~\citenum{PRA21_Hodgson_exact_Nc-eDFT_1D} and \citenum{Cernatic2022}. In this case, we only need to include into the ensemble the $N$-electron $\Psi_\nu =\Psi_0$, $(N+1)$-electron $\Psi_\kappa =\Psi_0^{N+1}$ (with weight $\xi_+$), and  $(N-1)$-electron $\Psi_{\kappa'} =\Psi_0^{N-1}$ (with weight $\xi_-$) ground states, so that Eq.~(\ref{eq:vHXC_dxiE}) now reads 
\begin{equation}\label{eq:compar_pots_fund_gap_pb} 
    \begin{split}
        \int d {\bf r} \, & \Big( v_{\rm Hxc}^{\bxi[\kappa\rightarrow\nu]}({\bf r})  - v_{\rm Hxc}^{\bxi[\kappa'\rightarrow\nu]}({\bf r}) \Big) \, \frac{n^{\bxi}({\bf r})}{N} 
        \\
        & =  \frac{\partial E_{\rm Hxc}^{\bxi}[n]}{\partial \xi_{+}}\Big|_{n=n^{\bxi}} + \frac{\partial E_{\rm Hxc}^{\bxi}[n]}{\partial \xi_{-}}\Big|_{n=n^{\bxi}}
    \end{split}
\end{equation}
for any weight values $\bxi\equiv\left\{\xi_+,\xi_-\right\}$. We recall that the above relation is a necessary condition for having the eKS system reproducing the fundamental gap exactly, \ie,  
\begin{equation} \label{eq:f_gap}
    I_0^{N}-I_0^{N+1} = \varepsilon^{\bxi[\kappa\rightarrow\nu]}_{N+1} -\varepsilon^{\bxi[\kappa'\rightarrow\nu]}_{N} \,.
\end{equation}
If we now take the zero-weight limit of Eq.~(\ref{eq:compar_pots_fund_gap_pb}), the difference in Hxc potential survives through the ensemble weight derivatives, evaluated at $\bxi=0$:
\begin{equation}  \label{eq:disc_shift_fund_0weights}
    \begin{split}
        \int d {\bf r} \, & \Big( v_{\rm Hxc}^{\bxi[\kappa\rightarrow\nu]}({\bf r})  - v_{\rm Hxc}^{\bxi[\kappa'\rightarrow\nu]}({\bf r}) \Big) \, \frac{n_{\Psi_{0}}({\bf r})}{N} 
        \\
        & \underset{\bxi\rightarrow 0}{=}  \left[\frac{\partial E_{\rm Hxc}^{\bxi}[n_{\Psi_{0}]}}{\partial \xi_{+}} + \frac{\partial E_{\rm Hxc}^{\bxi}[n_{\Psi_{0}}]}{\partial \xi_{-}}\right]_{\bxi=0}.
    \end{split}
\end{equation}
In fact, the $N$-electron ground-state limit can be reached in different ways. We can either consider the smaller ensemble $\left\{\Psi_0,\Psi_0^{N-1}\right\}$ for which $\xi_+=0$ and, in this case, the exactification of the IP theorem implies, in the $\xi_-\rightarrow 0$ limit (see Eqs.~\eqref{eq:shift_IP_Koopmans} and \eqref{eq:d_IP_Koopmans}),
\be\label{eq:fund_gap_pb_Hxc_pot_exact_IPN}
\begin{split}
&\int d\br\, v^{\xi_+=0}_{\rm Hxc}(\br)n_{\Psi_0}(\br)
\\
&=E_{\rm Hxc}[n_{\Psi_0}]
-N\left.\frac{\partial E_{\rm Hxc}^{\bxi}[n_{\Psi_{0}]}}{\partial \xi_{-}}\right|_{\bxi=0},
\end{split}
\ee
where $E_{\rm Hxc}[n]=E^{\bxi=0}_{\rm Hxc}[n]$ is the regular ground-state Hxc functional, or we consider the other smaller ensemble $\left\{\Psi_0,\Psi_0^{N+1}\right\}$ (for which $\xi_-=0$) whose corresponding IP theorem implies that, in the $\xi_+\rightarrow 0$ limit,
\be
\begin{split}
&\int d\br\, v^{\xi_+\rightarrow 0}_{\rm Hxc}(\br)n_{\Psi_0}(\br)
\\
&=E_{\rm Hxc}[n_{\Psi_0}]
+N\left.\frac{\partial E_{\rm Hxc}^{\bxi}[n_{\Psi_{0}]}}{\partial \xi_{+}}\right|_{\bxi=0}.
\end{split}
\ee
By taking the difference of the two expressions, we obtain a relation analogous to that of Eq.~(\ref{eq:disc_shift_fund_0weights}),
\begin{equation} 
    \begin{split}
        \int d {\bf r} \, & \Big(v^{\xi_+\rightarrow 0}_{\rm Hxc}(\br)-v^{\xi_+=0}_{\rm Hxc}(\br)\Big) \, \frac{n_{\Psi_{0}}({\bf r})}{N} 
        \\
        & {=}  \left[\frac{\partial E_{\rm Hxc}^{\bxi}[n_{\Psi_{0}]}}{\partial \xi_{+}} + \frac{\partial E_{\rm Hxc}^{\bxi}[n_{\Psi_{0}}]}{\partial \xi_{-}}\right]_{\bxi=0},
    \end{split}
\end{equation}
where the jump in Hxc potential upon the infinitesimal addition of an electron, which consists in switching from $\xi_+=0$ to $\xi_+\rightarrow 0$ in the language of Nc-eDFT (or crossing an integer number of electrons in PPLB DFT), is fully described by the ensemble weight derivatives of the Hxc functional evaluated for the (fixed) ground-state density.\\  

We can similarly revisit the seminal work of Levy on the optical gap~\cite{levy1995excitation} by considering another ensemble consisting of the $(N-1)$-electron ground state $\Psi_\nu =\Psi_0^{N-1}$ (with weight $\xi_{-}$), the reference $N$-electron ground state $\Psi_{\kappa'} =\Psi_0$, and the first (singly) excited state $\Psi_{\kappa} =\Psi_1^{N}$ (with weight $\xi$). In this case, the exactification of Koopmans' theorem for the evaluation of the neutral excitation energy
\begin{equation}
\begin{split}
    \omega_1 = I_0^N-I_1^N=\varepsilon^{\bxi[\kappa\rightarrow\nu]}_{N+1} -\varepsilon^{\bxi[\kappa'\rightarrow\nu]}_{N}
\end{split}
\end{equation}
implies, according to Eq.~(\ref{eq:vHXC_dxiE}), 
\begin{equation} 
\begin{split}
        &\int d {\bf r} \,  \Big( v_{\rm Hxc}^{\bxi[\kappa\rightarrow\nu]}({\bf r})  - v_{\rm Hxc}^{\bxi[\kappa'\rightarrow\nu]}({\bf r}) \Big) \, \frac{n^{\bxi}({\bf r})}{N} 
        \\
        &=  \frac{\partial E_{\rm Hxc}^{\bxi}[n]}{\partial \xi}\Big|_{n=n^{\bxi}}, 
\end{split}
\end{equation}
which gives in the zero-weight limit
\begin{equation}  \label{eq:disc_shift_opt_0weights}
    \begin{split}
        \int d {\bf r} \, & \Big( v_{\rm Hxc}^{\bxi[\kappa\rightarrow\nu]}({\bf r})  - v_{\rm Hxc}^{\bxi[\kappa'\rightarrow\nu]}({\bf r}) \Big) \, \frac{n_{\Psi_{0}}({\bf r})}{N} 
        \\
        & \underset{\bxi\rightarrow 0}{=}  \left[\frac{\partial E_{\rm Hxc}^{\bxi}[n_{\Psi_{0}]}}{\partial \xi}\right]_{\bxi=0}.
    \end{split}
\end{equation}

In that limit, $v_{\rm Hxc}^{\bxi[\kappa'\rightarrow\nu]}$ does not involve the neutral excited state and is identical to the Hxc potential of Eq.~(\ref{eq:fund_gap_pb_Hxc_pot_exact_IPN}). For that reason, it can simply be denoted as $v^{\xi=0}_{\rm Hxc}$, while $v_{\rm Hxc}^{\bxi[\kappa\rightarrow\nu]}$, that we simply denote $v^{\xi\rightarrow 0}_{\rm Hxc}$ in the same limit, is such that (see Eqs.~\eqref{eq:shift_IP_Koopmans} and \eqref{eq:d_IP_Koopmans})
\be
\begin{split}
&\int d\br\, v^{\xi\rightarrow 0}_{\rm Hxc}(\br)n_{\Psi_0}(\br)
\\
&=E_{\rm Hxc}[n_{\Psi_0}]
+N\left(\frac{\partial E_{\rm Hxc}^{\bxi}[n_{\Psi_{0}]}}{\partial \xi}-\frac{\partial E_{\rm Hxc}^{\bxi}[n_{\Psi_{0}]}}{\partial \xi_-}\right)_{\bxi=0}
\\
&=\int d\br\, v^{\xi=0}_{\rm Hxc}(\br)n_{\Psi_0}(\br)+N\left.\frac{\partial E_{\rm Hxc}^{\bxi}[n_{\Psi_{0}]}}{\partial \xi}\right|_{\bxi=0}.
\end{split}
\ee
As a result, Eq.~(\ref{eq:disc_shift_opt_0weights}) can be rewritten as follows,
\begin{equation} 
    \begin{split}
        &\int d {\bf r} \,\Big(v^{\xi\rightarrow 0}_{\rm Hxc}(\br)-v^{\xi=0}_{\rm Hxc}(\br)\Big) \, \frac{n_{\Psi_{0}}({\bf r})}{N} 
        \\
        & {=}\left[\frac{\partial E_{\rm Hxc}^{\bxi}[n_{\Psi_{0}]}}{\partial \xi}\right]_{\bxi=0},
    \end{split}
\end{equation}
where, {\it via} the Nc ensemble formalism, the equivalence between derivative discontinuities and weight derivatives is now made transparent for neutral excitations too.

\section{eDFT beyond the calculation of energies}\label{sec:densities_and_lrfunc_from_eDFT}

We will now show in this section how excited-state properties such as densities and static density-density linear response functions can be extracted, in principle exactly, from a single eKS calculation, along the lines of the strategy adopted previously for evaluating energy levels. For simplicity, from now on, we will restrict the discussion to TGOK ensembles, without any loss of generality. We refer the interested reader to a recent work~\cite{dupuy2026fukui} where an alternative DFT of Fukui functions has been derived through an extension of the present formalism to Nc ensembles.   

\subsection{Extraction of individual-state densities}\label{sec:extraction_ind_densities}

Extraction of excited-state densities from frameworks first designed for energy level calculations is gaining interest, as shown in a recent work using TDDFT~\cite{baranova2025_excited-state_densities}. We presently consider how to do so from an ensemble perspective. The philosophy of excited-state energies extraction from an ensemble calculation carries to individual densities. Indeed, the physical ensemble density also varies linearly with the weights (see Eq.~\eqref{eq:nc_dens}, where $N_\lambda=N$ in the present case of TGOK ensembles), \ie,
\begin{equation}
    \frac{\partial n^{\bxi}}{\partial \xi_{\lambda}} \underset{\lambda>0}{=} n_{\Psi_{\lambda}} - n_{\Psi_0},
\end{equation}
so that one can isolate any (ground- or excited-) state density by considering appropriate weight derivatives of the ensemble:
\begin{equation} \label{eq:indv_n_dnxi}
\begin{split}
    n_{\Psi_{\nu}} 
    &\underset{\nu\geq 0}{=}\left(n_{\Psi_{\nu}}-n_{\Psi_0}\right)+n_{\Psi_0}
    \\
    &=\sum_{\lambda>0}\delta_{\lambda\nu}\frac{\partial n^{\bxi}}{\partial \xi_{\lambda}}+\left(n^{\bxi}-\sum_{\lambda>0}\xi_{\lambda}\frac{\partial n^{\bxi}}{\partial \xi_{\lambda}}\right)
    \\
    &= n^{\bxi} + \sum_{\lambda>0} \big( \delta_{\lambda\nu} - \xi_{\lambda} \big) \frac{\partial n^{\bxi}}{\partial \xi_{\lambda}} 
     \,.
\end{split}
\end{equation}
To turn this into an exact working eDFT equation, weight derivatives of the ensemble density have been related in Refs.~\onlinecite{Fromager_2020,Fromager2025indvElevel} to the individual KS densities as follows (see Eq.~\eqref{eq:nc_KS_dens}),
\begin{equation} \label{eq:dndxi_ks}
    \frac{\partial n^{\bxi}}{\partial \xi_{\lambda}} = n_{\Phi^{\bxi}_{\lambda}} - n_{\Phi^{\bxi}_0} + \sum_{\nu\geq 0} \xi_{\nu} \frac{\partial n_{\Phi^{\bxi}_{\nu}}}{\partial \xi_{\lambda}},
\end{equation}
where, unlike in the true physical system, the weight derivatives of the KS states do not vanish~\cite{Cernatic2022}. Let us recall that all density functional quantities here are evaluated at the exact ensemble density $n^{\bxi}$, with the shorthand notation $\displaystyle n_{\Phi^{\bxi}_{\nu}[n^{\bxi}]}\equiv n_{\Phi^{\bxi}_{\nu}}$ used throughout.  
We thus need to unravel how individual KS densities vary with weights to match the (also weight-dependent) variationally-determined ensemble density $n^{\bxi}$. This matching is performed by the eKS potential $v^{\bxi}_{\rm s}$ through Eqs.~\eqref{eq:KS_Schrodinger} and \eqref{eq:ks_pot}, so that the KS densities can be equivalently seen as functionals of $v^{\bxi}_{\rm s}$. By chain rule, we rewrite Eq.~\eqref{eq:dndxi_ks} using variations of $v^{\bxi}_{\rm s}$ with respect to the weights:
\begin{equation}
    \begin{split}
            \sum_{\nu\geq 0} \xi_{\nu} \frac{\partial n_{\Phi^{\bxi}_{\nu}}}{\partial \xi_{\lambda}} & = \sum_{\nu\geq 0} \xi_{\nu} \frac{\delta n_{\Phi^{\bxi}_{\nu}}}{\delta v^{\bxi}_{\rm s}} \star \frac{\partial v^{\bxi}_{\rm s}}{\partial \xi_{\lambda}} 
            \\
            & = \frac{\delta n^{\bxi}}{\delta v^{\bxi}_{\rm s}} \star \frac{\partial v^{\bxi}_{\rm s}}{\partial \xi_{\lambda}}  := \chi^{\bxi}_{\rm s} \star \frac{\partial v^{\bxi}_{\rm s}}{\partial \xi_{\lambda}}
            ,
    \end{split}
\end{equation}
where the KS ensemble (static density-density linear) response function $\chi^{\bxi}_{\rm s}$ naturally emerges from Eq.~(\ref{eq:nc_KS_dens}), and we introduced the shorthand notation
\begin{equation}
    \Big(f\star g\Big) ({\bf r},{\bf r}'') = \int d {\bf r}' f({\bf r},{\bf r}') g({\bf r}',{\bf r}'') \,.
\end{equation}
As the exact eKS potential varies with the weights only through the ensemble density-functional Hxc potential, according to Eq.~\eqref{eq:ks_pot}, we can proceed with the following simplification,
\begin{equation} \label{eq:vKSwd}
    \frac{\partial v^{\bxi}_{\rm s}}{\partial \xi_{\lambda}} =   \frac{\partial v^{\bxi}_{\rm Hxc}}{\partial \xi_{\lambda}} =  \frac{\partial v^{\bxi}_{\rm Hxc}[n]}{\partial\xi_\lambda}\bigg|_{n=n^{\bxi}}   + f^{\bxi}_{\rm Hxc} \star \frac{\partial n^{\bxi}}{\partial \xi_{\lambda}}
    ,
\end{equation}
where 
\begin{equation}
    f^{\bxi}_{\rm Hxc}\equiv \left.f^{\bxi}_{\rm Hxc}[n]({\bf r},{\bf r}')\right|_{n=n^{\bxi}} = \frac{\delta^2 E^{\bxi}_{\rm Hxc}[n]}{\delta n({\bf r}) \delta n({\bf r}')}\bigg|_{n=n^{\bxi}} \,
\end{equation}
is the analog for ensembles of the Hxc kernel. We now deduce from Eq.~(\ref{eq:dndxi_ks}) the working integral expression for the weight derivative of the ensemble density,
\be
\begin{split}
\frac{\partial n^{\bxi}}{\partial \xi_{\lambda}} &= n_{\Phi^{\bxi}_{\lambda}} - n_{\Phi^{\bxi}_0}
\\
&\quad+\chi^{\bxi}_{\rm s}\star\left(\frac{\partial v^{\bxi}_{\rm Hxc}[n]}{\partial\xi_\lambda}\bigg|_{n=n^{\bxi}}   + f^{\bxi}_{\rm Hxc} \star \frac{\partial n^{\bxi}}{\partial \xi_{\lambda}}\right)
,
\end{split}
\ee
or, equivalently,
\begin{equation}
\begin{split}
    \frac{\partial n^{\bxi}}{\partial \xi_{\lambda}} 
    &=\left[ 1 - \chi^{\bxi}_{\rm s} \star f^{\bxi}_{\rm Hxc} \right]^{-1}
    \\
    &\quad\star
    \left(n_{\Phi^{\bxi}_{\lambda}} - n_{\Phi^{\bxi}_0} 
    +\chi^{\bxi}_{\rm s} \star \frac{\partial v^{\bxi}_{\rm Hxc}[n]}{\partial\xi_\lambda}\bigg|_{n=n^{\bxi}}\right)  
    ,
\end{split}
\end{equation}
which, as readily seen, can be deduced from the eKS system and its linear response, in principle exactly. As a result, any individual-state density can be expressed as follows, according to Eq.~(\ref{eq:indv_n_dnxi}),
\begin{equation}\label{eq:ind_dens_only_from_KS_quantities}
    \begin{split}
             &n_{\Psi_\nu} =n^{\bxi} 
             + \left[ 1 - \chi^{\bxi}_{\rm s} \star f^{\bxi}_{\rm Hxc} \right]^{-1}\left(n_{\Phi_\nu^{\bxi}}-n^{\bxi}\right)
             \\
             &\left[ 1 - \chi^{\bxi}_{\rm s} \star f^{\bxi}_{\rm Hxc} \right]^{-1}\chi^{\bxi}_{\rm s}\star \sum_{\lambda>0} (\delta_{\lambda\nu}-\xi_\lambda) \frac{\partial v^{\bxi}_{\rm Hxc}[n]}{\partial\xi_\lambda}\bigg|_{n=n^{\bxi}}.  
    \end{split}
\end{equation}
An alternative and more compact expression is finally obtained from the ensemble Dyson equation~\cite{Fromager2025indvElevel} that relates the true ensemble response function $\chi^{\bxi}$ to the eKS one,
\be
\begin{split}
\chi^{\bxi} &\equiv \dfrac{\delta n^{\bxi}}{\delta v_{\rm ext}}
=\chi^{\bxi}_{\rm s}\star\dfrac{\delta v^{\bxi}_{\rm s}}{\delta  v_{\rm ext}}
\\
&=\chi^{\bxi}_{\rm s}+\chi^{\bxi}_{\rm s}\star f^{\bxi}_{\rm Hxc}\star \chi^{\bxi}, 
\end{split}
\ee
or, equivalently,
\begin{equation} \label{eq:ens_dyson}
    \big(\chi^{\bxi} \big)^{-1} = \big(\chi^{\bxi}_{\rm s} \big)^{-1} - f^{\bxi}_{\rm Hxc},
\end{equation}
so that the prefactor in both second and third terms on the right-hand side of Eq.~\eqref{eq:ind_dens_only_from_KS_quantities} simply reads
\begin{equation}
    \left[ 1 - \chi^{\bxi}_{\rm s} \star f^{\bxi}_{\rm Hxc} \right]^{-1} = \chi^{\bxi} \star \big(\chi^{\bxi}_{\rm s} \big)^{-1}
    =1+\chi^{\bxi} \star f^{\bxi}_{\rm Hxc},
\end{equation}
thus leading to the following exact relation between physical and KS individual densities:
\begin{equation} \label{eq:diffpsifull}
    \begin{split}
             n_{\Psi_\nu} = & \big(1+\chi^{\bxi} \star  f^{\bxi}_{\rm Hxc} \big) \star n_{\Phi_\nu^{\bxi}}
            -\chi^{\bxi} \star  f^{\bxi}_{\rm Hxc} \star n^{\bxi} 
            \\
            & + \sum_{\lambda>0} (\delta_{\lambda\nu}-\xi_\lambda)\, \chi^{\bxi} \star\frac{\partial v^{\bxi}_{\rm Hxc}[n]}{\partial\xi_\lambda}\bigg|_{n=n^{\bxi}}.  
    \end{split}
\end{equation}
This is a key result of Refs.~\citenum{Fromager2025indvElevel} and~\citenum{dupuy2025_exact_static} that we recall in this section. Let us briefly discuss how it would be used in practice: From a single eKS calculation (performed with a given choice of weights $\bxi$), one can obtain the ensemble density $n^{\bxi}$, the individual KS densities, and the eKS linear response function. One then needs to solve the ensemble Dyson Eq.~\eqref{eq:ens_dyson} in order to determine the true ensemble response function, thus giving access to the true ground- and excited-state densities. Note that, in the exact theory, the result should not depend on the choice of weights. This can be used as a diagnostic for assessing eDFAs, along the lines of Sec.~\ref{sec:scaling_gs_DFAs}.\\ 

While the above derivation is the most compact, and easily relatable to Sec.~\ref{sec:extraction_ind_ener_levels_NceDFT}, there is an alternative route to Eq.~\eqref{eq:diffpsifull} with several practical implications for which we only summarize the conceptually crucial points, and refer the reader to Ref. \onlinecite{Fromager2025indvElevel} for a detailed proof. The important observation to make is that, even though individual energy levels are not (even local) minima of the energy, they correspond, individually, to the {\it stationary points} (\ie, they form saddle points) of an exact (state-dependent) ensemble density functional that reads, for any external potential $v$~\cite{Fromager2025indvElevel} (see also Eqs.~(\ref{eq:KS_ens_non-int-limit}) and (\ref{eq:int_ens_dens_func_dens_matrix_op_exp})),
\begin{equation} \label{eq:indvElvl_in_terms_of_pots}
    \begin{split}    
        E^{\bxi}_{v,\nu}[n] &\underset{\nu\geq 0}{=} \left\langle\Phi^{\bxi}_\nu[n]\middle\vert\hat{T}\middle\vert\Phi^{\bxi}_\nu[n]\right\rangle + E^{\bxi}_{\rm Hxc}[n] 
        \\
        &\quad + \sum_{\lambda>0}(\delta_{\lambda\nu}-\xi_{\lambda}) \frac{\partial E^{\bxi}_{\rm Hxc}[n]}{\partial\xi_\lambda}
        - \contract{v^{\bxi}_{\rm{Hxc}}[n]}{n} 
        \\
        &
        \quad+ \contract{v+\delta T_{\rm s}^{\bxi}[n]/\delta n+v^{\bxi}_{\rm{Hxc}}[n]}{n_{\Psi^{\bxi}_{\nu}}[n]}
        \\
        &\quad  -\contract{\delta T_{\rm s}^{\bxi}[n]/\delta n}{n_{\Phi^{\bxi}_{\nu}}[n]}.
    \end{split}
\end{equation}

As shown in Ref.~\citenum{Fromager2025indvElevel}, combining the stationarity condition 
\begin{equation} \label{eq:stindvElvl}
    \frac{\delta E^{\bxi}_{v_{\rm ext},\nu}[n]}{\delta n} \Big|_{n=n^{\bxi}} = 0
\end{equation}
with the eKS Eq.~(\ref{eq:KS_Schrodinger}) leads to the exact same individual-state density expression of Eq.~\eqref{eq:diffpsifull}, {\it without} relying explicitly on the linearity-in-weights of the ensemble density. This offers an alternative path for practical eDFT calculations of excited-state energy levels and densities, as eDFAs, which can be introduced into the exact expression of Eq.~(\ref{eq:indvElvl_in_terms_of_pots}), may induce weight curvatures that can be consistently tackled through stationarity conditions~\cite{senjean2015linear}. Moreover, as discussed further in the following, it also paves the way towards the eDFT of excited-state linear response properties.

\subsection{Static linear response of excited states}\label{sec:static_lr-eDFT}

In addition to accessing individual energy levels and densities, there is interest in extracting excited-state linear responses from an eDFT calculation~\cite{dupuy2025_exact_static}. Notably, the computation of polarizabilities in both ground and excited states, from which the change of solute-solvent interactions upon electronic excitations can be quantified~\cite{C7CP08549D}, for example, requires the knowledge of individual-state responses to infinitesimal external electric field variations. From a more fundamental viewpoint, a state-specific analog to the ensemble Dyson Eq.~\eqref{eq:ens_dyson} would point to an explicit definition of individual (ground- or excited-) state Hxc kernels within the ensemble.

In Ref.~\onlinecite{dupuy2025_exact_static}, some of the authors showed how to evaluate exactly (in principle) the individual linear response functions of the physical states from the responses of their KS analogs. Again, the stationarity of the ensemble density-functional energy levels can be exploited. More precisely, the fact that Eq.~\eqref{eq:stindvElvl} holds for any external potential $v$ implies
\begin{equation} \label{eq:dblevarEnu}
    \frac{\delta}{\delta v}\left[ \frac{\delta E^{\bxi}_{v,\nu}[n]}{\delta n}\Bigg|_{n=n^{\bxi}_v}  \right] \Bigg|_{v=v_{\text{ext}}} =  0,   
\end{equation}
with the ensuing derivation detailed in Ref.~\onlinecite{dupuy2025_exact_static}, thus leading to the following exact relation between the physical individual-state (density-density linear) response function $\chi_\nu\equiv \delta n_{\Psi_\nu}/\delta v_{\rm ext}$, its KS analog $\chi^{\bxi}_{{\rm s},\nu}\equiv \delta n_{\Phi^{\bxi}_\nu}/\delta v^{\bxi}_{\rm s}$, as well as both physical ensemble and eKS response functions:
\begin{equation} \label{eq:dblsta2}
    \begin{split}
    &\big(\chi^{\bxi}\big)^{-1}\chi_\nu =  \bigg[\sum_{\lambda>0} (\delta_{\lambda\nu}-\xi_\lambda) \frac{\partial f^{\bxi}_{\rm Hxc}[n]}{\partial\xi_\lambda}\Bigg|_{n=n^{\bxi}}  
    \\ & + \left(g^{\bxi}_{\rm Hxc} \star (n_{\Psi_\nu}-n^{\bxi})\right)  - f^{\bxi}_{\rm Hxc} \bigg] \star \chi^{\bxi}
    \\
    & + \big(\chi^{\bxi}_{\rm s}\big)^{-1} \star \bigg[ \chi^{\bxi}_{{\rm s},\nu}  
    \\
    & +  \left(\chi^{[2]\bxi}_{\rm s} \star \big(\chi^{\bxi}_{\rm s} \big)^{-1} \star (n_{\Psi_\nu}-n_{\Phi_\nu^{\bxi}})\right)  \bigg] \star \big(\chi^{\bxi}_{\rm s}\big)^{-1}  \star \chi^{\bxi},
    \end{split}
\end{equation}
where $\chi^{[2]\bxi}_{\rm s}\equiv {\delta \chi^{\bxi}_{\rm s}/{\delta v^{\bxi}_{\rm s}}}$ is the quadratic eKS response function and $g_{\rm Hxc}^{\bxi}\equiv \left.{\delta f^{\bxi}_{\rm Hxc}[n]}/{\delta n}\right|_{n=n^{\bxi}}$ denotes the (static) ensemble Hxc quadratic kernel.

As readily seen from Eq.~\eqref{eq:dblsta2}, the linear response functions of states belonging to a density-functional ensemble do not obey, individually, a simple Dyson equation similar to that of Eq.~(\ref{eq:ens_dyson}). 
Nonetheless, if we consider the individual components of the inverse ensemble linear response function instead,
\begin{equation} \label{eq:Rnu}
    (R^{\bxi}_\nu)^{-1} :=  \big(\chi^{\bxi}\big)^{-1} \star \chi_\nu \star \big(\chi^{\bxi}\big)^{-1},
\end{equation}
as well as their KS analogs
\begin{equation}\label{eq:Rnu_KS}
    (R^{\bxi}_{{\rm s},\nu})^{-1} =  \big(\chi^{\bxi}_{\rm s}\big)^{-1} \star \chi^{\bxi}_{{\rm s},\nu} \star \big(\chi^{\bxi}_{\rm s}\big)^{-1},
\end{equation}
for which we easily verify that
\be\label{eq:weighted_sum_Rnu}
\sum_{\nu\geq0} \xi_\nu (R^{\bxi}_{\nu})^{-1} = \big(\chi^{\bxi}\big)^{-1} \star \chi^{\bxi} \star (\chi^{\bxi})^{-1}=\big(\chi^{\bxi}\big)^{-1}
\ee
and
\begin{equation}\label{eq:weighted_sum_Rnu_KS}
        \sum_{\nu\geq0} \xi_\nu (R^{\bxi}_{{\rm s},\nu})^{-1} =  (\chi^{\bxi}_{\rm s})^{-1},
\end{equation}
hence the name ``component'' given to both quantities, then the following Dyson equation emerges from Eq.~\eqref{eq:dblsta2},
\begin{equation} \label{eq:dysonlike}
    (R^{\bxi}_{\nu})^{-1} = (R^{\bxi}_{{\rm s},\nu})^{-1} -\Xi^{\bxi}_{{\rm Hxc},\nu},
\end{equation}
with the (weight-dependent) individual-state Hxc kernel $\Xi^{\bxi}_{{\rm Hxc},\nu}$ being identified as
\begin{equation} \label{eq:Xinu}
    \begin{split}
        &\Xi^{\bxi}_{{\rm Hxc},\nu} = f^{\bxi}_{\rm Hxc} -\sum_{\lambda>0} (\delta_{\lambda\nu}-\xi_\lambda) \frac{\partial f^{\bxi}_{\rm Hxc}[n]}{\partial\xi_\lambda}\Bigg|_{n=n^{\bxi}} 
        \\
        & - \left(g^{\bxi}_{\rm Hxc} \star (n_{\Psi_\nu}-n^{\bxi})\right) 
        \\
         & -  \big(\chi^{\bxi}_{\rm s}\big)^{-1}  \star \bigg(  \chi^{[2]\bxi}_{\rm s} \star \big(\chi^{\bxi}_{\rm s} \big)^{-1} 
        \star (n_{\Psi_\nu}-n_{\Phi_\nu^{\bxi}})  \bigg) \star \big(\chi^{\bxi}_{\rm s}\big)^{-1}. 
    \end{split}
\end{equation}
Note that, by construction (see Eqs.~(\ref{eq:weighted_sum_Rnu}) and (\ref{eq:weighted_sum_Rnu_KS})), the above Hxc kernel is the $\nu$th component of the total ensemble Hxc kernel, \ie,
\be\label{eq:average_ind_kernels_Xi_def}
\begin{split}
\sum_{\nu\geq0} \xi_\nu \Xi^{\bxi}_{{\rm Hxc},\nu}
&=\sum_{\nu\geq0} \xi_\nu (R^{\bxi}_{{\rm s},\nu})^{-1}- \sum_{\nu\geq0} \xi_\nu (R^{\bxi}_{\nu})^{-1}
\\
&= (\chi^{\bxi}_{\rm s})^{-1}-\big(\chi^{\bxi}\big)^{-1}
\\
&=f^{\bxi}_{\rm Hxc},
\end{split}
\ee
from which any individual linear response within the ensemble can be obtained, in principle exactly, as follows, 
\be 
\begin{split}
&\chi_\nu
=\chi^{\bxi}\star (R^{\bxi}_{\nu})^{-1}\star \chi^{\bxi} 
\\
&=\chi^{\bxi}\star\left(\big(\chi^{\bxi}_{\rm s}\big)^{-1} \star \chi^{\bxi}_{{\rm s},\nu} \star \big(\chi^{\bxi}_{\rm s}\big)^{-1}-\Xi^{\bxi}_{{\rm Hxc},\nu}\right)\star\chi^{\bxi},
\end{split}
\ee
or, equivalently,
\be\label{eq:chinu}
\begin{split}
\chi_\nu &=\left(1+\chi^{\bxi}\star f_{\rm Hxc}^{\bxi}\right)\star \chi^{\bxi}_{{\rm s},\nu}\star\left(1+f_{\rm Hxc}^{\bxi}\star \chi^{\bxi}\right)
\\
&\quad -\chi^{\bxi}\star \Xi^{\bxi}_{{\rm Hxc},\nu}\star\chi^{\bxi}
.
\end{split}
\ee
While we have shown that it is technically possible to extract static linear response functions of ground and excited states from the eKS system, we refer the reader to Ref.~\citenum{dupuy2025_exact_static} for further discussion about how to turn the approach into a practical computational tool. Let us finally mention that a time-dependent extension of the theory has recently been explored in Ref.~\citenum{daas2025ensembletimedependentdensityfunctional}.


\section{$\Delta$-SCF as an approximation to EDFT}\label{sec:Delta_scf_from_eDFT}

The $\Delta$-SCF approach to excited states, relying on orbital optimization with non-aufbau occupation numbers, has recently received a lot of attention and new developments\cite{Levi20_Variational,Levi_DOMOM_2020,Ivanov21_Method,Hait21_Orbital,Schmerwitz22_Variational,Schmerwitz26_Freeze-and-Release} thanks to its ability to accurately predict double excitations, but also Rydberg\cite{OODFTRydberg_Levi2023} and charge-transfer\cite{Levi_OODFTvsTDDFT2024} states. Seminal~\cite{Gorling95dftexc} and more recent~\cite{yang2024foundationdeltascfapproachdensity,luber25DSCF} works have drawn connections between the {\it ad hoc} $\Delta$-SCF and more clearly-derived theories. Here we will show how $\Delta$-SCF can be recovered from eDFT when well-identified density-functional simplifying assumptions are made~\cite{Fromager2025indvElevel} (see also Ref.~\citenum{gould2025stationaryconditionsexcitedstates}). Indeed, starting from the exact ensemble density-functional energy level expression of Eq.~\eqref{eq:indvElvl_in_terms_of_pots} and proceeding with the (rather crude) density approximations
\begin{equation} \label{eq:appnodd}
   n_{\Psi_{\nu}^{\bxi}[n]}({\bf r}) \approx n({\bf r}) \approx n_{\Phi_{\nu}^{\bxi}[n]}({\bf r}),
\end{equation}
with, on top of it, the following eDFA 
\begin{equation}\label{eq:ensemblization_from_GS_func_ind_densities}
    E^{\bxi}_{\rm Hxc}[n] \approx \sum_{\nu\geq0} \xi_{\nu} E_{\rm Hxc}[n_{\Phi_{\nu}^{\bxi}[n]}]
\end{equation}
based on the regular ground-state Hxc functional $E_{\rm Hxc}[n] = E^{\bxi=0}_{\rm Hxc}[n]$,
which implies, according to Eq.~(\ref{eq:appnodd}),
\be
\begin{split}
&E^{\bxi}_{\rm Hxc}[n] 
+ \sum_{\lambda>0}(\delta_{\lambda\nu}-\xi_{\lambda}) \frac{\partial E^{\bxi}_{\rm Hxc}[n]}{\partial\xi_\lambda}
\\
&\approx E^{\bxi}_{\rm Hxc}[n] 
\\
&\quad+ \sum_{\lambda>0}(\delta_{\lambda\nu}-\xi_{\lambda})\left(E_{\rm Hxc}[n_{\Phi_{\lambda}^{\bxi}[n]}]-E_{\rm Hxc}[n_{\Phi_{0}^{\bxi}[n]}]\right)
\\
&=E_{\rm Hxc}[n_{\Phi_{\nu}^{\bxi}[n]}],
\end{split}
\ee
we obtain the (drastically) simplified energy expression, for $v=v_{\rm ext}$,
\begin{equation}
    \begin{split}
            E^{\bxi}_\nu[n] \approx & \;\bra{\Phi_{\nu}^{\bxi}[n]} \hat{T} + \hat{V}_{\rm ext} \ket{\Phi_{\nu}^{\bxi}[n]} + E_{\rm Hxc}[n_{\Phi_{\nu}^{\bxi}[n]}]
            \\
            &\equiv \; E^{\rm KS-DFT}\big[\Phi_{\nu}^{\bxi}[n]\big],
    \end{split}
\end{equation}
which is essentially that of the $\Delta$-SCF method. The remaining connection with eDFT is the ensemble density $n$ which is, in the exact theory, the proper variational variable. In this context, the exact stationarity condition of Eq.~(\ref{eq:stindvElvl}) becomes
\begin{equation}\label{eq:stat_cond_ens_dens_like_Delta-SCF}
    0\approx \bra{\delta \Phi_{\nu}^{\bxi}[n]} \hat{T} + \hat{V}_{\rm ext} + \hat{V}_{\rm Hxc}[n_{\Phi_{\nu}^{\bxi}[n]}] \ket{\Phi_{\nu}^{\bxi}[n]}|_{n=n^{\bxi}}, 
\end{equation}
where $\hat{V}_{\rm Hxc}[n] = \int d{\bf r}\,{\delta E_{\rm Hxc}[n]}/{\delta n({\bf r})} \, \hat{n}({\bf r})$ is the regular ground-state density-functional Hxc potential operator. While a $\Delta$-SCF calculation differs from the above eDFT-based construction, strictly speaking, because the former searches directly for saddle points of $E^{\rm KS-DFT}\big[\Phi]$ with respect to single-configuration wavefunctions $\Phi$, its stationary solutions will always satisfy Eq.~(\ref{eq:stat_cond_ens_dens_like_Delta-SCF}), thus establishing a clear connection between the two approaches.  
In order to improve on the method, from the perspective of eDFT, one should realize, as pointed out recently by Gould~\cite{gould2025stationaryconditionsexcitedstates}, that the key approximation of Eq.~\eqref{eq:appnodd} completely neglects density-driven correlations, \ie, correlations induced by the fact that, unlike in ground-state DFT, physical and KS states do not share the same density within an ensemble, individually, even though the total ensemble densities match~\cite{PRL19_Gould_DD_correlation,Fromager_2020,Cernatic2022,gould2026ensemblization}. Including these correlations by referring to the fluctuation-dissipation theorem (FDT) and the adiabatic connection formalism, together with the low-density regime, is a promising strategy~\cite{gould2025stationaryconditionsexcitedstates,gould2026ensemblization}. One could also replace the conventional eKS equations by the (more involved~\cite{Gould2021_Ensemble_ugly,Cernatic2022} but free from the ghost-interaction errors~\cite{ensemble_ghost_interaction,Cernatic2022} that are further discussed in Sec.~\ref{sec:finding_scaling_function}) self-consistent one-electron equations deduced from the eDFA of Eq.~(\ref{eq:ensemblization_from_GS_func_ind_densities}). The connection with eDFT is also expected to be useful for improving on the description of excited-state properties beyond excitation energies, in the light of Sec.~\ref{sec:static_lr-eDFT}. Work is currently in progress in these directions.


\section{Exploring alternative routes toward practical eDFT calculations}\label{sec:routes_practice}

Several strategies are currently pursued to device practical computational tools based on eDFT. Orbital-based eDFA finds its appeal in bypassing development of functionals that depend explicitly on both the weights and the density. At the ensemble Hx-only level of the theory, it has been applied extensively to excitation energies~\cite{yang2014exact,pribramjones2014excitations,yang2017direct,gould2018charge,sagredo2018can,deur2018exploring,senjean2018unified,deur2019ground,Senjean_2020} in difficult situations for TDDFT such as states with a double excitation character. It has recently been extended to include correlation at the G\"{o}rling--Levy second-order perturbation theory (GLPT2) level~\cite{Yang2021_Second}. Turning to development of full-fledged eDFAs, crucial work has been done to understand the issues plaguing a naive use of regular KS-DFT principles, in particular regarding the definition of the proper ensemble analog to the Hartree and exchange terms\cite{gould2017hartree}. This avoids crippling self-interaction errors and the non-uniqueness disaster in presence of degeneracies.
Extension of the FDT to ensembles\cite{PRL20_Gould_Hartree_def_from_ACDF_th,Gould2021_Double} yielded separate expression for Hartree and exchange components, allowing use of hybrid functionals.
Finally, exploiting the ensemble FDT together with the adiabatic connection formula for ensembles yielded a general expression for the correlation energy\cite{PRL20_Gould_Hartree_def_from_ACDF_th}. Interestingly, in addition to state-driven contributions which are direct analogs of the correlation in KS-DFT, it presents an additional term bearing density-driven contributions specific to eDFT. Note existence of these terms was reported\cite{PRL19_Gould_DD_correlation} and refined\cite{Fromager_2020} shortly before the FDT argument was presented. Interpolation between known low and high correlation limits of the theory~\cite{Gould2023_Electronic,scott2024exact}, exploiting their correspondence to low and high density limits respectively through the adiabatic connection formalism, proved to be an effective strategy to build eDFAs including correlation energy. It was also used to specifically derive an approximation of the density-driven contributions to the ensemble correlation energy\cite{Gould2025_PRL_tate-Specific}. We refer the reader to Ref.~\citenum{gould2026ensemblization} for a recent overview.

In the following, we will explore three (very) different paths to the same general aim as the work mentioned above, albeit with noticeably different strategies. The first one (discussed in Sec.~\ref{sec:scaling_gs_DFAs}) consists in recycling regular ground-state DFAs through a weight-dependent scaling function. The second route (discussed in Sec.~\ref{sec:mrpt_eDFT}) consists in applying 
many-body perturbation theory in the context of eDFT, with a particular focus on its multi-reference formulation and the implications in the definition of ensemble Hartree, exchange, and correlation energies, individually. Finally, we discuss in Sec.~\ref{sec:ensemble_LPFET} the use of the eDFT formalism for designing a rigorous quantum embedding theory of excited states, by analogy with the recently proposed LPFET of pure ground states~\cite{makhlouf2025_local_potential,makhlouf2026generalizedlocalpotentialfunctional}.

\subsection{Recycling regular DFAs through a weight-dependent scaling function}\label{sec:scaling_gs_DFAs} 

\subsubsection{Motivation}

Regular KS-DFT is neatly recovered from eDFT by taking all ensemble weights to be zero. Yet, there is more to the zero-weight limit than the ensemble falling back to just being the ground state. Let us make it clear that what we refer to as the zero-weight limit of eDFT includes infinitesimal deviations from $\bxi=0$, so that differentiations with respect to ensemble weights allowing to extract individual-state properties (the energies in Eq.~(\ref{eq:ind_e_nc}), the densities in Eq.~(\ref{eq:diffpsifull}), and the static density-density linear response functions in Eq.~(\ref{eq:chinu})) can be performed. This is not possible if we immediately set $\bxi=0$ in the variational ensemble energy expression of Eq.~(\ref{eq:var_principle}), which gives back ground-state KS-DFT and nothing more. With this subtlety in mind, all relations discussed previously between individual (ground- and excited-) state properties and those of the ensemble remain valid when the weight derivatives are evaluated at $\bxi=0$. This brings a considerable simplification, as all ensemble properties and Hxc ``fields'' (energy, potential, and kernel) become their comparatively simpler and more easily approximated ground-state analogs. The remaining ensemble correction to KS-DFT that needs to be modeled is brought by ensemble weight derivatives of the relevant ensemble Hxc quantity, for example ${\partial E^{\bxi}_{\rm Hxc}[n_{\Psi_0}]}/{\partial \xi_\lambda}|_{\bxi=0}$ for the energies (see Eq.~(\ref{eq:ind_e_nc})). Approximating the latter weight derivatives reduces the effort to modeling the weight dependence of the ensemble Hxc functional in the vicinity of the zero-weight limit and the ground-state density $n_{\Psi_0}$, which also is considerably simpler than modeling the full density and weight dependence of the ensemble Hxc functional.\\ 

To that aim, we explore in this section the following ensemble scaled DFA,
\be \label{eq:ehxc_scaling}
E_{\rm Hxc}^{\bxi}[n]\approx s(\bxi)E_{\rm
Hxc}^{{\bxi}=0}[n]=s(\bxi)E_{\rm Hxc}[n],
\ee
where the (to-be-determined) weight-dependent scaling function $s(\bxi)$ ``dresses'' the regular ground-state Hxc functional. The above ansatz implies those of the ensemble Hxc potential and the Hxc weight derivatives, \ie, 
\begin{equation} \label{eq:pot_ansatz}
    v^{\bxi}_{\rm Hxc}[n] \approx s(\bxi) v_{\rm Hxc}[n],
\end{equation}
and
\be
\dfrac{\partial E^{\bxi}_{\rm Hxc}[n]}{\partial \xi_\lambda}\approx E_{\rm Hxc}[n]\dfrac{\partial s(\bxi)}{\partial \xi_\lambda}, 
\ee
respectively. The ansatz of Eq.~\eqref{eq:ehxc_scaling} seems like a very simple form building weight dependence on top of a regular (semi)-local functional, for example. The motivation for scaling both Hartree, exchange, and correlation contributions relies on the assumption that ghost-interaction errors (which originate from unphysical energy couplings between different states in the approximate Hartree energy~\cite{ensemble_ghost_interaction,Cernatic2022}, for example, when evaluated for the ensemble density) might be adequately corrected. This is actually confirmed later in Sec.~\ref{subsec:scaled_eDFA_Hubbard_dimer} in the prototypical Hubbard dimer model. This form also has the appeal to preserve the exact uniform coordinate scaling relation of the ground-state Hxc functional that was shown to hold also for TGOK ensembles~\cite{Nagy_ensAC}. Recent investigations of its consequences and its link to the coupling constant of the adiabatic connection~\cite{Nagy_ensAC} have proved to be fruitful in generating eDFAs~\cite{Gould2023_Electronic,scott2024exact}. As discussed in Appendix~\ref{appendix:details_scaled_eDFA}, where the ansatz of Eq.~\eqref{eq:ehxc_scaling} is further rationalized, the exact uniform coordinate scaling constraint actually holds in the more general case of Nc ensembles. In the following, we will show how the scaling function can be determined, for a given system, by exploiting exact properties of eDFT, thus illustrating its relevance and potential impact in further developments of eDFAs. 

\subsubsection{Finding the appropriate scaling function}\label{sec:finding_scaling_function}

From its definition in Eq.~(\ref{eq:ehxc_scaling}), it follows immediately that $s(\bxi=0)=1$. In order to assess the relevance of the ensemble scaled ansatz, we need to determine the scaling function (as a function of the weights) from that ``initial condition'', for example, from working differential equations that need to be identified. For that purpose, we can use the fact that each energy level, as determined from exact eDFT, according to Eq.~\eqref{eq:ind_e_nc}, should be the physical {\it weight-independent} one, even though it is evaluated from weight-dependent quantities. By inserting the ensemble scaled ansatz of Eq.~\eqref{eq:ehxc_scaling} into Eq.~\eqref{eq:ind_e_nc}, \ie,
\begin{equation} 
\begin{split}
        E_\nu & \underset{\nu\geq0}{=} \mathcal{E}^{\bxi}_{\nu} + \frac{N_\nu}{N}  \Big( s(\bxi) E_{\rm Hxc}[n^{\bxi}] - \big( v^{\bxi}_{\rm Hxc} \big| n^{\bxi} \big) \Big)
        \\
        & \quad + \sum_{\lambda>0} \left(\delta_{\lambda\nu}-\frac{N_\nu}{N} \xi_{\lambda}\right) E_{\rm Hxc}[n^{\bxi}]\frac{\partial s({\bm \bxi})}{\partial \xi_{\lambda}} ,
\end{split}
\end{equation}
we immediately obtain the desired equations,
\begin{equation} \label{eq:dsx}
    \sum_{\lambda>0} \left(\delta_{\lambda\nu}-\frac{N_\nu}{N} \xi_{\lambda}\right) \frac{\partial s({\bm \bxi})}{\partial \xi_{\lambda}} \underset{\nu\geq 0}{=} c_{\nu}({\bxi}) -\frac{N_\nu}{N} s(\bxi),
\end{equation}
where the weight evolution of the scaling function is driven by the following state-dependent coefficient functions,
\begin{equation} \label{eq:cnu}
    c_{\nu}({\bxi}) := \frac{E_\nu-\mathcal{E}^{\bxi}_{\nu} + \frac{N_\nu}{N} \big( v^{\bxi}_{\rm Hxc} \big| n^{\bxi} \big)}{E_{\rm Hxc}[n^{\bxi}]}
    .
\end{equation}
In the above construction, it is assumed that the true energies are known, which is of course not the case in practice, otherwise the approach would be pointless. In this paper, we are only discussing the relevance of the ansatz. We refer the reader to a recent work on the calculation of Fukui functions within Nc-eDFT~\cite{dupuy2026fukui}, where more practical strategies for the design of scaling functions are presented.\\    

A natural question to be asked at this point is from which state should we actually determine the scaling function? Being able to rely exclusively on the ground state, for which
\be\label{eq:C0_equal_one_zero_weight}
c_{0}({\bxi}=0)=1,
\ee
since
\be\label{eq:zero_weight_limit_GS_dens}
n_{\Phi_0^{\bxi=0}}=n^{\bxi=0}:=n^0=n_{\Psi_0}
\ee
in the zero-weight limit, is of course extremely appealing but, unfortunately, not very instructive if we remain close to that limit. Indeed, if we consider the simplest case of a TGOK bi-ensemble, with $\bxi \equiv \xi_1=\xi$, Eq.~(\ref{eq:dsx}) reads as follows for the ground state ($\nu=0$),
\be
-\xi\frac{d s(\xi)}{d\xi}=c_0(\xi)-s(\xi),
\ee
or, equivalently,
\be
\label{eq:scaling_from_GS_deadend}
c_0(\xi)=s(\xi)-\xi\frac{d s(\xi)}{d\xi}.
\ee
As readily seen from the above equation, the first-order derivative at $\xi=0$ of the scaling function, which is the key information we need to evaluate excitation energies from a  regular KS-DFT calculation (see the zero-weight limit of Eq.~(\ref{eq:eb_gap_nc})), is actually inaccessible from the ground state, the first-order terms in $\xi$ disappearing in the right-hand side of Eq.~(\ref{eq:scaling_from_GS_deadend}). This is consistent with the fact that the derivative of $c_0(\xi)$ vanishes in the zero-weight limit, according to the Hellmann--Feynman theorem:
\be\label{eq:scaling_E0_nothing_to_learn}
\begin{split}
\left.\dfrac{d c_0(\xi)}{d\xi}\right|_{\xi=0}
&=
\dfrac{1}{E_{\rm Hxc}[n^{0}]}
\left[
-\left.\dfrac{\partial (v_{\rm Hxc}^{\xi}\vert n)}{\partial \xi}
\right|_{n=n^0}
\right.
\\
&\quad+\left.\dfrac{\partial \left(v_{\rm Hxc}^{\xi}\vert n^\xi\right)}{\partial \xi}\right]_{\xi=0}
\\
&
\quad
-
\dfrac{1}{E_{\rm Hxc}[n^{0}]}(v_{\rm Hxc}^{\xi}\vert \partial n^\xi/\partial\xi)_{\xi=0}
\\
&=0.
\end{split}
\ee

Therefore, we need to rely on the excited state instead, for which the working Eq.~(\ref{eq:dsx}) reads
\begin{equation}
    \frac{ds(\xi)}{d \xi}= \frac{c_{1}(\xi) -s(\xi)}{1-\xi },
\end{equation}
where we can easily verify that the derivative of the scaling function can now be extracted and it is non-zero,
\begin{equation}
\begin{split}
    \frac{ds({\bf \xi})}{d\xi}\Big|_{\xi=0}
    &= c_{1}(0)-1
    \\
    &=\dfrac{E_1-\mathcal{E}_1^{0}+(v_{\rm Hxc}\vert n^0) -E_{\rm Hxc}[n^{0}]}{E_{\rm Hxc}[n^{0}]}
    \\
    &\neq 0,
\end{split}
\end{equation}
unlike in the case of the ground state, for which regular KS-DFT is exclusively designed (see Eqs.~(\ref{eq:C0_equal_one_zero_weight}) and (\ref{eq:zero_weight_limit_GS_dens})).


\subsubsection{Illustrative example}\label{subsec:scaled_eDFA_Hubbard_dimer}

As a proof of principle, we apply the procedure to the bi-ensemble of ground and first-excited singlet states of the two-electron Hubbard dimer whose Hamiltonian reads in second quantization~\cite{carrascal2015hubbard}
\be
\begin{split}
\hat{H}=\hat{\mathcal{T}}+\hat{\mathcal{U}}+\hat{\mathcal{V}},
\end{split}
\ee
where
\begin{subequations}\label{eq:Hamiltonian_Hdim}
\begin{align}
\hat{\mathcal{T}}&=-t\sum_{\sigma = \uparrow,\downarrow}
(\hat{c}^{\dagger}_{0\sigma}\hat{c}_{1\sigma} + \hat{c}^{\dagger}_{1\sigma}\hat{c}_{0\sigma}),
\label{eq:Hamiltonian_Hdim_hopping}
\\
\hat{\mathcal{U}}&=U\sum_{i=0}^{1}\hat{n}_{i\uparrow}\hat{n}_{i\downarrow},
\label{eq:Hamiltonian_Hdim_U}
\\
\hat{\mathcal{V}}&=\frac{\Delta v}{2}(\hat{n}_{1} - \hat{n}_{0})
\label{eq:Hamiltonian_Hdim_vext}
\end{align}
\end{subequations}
are the hopping (analog of the kinetic energy for a lattice model), the on-site electronic repulsion, and local potential operators, respectively. In Eq.~(\ref{eq:Hamiltonian_Hdim}), $i \in \qty{0,1}$ labels the two atomic sites,  $\hat{n}_{i\sigma} = \hat{c}^{\dagger}_{i\sigma}\hat{c}_{i\sigma}$
is the spin-site occupation operator, and $\hat{n}_{i} = \sum_{\sigma =
\uparrow,\downarrow}\hat{n}_{i\sigma}$ plays the role of
the density operator. While the difference in local potential $\Delta
v$ controls the asymmetry of the model, the ratio $U/t$ can be used to tune the strength of electron correlation. In this model, the ensemble electronic density
$\left\{n_i=\langle\hat{n}_i\rangle\right\}_{i=0,1}$ reduces to a single
number $n$ with site
occupations equal to $n_{0}=n$ and $n_{1} = 2-n$. The system's parameters are set to $U=1$, $t=1$, and $\Delta v=0.5$ in the following, which corresponds to a standard moderately correlated regime.\\

\begin{figure}[h]
    \centering
    \includegraphics[width=\linewidth]{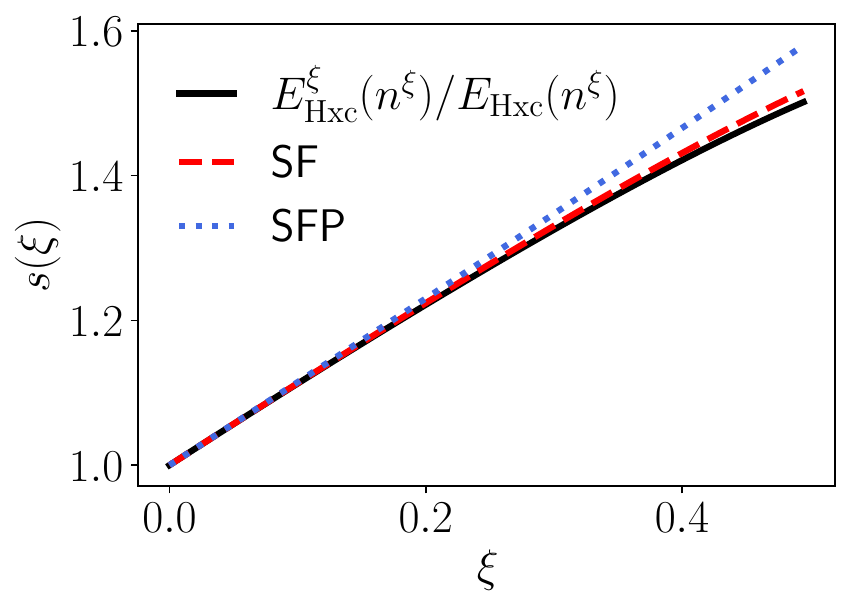}
    \caption{Scaling function obtained in two ways (SF and SFP) and plotted as a function of the bi-ensemble weight. Comparison is made with the exact scaling (in black). See text for further details.}
    \label{fig:scalingfcts}
\end{figure}

\begin{figure}[h]
    \centering
    \includegraphics[width=\linewidth]{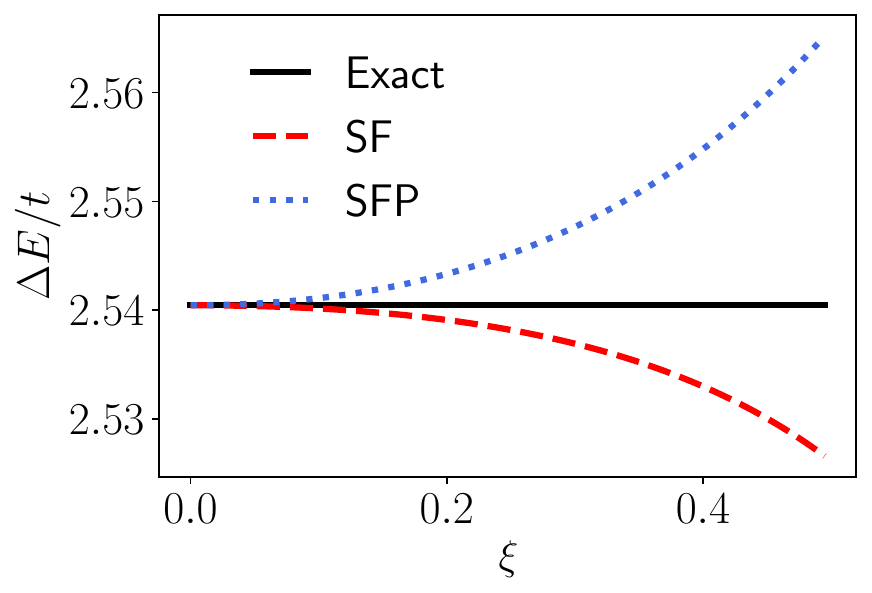}
    \caption{Prediction of the energy gap from the scaling functions as a function of the bi-ensemble weight. See text for further details.}
    \label{fig:Egap}
\end{figure}

\begin{figure*}
    \centering
    \begin{tabular}{cc}
        \includegraphics[width=0.45\linewidth]{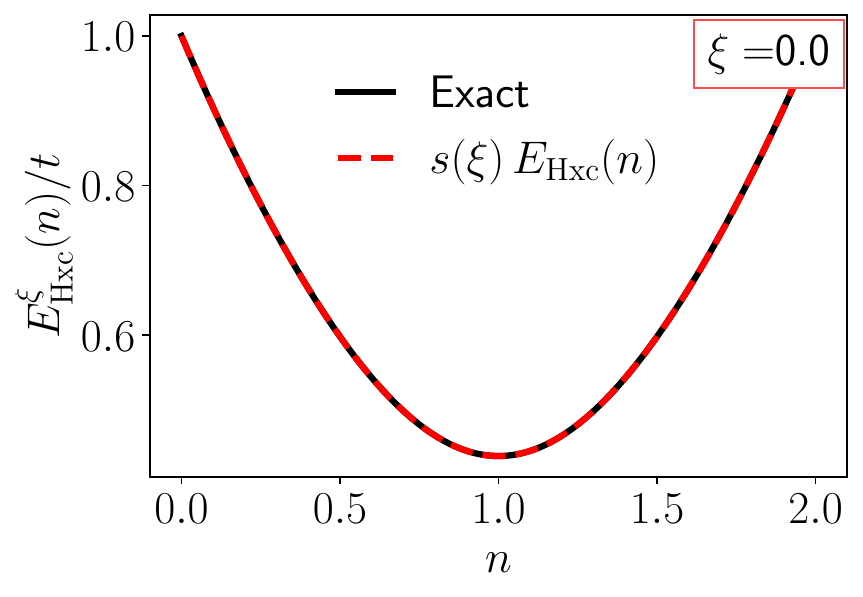}&
        \includegraphics[width=0.45\linewidth]{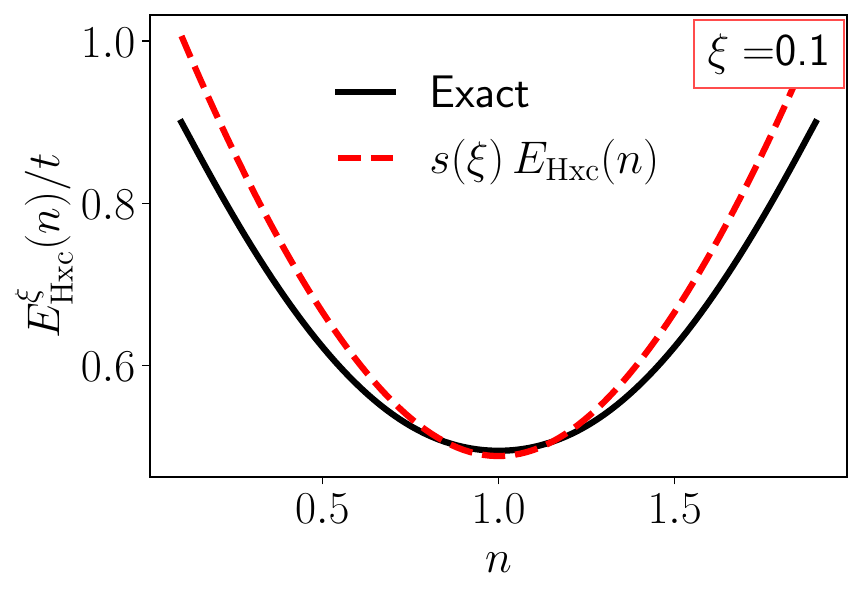}\\[2\tabcolsep]
        \includegraphics[width=0.45\linewidth]{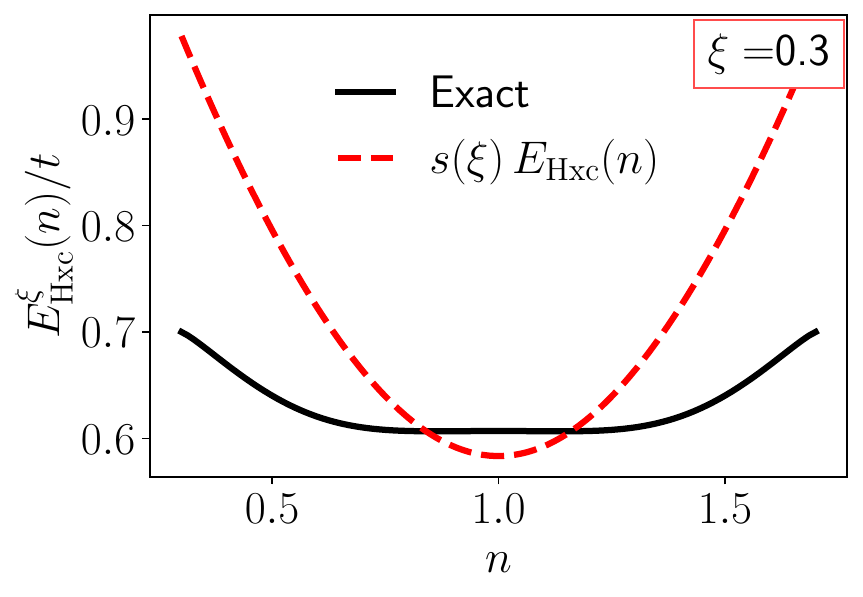}&
        \includegraphics[width=0.45\linewidth]{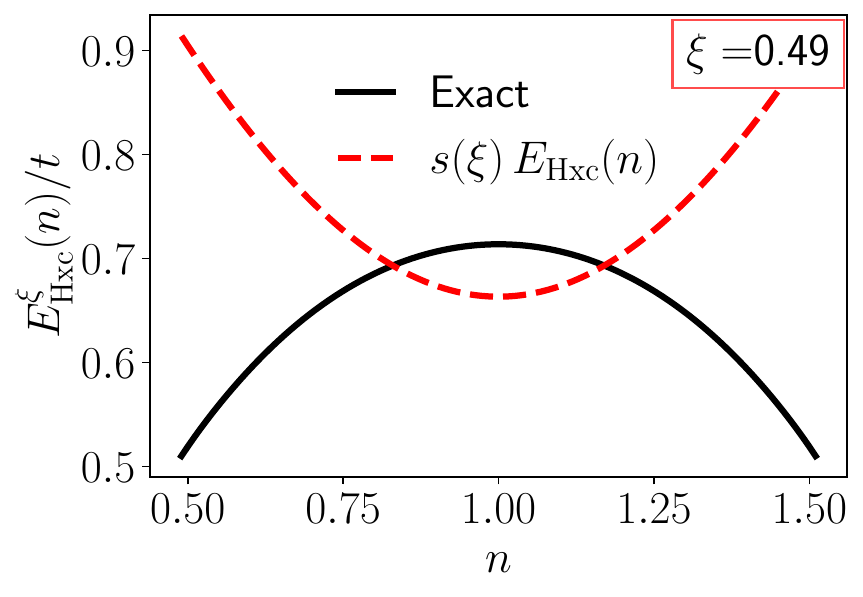}
    \end{tabular}
    \caption{Comparison of the exact ensemble Hxc density functional with the ensemble scaled one (SF) for different bi-ensemble weight values.}
    \label{fig:Ehxc}
\end{figure*}
We will use the exact $E_1$ which can be determined analytically in the model~\cite{deur2017exact} and compute the scaling function in two ways. The latter will differ in the ensemble Hxc potential that is used in Eq.~\eqref{eq:cnu}. In the first approach, referred to as scaled functional (SF), the exact potential is used (see Ref.~\citenum{deur2017exact} where the exact numerical evaluation of ensemble density-functional Hxc energies and potentials is described in detail) while in the second approach, referred to as scaled functional-potential (SFP), the approximation of Eq.~\eqref{eq:pot_ansatz} will be used instead. The resulting scaling functions are compared on Fig.~\ref{fig:scalingfcts}, with SF (in red) and SFP (in blue), and for comparison the ratio of the exact ensemble and ground-state Hxc functionals evaluated at the exact ensemble density (in black). The SF scaling function is very close to the exact ratio of functionals. In a way, this is remarkable as the scaling function is exclusively determined from the excited-state energy, not the ensemble one. The error that occurs at larger weight values within the SFP approximation already suggests that the scaling function should be made density-dependent.

Both SF and SFP scaling functions predict the correct first excited-state energy, by construction, provided they are used with the appropriate Hxc potential (exact for SF and Eq.~\eqref{eq:pot_ansatz} for SFP). We now turn to the energy gap $\Delta E = E_1-E_0$ prediction as a function of the weight $\xi$, which is reported on Fig.~\ref{fig:Egap}. While remaining exact through first-order in $\xi$ as we deviate from the ground-state limit of the theory, by construction (see Eq.~(\ref{eq:scaling_E0_nothing_to_learn})), an error is accumulated on the ground-state energy prediction as the weight increases further. Nonetheless, comparison of the error on the gap evaluated at $\xi=0.5$ (of about $2.5 \times 10^{-2} \, t$ for SFP and half of that for SF) to the variation of the KS gap over the same weight variation (adding up to $1.24 \times 10^{-1} \, t$) shows the positive impact of the scaling function.

\begin{figure}[h]
    \centering
    \includegraphics[width=\linewidth]{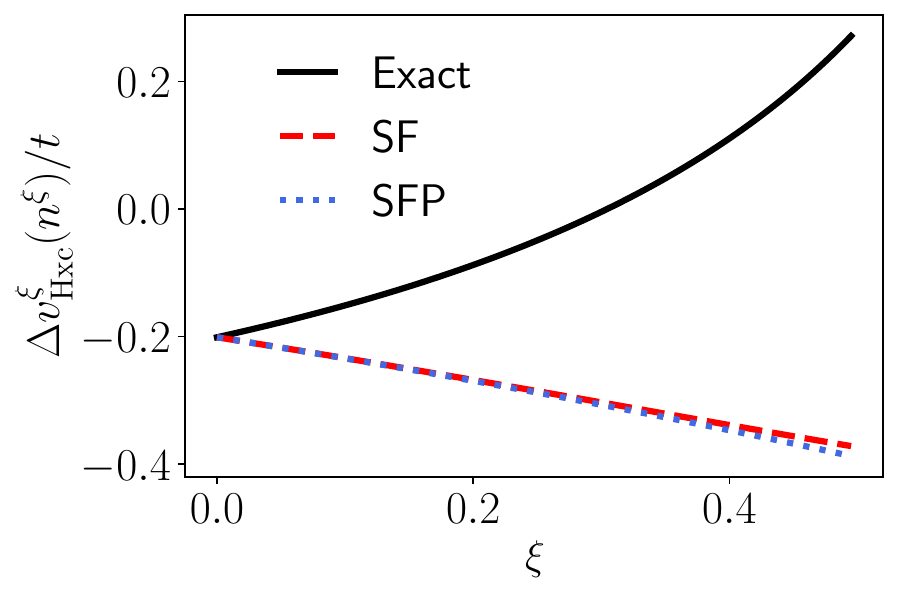}
    \caption{Comparison of the exact ensemble Hxc potential and the ensemble scaled ones (SF and SFP), both evaluated at the exact ensemble density $n^\xi$. See text for further details.}
    \label{fig:vhxc}
\end{figure}

Finally, we turn to the comparison of the ensemble scaled Hxc density functional (obtained from the most accurate SF scheme) with the exact one. On Fig.~\ref{fig:Ehxc}, they are both plotted (in red and black, respectively) on four panels corresponding to increasing weight values. At low weights (top panels) the match is pretty good, as expected. On the other hand, the exact ensemble Hxc functional adopts at higher weights a density dependence that is too starkly different from that of the ground-state functional (see the bottom panels as well as Refs.~\citenum{deur2017exact} and \citenum{deur2018exploring}) to allow a reasonable approximation by a density-independent scaling function. As already shown in Fig.~\ref{fig:scalingfcts}, in the latter case, the scaled functional is still very accurate at the exact ensemble density values of the system under study (where the curves cross), \ie, the values of the density for which the scaling function has been determined (from the excited-state energy). In practice, one needs not the approximation of $E^{\bxi}_{\rm Hxc}[n]$ to match the exact one everywhere both density- and weight-wise. As explained above, a practical approximation needs only to be valid around the ensemble density $n^{\bxi}$ for a given set of weights, and the vicinity of the zero-weight limit is all we need for an ensemble correction to a ground-state KS-DFT calculation. Nonetheless, the absence of density dependence in the scaling function is expected to be problematic for solving the eKS equations. Indeed, as seen on Fig.~\ref{fig:vhxc}, the ensemble scaled Hxc potential evaluated from Eq.~\eqref{eq:pot_ansatz} with the exact ensemble density differs substantially from the exact ensemble Hxc potential as soon as the ensemble weight deviates from zero. Future work should explore how to design more flexible ensemble scaling functions, ideally without relying on accurate energies of a given system. A first promising investigation in this direction has recently been performed by some of the authors in connection with the calculation of Fukui functions~\cite{dupuy2026fukui}. Properties of the EEXX functional and the extension to ensembles of GLPT2 (see Sec.~\ref{subsec_motivation_MRPT}) have been used for that purpose.

\subsection{Multi-reference ensemble density-functional perturbation theory}\label{sec:mrpt_eDFT}

\subsubsection{Motivation}\label{subsec_motivation_MRPT}

While significant progress has been made in the practical implementation of orbital-dependent DFAs~\cite{Teale2022_DFT_exchange,Trushin25_Accurate}, this strategy has been mostly used in the context of eDFT for computing ensemble Hartree and exchange energies~\cite{yang2017direct,yang2014exact,sagredo2018can}. In the seminal work of Yang~\cite{Yang2021_Second}, GLPT2 has been extended to the calculation of ensemble correlation energies. In fact, employing multi-reference (quasi-degenerate) perturbation theory, where the projection of each low-lying electronic state onto the eKS space can be refined up to a given order of perturbation (a feature that ensemble GLPT2~\cite{Yang2021_Second} does not have), seems like a more natural approach to density-functional ensembles, as discussed further in the following.\\

We will start with a Rayleigh--Schr\"{o}dinger (RS) formulation of the theory, where the true interacting problem is projected onto the eKS space (see Sec.~\ref{eq:ensemble_RS_theory}). On that basis, the decomposition of the ensemble Hxc energy into separate Hartree, exchange, and correlation contributions will be revisited in Sec.~\ref{sec:discuss_H-x-c_defs}. Applying perturbation theory in this context (see Sec.~\ref{sec:PT_expansion_RS}) leads to a quasi-degenerate flavor of ensemble GLPT2. For completeness, we will explore in Sec.~\ref{sec:mrpt_eDFT-unitary_formulation} an alternative unitary formulation, along the lines of Van Vleck (VV) perturbation theory~\cite{VanVleck29_On,SOKOLOV2024121}. Formal and practical advantages/drawbacks of both approaches will be discussed.

\subsubsection{Projection-based formulation}\label{eq:ensemble_RS_theory}

Let us consider a $M$-state TGOK ensemble with weights $\bxi\equiv \left\{\xi_1,\ldots,\xi_{M-1}\right\}$, where the ground and low-lying excited states of interest ($M$ states in total) describe the same integer number $N$ of electrons. Following the philosophy of RS perturbation theory (RSPT)~\cite{Lindgren1986}, we propose to rewrite the physical Schr\"{o}dinger equation, which is fulfilled by the states that belong to the ensemble, in terms of their projections onto the eKS space:
\be\label{eq:tilde_projections_def}
\ket{\Psi_\nu}\rightarrow \ket{\tilde{\Phi}^{\bxi}_\nu}=\hat{P}^{\bxi}\ket{\Psi_\nu},\, 0\leq \nu<M,
\ee
where the projection operator can be expressed as follows in terms of the eKS wavefunctions (see Eq.~(\ref{eq:KS_Schrodinger})),
\be\label{eq:projector_ortho_basis_KS_model_space}
\hat{P}^{\bxi}=\sum_{0\leq \nu<M}\ket{\Phi^{\bxi}_\nu}\bra{\Phi^{\bxi}_\nu}.
\ee
We stress that the latter are not necessarily Slater determinants. Indeed, in the limit of non-interacting electrons, where the electronic repulsion is infinitesimal, they become configuration state functions (CSFs)~\cite{gould2017hartree}. We will return to this point later. While both KS and physical ensembles consist of orthonormal solutions, the so-called {\it model} eKS wavefunctions $\left\{\tilde{\Phi}^{\bxi}_\nu\right\}$ are obviously not orthonormal. If we proceed with their orthonormalization,
\be
\ket{\tilde{\Phi}^{\bxi}_\nu}\rightarrow \sum_{0\leq \nu<M}\left[\tilde{\bf S}^{-\frac{1}{2}}\right]_{\mu\nu}\ket{\tilde{\Phi}^{\bxi}_\nu},
\ee
where $\tilde{\bf S}\equiv \left\{\langle \tilde{\Phi}^{\bxi}_\mu\vert \tilde{\Phi}^{\bxi}_\nu \rangle\right\}$ is the overlap matrix, then the projector can be expressed equivalently as follows,   
\be\label{eq:projector_model_KS_space_non-ortho_basis}
\hat{P}^{\bxi}=\sum_{0\leq \mu,\nu<M} \left[\tilde{\bf S}^{-1}\right]_{\mu\nu}\ket{\tilde{\Phi}^{\bxi}_\mu}\bra{\tilde{\Phi}^{\bxi}_\nu}.
\ee
\\

The connection between the model eKS wavefunctions and the (non-normalized in the present theory) physical solutions relies on the ensemble wave operator $\hat{\Omega}^{\bxi}$ that reconstructs the latter solutions from the former:
\be\label{eq:def_wave_operator}
\ket{\Psi_\nu}=\hat{\Omega}^{\bxi}\ket{\tilde{\Phi}^{\bxi}_\nu},\, 0\leq \nu<M. 
\ee
It can be determined, in principle exactly, by solving the so-called generalized Bloch equation, which consists in reformulating the Schr\"{o}dinger equation by applying the wave operator on top of the projector, \ie, $\hat{P}^{\bxi}\rightarrow \hat{\Omega}^{\bxi}\hat{P}^{\bxi}$, so that, in the expression of Eq.~(\ref{eq:projector_model_KS_space_non-ortho_basis}), the model eKS wavefunctions in the kets are replaced by their physical analogs:
\be
\begin{split}
&\hat{H}\hat{\Omega}^{\bxi}\hat{P}^{\bxi}
\\
&
=\hat{H}\left(\sum_{0\leq \mu,\nu<M}\left[\tilde{\bf S}^{-1}\right]_{\mu\nu} \ket{\Psi_\mu}\bra{\tilde{\Phi}^{\bxi}_\nu}\right)
\\
&=
\sum_{0\leq \mu,\nu<M} \left[\tilde{\bf S}^{-1}\right]_{\mu\nu}E_\mu\ket{\Psi_\mu}\bra{\tilde{\Phi}^{\bxi}_\nu}
,
\end{split}
\ee
or, in a more compact and equivalent form, 
\be\label{eq:first_step_proof_GBE}
\hat{H}\hat{\Omega}^{\bxi}\hat{P}^{\bxi}=\hat{\Omega}^{\bxi}\hat{P}^{\bxi}\hat{H}\hat{\Omega}^{\bxi}\hat{P}^{\bxi},
\ee
where we used the expression $E_\mu\ket{\Psi_\mu}=\hat{\Omega}^{\bxi}\hat{P}^{\bxi}\hat{H}\hat{\Omega}^{\bxi}\ket{\tilde{\Phi}^{\bxi}_\mu}$. 
As this formalism will ultimately be used for a perturbative treatment of electron correlation (see Sec.~\ref{sec:PT_expansion_RS}), where the KS wavefunctions are used as perturbers, it is natural to split the true Hamiltonian into the eKS one of Eq.~(\ref{eq:KS_Schrodinger}) and a complementary part,
\be\label{eq:decomp_true_Hamil_from_HEKS}
\begin{split}
\hat{H}&=\hat{H}_{\rm KS}^{\bxi}+\left(\hat{W}_{\rm ee}-\hat{V}^{\bxi}_{\rm Hxc}\right)
\\\label{eq:split_true_Hamil_into_KS_plus_W}
&\equiv \hat{H}_{\rm KS}^{\bxi}+\hat{\mathcal{W}}^{\bxi}, 
\end{split}
\ee
with $\hat{V}^{\bxi}_{\rm Hxc}=\int d\br\,v^{\bxi}_{\rm Hxc}(\br)\hat{n}(\br)$, so that we can rewrite Eq.~(\ref{eq:first_step_proof_GBE}) as
\be
\hat{H}\hat{\Omega}^{\bxi}\hat{P}^{\bxi}
=\hat{\Omega}^{\bxi}\hat{H}_{\rm KS}^{\bxi}\hat{P}^{\bxi}+\hat{\Omega}^{\bxi}\hat{P}^{\bxi}\hat{\mathcal{W}}^{\bxi}\hat{\Omega}^{\bxi}\hat{P}^{\bxi},
\ee
where we made the following simplification, according to Eqs.~(\ref{eq:KS_Schrodinger}), (\ref{eq:tilde_projections_def}), (\ref{eq:projector_ortho_basis_KS_model_space}), (\ref{eq:projector_model_KS_space_non-ortho_basis}), and (\ref{eq:def_wave_operator}), 
\be
\hat{P}^{\bxi}\hat{H}_{\rm KS}^{\bxi}\hat{\Omega}^{\bxi}\ket{\tilde{\Phi}^{\bxi}_\mu}=\hat{H}_{\rm KS}^{\bxi}\hat{P}^{\bxi}\hat{\Omega}^{\bxi}\ket{\tilde{\Phi}^{\bxi}_\mu}=\hat{H}_{\rm KS}^{\bxi}\ket{\tilde{\Phi}^{\bxi}_\mu},
\ee
thus leading to the final and standard form of the generalized Bloch equation:
\be\label{eq:GBE_RS-EDFT}
\left[\hat{\Omega}^{\bxi},\hat{H}_{\rm KS}^{\bxi}\right]
\hat{P}^{\bxi}=\left(\hat{\mathcal{W}}^{\bxi}\hat{\Omega}^{\bxi}-\hat{\Omega}^{\bxi}\hat{P}^{\bxi}\hat{\mathcal{W}}^{\bxi}\hat{\Omega}^{\bxi}\right)\hat{P}^{\bxi}.
\ee
Once the above equation has been solved, it does not give immediate access to the true physical (individual-state) wavefunctions because their projections onto the eKS space are still unknown. Nevertheless, if we know the wave operator then we can determine the exact projection of the Schr\"{o}dinger equation onto the eKS space, which then reads as an effective (non-Hermitian) Schr\"{o}dinger equation, according to Eqs.~(\ref{eq:tilde_projections_def}) and (\ref{eq:def_wave_operator}),  
\begin{subequations}
\begin{align}
&\hat{P}^{\bxi}\hat{H}\ket{\Psi_\nu}=E_\nu\hat{P}^{\bxi}\ket{\Psi_\nu}
\\
\label{eq:eff_Hamil_SE}
&\Leftrightarrow 
\hat{H}_{\rm eff}^{\bxi}\ket{\tilde{\Phi}^{\bxi}_\nu}=E_\nu\ket{\tilde{\Phi}^{\bxi}_\nu}
,
\end{align}
\end{subequations}
whose solutions are the desired projections. The effective Hamiltonian $\hat{H}_{\rm eff}^{\bxi}=\hat{P}^{\bxi}\hat{H}\hat{\Omega}^{\bxi}\hat{P}^{\bxi}$, which is indeed fully determined from $\hat{\Omega}^{\bxi}$, can be further simplified as follows, if we use the decomposition of Eq.~(\ref{eq:split_true_Hamil_into_KS_plus_W}), 
\be\label{eq:final_exp_non-Hermitian_Heff}
\hat{H}_{\rm eff}^{\bxi}=\hat{H}_{\rm KS}^{\bxi}\hat{P}^{\bxi}+\hat{P}^{\bxi}\hat{\mathcal{W}}^{\bxi}\hat{\Omega}^{\bxi}\hat{P}^{\bxi},
\ee
or, equivalently,
\be\label{eq:final_working_exp_Heff}
\hat{H}_{\rm eff}^{\bxi}=\hat{P}^{\bxi}\hat{H}\hat{P}^{\bxi}+\hat{P}^{\bxi}\hat{\mathcal{W}}^{\bxi}\left(\hat{\Omega}^{\bxi}-\hat{P}^{\bxi}\right)\hat{P}^{\bxi},
\ee
where we used both the fact that $\hat{P}^{\bxi}$ and $\hat{H}_{\rm KS}^{\bxi}$ commute and the following simplification (see Eq.~(\ref{eq:projector_model_KS_space_non-ortho_basis})), 
\be
\begin{split}
\hat{P}^{\bxi}\hat{\Omega}^{\bxi}\hat{P}^{\bxi}
&=\hat{P}^{\bxi}\sum_{0\leq \mu,\nu<M} \left[\tilde{\bf S}^{-1}\right]_{\mu\nu}\ket{\Psi_\mu}\bra{\tilde{\Phi}^{\bxi}_\nu}
\\
&=\sum_{0\leq \mu,\nu<M} \left[\tilde{\bf S}^{-1}\right]_{\mu\nu}\ket{\tilde{\Phi}^{\bxi}_\mu}\bra{\tilde{\Phi}^{\bxi}_\nu}
\\
&=\hat{P}^{\bxi}
.
\end{split}
\ee
So far we have essentially reviewed the settings of standard RSPT with an unsual choice of (unperturbed) reference Hamiltonian, namely the eKS one. Actually, as further discussed in the following, this reformulation of eDFT sheds another light (different from that of Ref.~\citenum{gould2026ensemblization}, for example) on the decomposition of the ensemble Hxc energy into separate Hartree, exchange, and correlation terms.     

\subsubsection{Revisiting the definition of Hartree, exchange, and correlation ensemble energies}\label{sec:discuss_H-x-c_defs}

A nice feature of the RS-eDFT formulated previously is that it gives directly access to the individual energy levels, in principle exactly. If we adopt, for convenience, the intermediate normalization conditions,
\be
\langle \Psi_\nu\vert \hat{P}^{\bxi}\vert \Psi_\nu\rangle 
&=\langle \tilde{\Phi}^{\bxi}_\nu\vert \tilde{\Phi}^{\bxi}_\nu \rangle=\left[\tilde{{\bf S}}\right]_{\nu\nu}\overset{0\leq \nu <M}{=}1, 
\ee
then the energies simply read, according to Eqs.~(\ref{eq:eff_Hamil_SE}) and (\ref{eq:final_working_exp_Heff}), 
\be\label{eq:exact_ind_ener_exp_from_RS_theory}
\begin{split}
E_\nu&=\langle \tilde{\Phi}^{\bxi}_\nu \vert\hat{H}_{\rm eff}^{\bxi}\vert \tilde{\Phi}^{\bxi}_\nu\rangle
\\
&=\langle \tilde{\Phi}^{\bxi}_\nu \vert\hat{T}\vert \tilde{\Phi}^{\bxi}_\nu\rangle
+\int \dr\, v_{\rm ext}(\br)n_{\tilde{\Phi}^{\bxi}_\nu}(\br)
\\
&\quad+\langle \tilde{\Phi}^{\bxi}_\nu \vert \hat{W}_{\rm ee}\vert \tilde{\Phi}^{\bxi}_\nu\rangle+\langle \tilde{\Phi}^{\bxi}_\nu \vert\hat{\mathcal{W}}^{\bxi}\left(\hat{\Omega}^{\bxi}-\hat{P}^{\bxi}\right)\vert \tilde{\Phi}^{\bxi}_\nu\rangle.
\end{split}
\ee
Interestingly, in the non-interacting limit $\alpha\rightarrow 0^+$ of the theory ($\alpha$ being the two-electron repulsion strength introduced in Eq.~(\ref{eq:KS_ens_non-int-limit})), the correction $\hat{\mathcal{W}}^{\bxi}$ to the eKS Hamiltonian boils down to $\alpha(\hat{W}_{\rm ee}-\hat{V}^{\bxi}_{\rm Hx})$ and the effective Hamiltonian simplifies as follows (see Eq.~(\ref{eq:final_exp_non-Hermitian_Heff})),
\be
\hat{H}_{\rm eff}^{\bxi}\rightarrow \hat{H}_{\rm KS}^{\bxi}\hat{P}^{\bxi}+\alpha\hat{P}^{\bxi}(\hat{W}_{\rm ee}-\hat{V}^{\bxi}_{\rm Hx})\hat{P}^{\bxi},
\ee
thus providing, after diagonalization, the eKS wavefunctions $\left\{{\Phi}^{\bxi}_\nu\right\}$. It becomes clear from the above expression that the latter are not Slater determinants but CSFs instead~\cite{gould2017hartree}. From the standard definition of the ensemble Hx energy (see Eqs.~(\ref{eq:KS_ens_non-int-limit}) and (\ref{eq:_standard_def_ens_Hx})),
\be
{E}^{\bxi}_{\rm Hx}[n^{\bxi}]=\sum_{0\leq \nu<M}\xi_\nu \langle {\Phi}^{\bxi}_\nu \vert\hat{W}_{\rm ee}\vert {\Phi}^{\bxi}_\nu\rangle, 
\ee
and Eq.~(\ref{eq:exact_ind_ener_exp_from_RS_theory}), we finally deduce the following exact expression for the ensemble correlation energy,
\be\label{eq:ens_corr_ener_exp_final_RS-EDFT}
\begin{split}
&E^{\bxi}_{\rm c}[n^{\bxi}]=
\sum _{0\leq \nu<M}\xi_\nu \left(\langle \tilde{\Phi}^{\bxi}_\nu \vert\hat{T}\vert \tilde{\Phi}^{\bxi}_\nu\rangle-\langle {\Phi}^{\bxi}_\nu \vert\hat{T}\vert {\Phi}^{\bxi}_\nu\rangle\right)
\\
&\quad+
\int \dr\, v_{\rm ext}(\br)\left(\sum _{0\leq \nu<M}\xi_\nu n_{\tilde{\Phi}^{\bxi}_\nu}(\br)-n^{\bxi}(\br)\right)
\\
&\quad+
\sum _{0\leq \nu<M}\xi_\nu \left(\langle \tilde{\Phi}^{\bxi}_\nu \vert\hat{W}_{\rm ee}\vert \tilde{\Phi}^{\bxi}_\nu\rangle-\langle {\Phi}^{\bxi}_\nu \vert\hat{W}_{\rm ee}\vert {\Phi}^{\bxi}_\nu\rangle\right)
\\
&\quad+
\sum _{0\leq \nu<M}\xi_\nu \langle \tilde{\Phi}^{\bxi}_\nu \vert\hat{\mathcal{W}}^{\bxi}\left(\hat{\Omega}^{\bxi}-\hat{P}^{\bxi}\right)\vert \tilde{\Phi}^{\bxi}_\nu\rangle.
\end{split}
\ee
The novelty in the above expression lies in the introduction of the auxiliary model eKS wavefunctions $\{\tilde{\Phi}^{\bxi}_\nu\}$, which are the projections of the exact physical states onto the non-interacting eKS space. Thus, we can identify a novel type of ensemble correlation that we refer here to as ensemble {\it projection correlation}. Its contribution to the ensemble energy consists of kinetic, repulsion, and external potential energy corrections (the three first terms on the right-hand side of Eq.~(\ref{eq:ens_corr_ener_exp_final_RS-EDFT})), where the non-interacting eKS wavefunctions are replaced by the model eKS ones. Even though we expect the latter to be individually closer in density to the physical states, which means that density-driven correlations~\cite{PRL19_Gould_DD_correlation,Fromager_2020,gould2026ensemblization} are (partially) included into the ensemble projection correlation, the model eKS wavefunctions have no reason to reproduce the ensemble density exactly, by construction, hence the second term on the right-hand side of Eq.~(\ref{eq:ens_corr_ener_exp_final_RS-EDFT}). Therefore, in the present alternative decomposition of the ensemble correlation energy, the remaining (fourth) term that we refer to as {\it orthogonal correlation} (because it is evaluated, for each state, from the orthogonal complement to the model eKS one) should contain both state- and density-driven correlation energy contributions~\cite{PRL19_Gould_DD_correlation,Fromager_2020,gould2026ensemblization}.\\

At this point it is important to realize that, in the projection/orthogonal correlation terminology we have just introduced, the non-interacting KS ensemble is used as reference, which is standard. Going beyond this description means introducing correlation in this case. We may however question this choice, especially in the present context of the (non-variational) RS-eDFT. Indeed, in the pure ground-state limit of the theory, the eKS space consists of a single state and the KS determinant is the projection of the exact ground state onto that space. If we extend this definition to ensembles, thus considering that interactions within the eKS space relate to Hartree and exchange physics only, then the use of the (non-orthogonal) model eKS states as reference, in place of the non-interacting eKS ones, becomes natural. Interestingly, similar conceptual issues regarding the definition of electron correlation have been encountered in the context of multi-determinant range-separated (e)DFT~\cite{TousrXmd,gori2009range,alam2016ghost}, thus leading to the introduction of a multi-determinant exchange energy. The above choice being made, the ensemble density-functional KS kinetic energy (that we cannot refer to as non-interacting anymore) and the Hx energies are now defined (and denoted with a tilde symbol, to make a clear distinction with the usual definitions) as follows,  
\be
\tilde{T}_{\rm s}^{\bxi}[n^{\bxi}]=\sum_{0\leq \nu<M}\xi_\nu \langle \tilde{\Phi}^{\bxi}_\nu \vert\hat{T}\vert \tilde{\Phi}^{\bxi}_\nu\rangle
\ee
and
\be\label{eq:projected_def_ens_Hx_ener}
\tilde{E}^{\bxi}_{\rm Hx}[n^{\bxi}]=\sum_{0\leq \nu<M}\xi_\nu \langle \tilde{\Phi}^{\bxi}_\nu \vert\hat{W}_{\rm ee}\vert \tilde{\Phi}^{\bxi}_\nu\rangle, 
\ee
respectively, and the complementary ensemble correlation energy reads, according to Eq.~(\ref{eq:exact_ind_ener_exp_from_RS_theory}), 
\be
\begin{split}
\tilde{E}^{\bxi}_{\rm c}[n^{\bxi}]
&=\int \dr\, v_{\rm ext}(\br)\left(\sum _{0\leq \nu<M}\xi_\nu n_{\tilde{\Phi}^{\bxi}_\nu}(\br)-n^{\bxi}(\br)\right)
\\
&\quad+
\sum _{0\leq \nu<M}\xi_\nu \langle \tilde{\Phi}^{\bxi}_\nu \vert\hat{\mathcal{W}}^{\bxi}\left(\hat{\Omega}^{\bxi}-\hat{P}^{\bxi}\right)\vert \tilde{\Phi}^{\bxi}_\nu\rangle
,
\end{split}
\ee
where the discrepancy between the model eKS density and the true ensemble density induces ensemble density-driven correlations (first term on the right-hand side of the above equation) while the remaing contribution is the orthogonal correlation energy introduced previously. As discussed further in the next section, the latter correlation can be described in perturbation theory through a given order. The model eKS wavefunctions will be updated accordingly, through the diagonalization of the effective Hamiltonian.\\

We should finally note that, in the light of Eq.~(\ref{eq:projected_def_ens_Hx_ener}), we may reconsider the definition of the ensemble Hartree and exchange energies, separately. Indeed, if we assume that the model eKS densities are closer to the exact densities than those of the bare non-interacting eKS wavefunctions, because, unlike the latter, they incorporate all exact features of the true individual states within the eKS space, it seems quite natural to evaluate the ensemble Hartree energy as follows,     
\be\label{eq:new_def_ens_Hartree_from_model_EKS_wfs}
\tilde{E}^{\bxi}_{\rm H}[n^{\bxi}]:=\sum_{0\leq \nu<M}\xi_\nu E_{\rm H}\left[n_{\tilde{\Phi}^{\bxi}_\nu}\right],
\ee
where $E_{\rm H}[n]=(1/2)\int d\br\int d\br'\, n(\br)n(\br')/\vert\br -\br'\vert$ is the regular Hartree density functional, which automatically implies the following redefinition of the ensemble exchange energy, according to Eq.~(\ref{eq:projected_def_ens_Hx_ener}),
\be
\tilde{E}^{\bxi}_{\rm x}[n^{\bxi}]:=
\sum_{0\leq \nu<M}\xi_\nu\left(\langle \tilde{\Phi}^{\bxi}_\nu \vert\hat{W}_{\rm ee}\vert \tilde{\Phi}^{\bxi}_\nu\rangle-E_{\rm H}\left[n_{\tilde{\Phi}^{\bxi}_\nu}\right]\right).
\ee
Even though the separation of ensemble Hartree and exchange energies proposed by Gould and coworkers~\cite{gould2026ensemblization} is different for excited states corresponding to neutral excitations, in particular because it relies on the FDT~\cite{PRL20_Gould_Hartree_def_from_ACDF_th}, it is interesting to note that their definition of the ensemble Hartree energy and that of Eq.~(\ref{eq:new_def_ens_Hartree_from_model_EKS_wfs}) have a key feature in common. Both can be expressed in terms of the individual non-interacting eKS wavefunctions or, to be more precise, in terms of their densities {\it and} their couplings through the density operator. The above redefinitions are expected to be useful if the model eKS wavefunctions can be evaluated in practice. Perturbation theory is invoked in the following for that purpose.   

\subsubsection{Perturbation expansions and ensemble density constraint}\label{sec:PT_expansion_RS}

Following regular RSPT~\cite{Lindgren1986}, we consider the RS-eDFT derived previously in the particular case of partially-interacting electrons, \ie,   
\be\label{eq:splitting_true_Hamilt_into_KS_plus_pert}
\hat{H}\rightarrow \hat{H}(\alpha)=\hat{H}_{\rm KS}^{\bxi}+\hat{\mathcal{W}}^{\bxi}(\alpha),
\ee
where the interaction strength $\alpha$ is assumed to be close to zero, such that the perturbation operator can be Taylor-expanded as follows,
\be\label{eq:perturb_in_RSEDFPT}
\hat{\mathcal{W}}^{\bxi}(\alpha)=\alpha\left(\hat{W}_{\rm ee}-\hat{V}_{\rm Hx}^{\bxi}\right)-\hat{V}_{\rm c}^{\bxi}(\alpha),
\ee
the ensemble correlation potential being itself expanded from second order:
\be
\hat{V}_{\rm c}^{\bxi}(\alpha)
=\alpha^2\hat{V}_{\rm c}^{\bxi(2)}+\alpha^3\hat{V}_{\rm c}^{\bxi(3)}+\ldots
\ee
The perturbation expansion of the wave operator 
\be
\hat{\Omega}^{\bxi}\rightarrow \hat{\Omega}^{\bxi}(\alpha)=\hat{P}^{\bxi}+\alpha\hat{\Omega}^{\bxi(1)}+
\alpha^2\hat{\Omega}^{\bxi(2)}+\ldots
\ee
is determined order by order from the generalized Bloch Eq.~(\ref{eq:GBE_RS-EDFT}), thus giving access to the model eKS wavefunctions $\tilde{\Phi}^{\bxi}_\nu(\alpha)$ through the diagonalization of the effective Hamiltonian (see Eq.~(\ref{eq:final_exp_non-Hermitian_Heff}))
\be
\hat{H}_{\rm eff}^{\bxi}(\alpha)=\hat{H}_{\rm KS}^{\bxi}\hat{P}^{\bxi}+\hat{P}^{\bxi}\hat{\mathcal{W}}^{\bxi}(\alpha)\hat{\Omega}^{\bxi}(\alpha)\hat{P}^{\bxi}.
\ee
Thus, we can obtain the perturbation expansion of the true wavefunctions, individually, 
\be
\ket{\Psi_\nu}\rightarrow \ket{\Psi_\nu(\alpha)}=\hat{\Omega}^{\bxi}(\alpha)\ket{\tilde{\Phi}^{\bxi}_\nu(\alpha)},
\ee
from which the individual densities can be evaluated as follows,
\be\label{eq:ind_density_expression_with_normalization}
\begin{split}
n_{\Psi_\nu}(\br)&\rightarrow n_{\Psi_\nu(\alpha)}(\br)=
\dfrac{\langle \Psi_\nu(\alpha)\vert \hat{n}(\br)\vert \Psi_\nu(\alpha)\rangle}{\langle \Psi_\nu(\alpha)\vert \Psi_\nu(\alpha)\rangle}
\\
&\quad =\dfrac{\langle \tilde{\Phi}^{\bxi}_\nu(\alpha) \vert \hat{\Omega}^{\bxi}(\alpha)^\dagger \hat{n}(\br)\hat{\Omega}^{\bxi}(\alpha)\vert \tilde{\Phi}^{\bxi}_\nu(\alpha)\rangle}{\langle \tilde{\Phi}^{\bxi}_\nu(\alpha) \vert \hat{\Omega}^{\bxi}(\alpha)^\dagger\hat{\Omega}^{\bxi}(\alpha)\vert \tilde{\Phi}^{\bxi}_\nu(\alpha)\rangle}
.
\end{split}
\ee
A major difference with regular RSPT is that the perturbation operator or, more precisely, the perturbative expansion of the ensemble Hxc potential, is generally unknown. Nonetheless, the latter can in principle be determined order by order from the {\it ensemble} density constraint,
\be
\sum_{0\leq \nu<M}\xi_\nu n_{\Psi_\nu(\alpha)}(\br)=\sum_{0\leq \nu<M}\xi_\nu n_{\Phi^{\bxi}_\nu}(\br), \,\forall \alpha,
\ee
thus reminding us that the non-interacting eKS wavefunctions still play a key role in the theory, as they define the model eKS space from which all physical properties are evaluated. Those are recovered (approximately) through a given order of perturbation, by setting $\alpha=1$, ultimately.\\ 

The above quasi-degenerate flavor of ensemble GLPT~\cite{Yang2021_Second} raises two technical questions. The first one relates to the non-orthogonality of the model eKS solutions $\{\tilde{\Phi}_\nu^{\bxi}\equiv\tilde{\Phi}_\nu^{\bxi}(\alpha=1)\}$ that are our new reference eKS wavefunctions. Indeed, these {\it effective} KS wavefunctions are, by construction, the projections of the exact solutions to the Schr\"{o}dinger equation onto the eKS space (see Eq.~(\ref{eq:tilde_projections_def})). Unlike the original non-interacting eKS wavefunctions, they are eigenfunctions of a non-Hermitian effective Hamiltonian (see Eqs.~(\ref{eq:eff_Hamil_SE}) and (\ref{eq:final_exp_non-Hermitian_Heff})). Even though we expect the latter to be ``closer'' to the true solutions than the former, the non-orthogonality of these effective KS wavefunctions might be questionable. The second (related) question that arises is about the non-unitary character of the wave operator, which implies dealing with the perturbation expansion of the normalization factor in Eq.~(\ref{eq:ind_density_expression_with_normalization}). Even though this is manageable (see Ref.~\citenum{srDFT_densitymatrixformulation}, for example), turning to a unitary theory, in the spirit of VV perturbation theory~\cite{VanVleck29_On,SOKOLOV2024121}, could possibly ease its formulation and implementation. These points are discussed in the next section.     

\subsubsection{Unitary formulation}\label{sec:mrpt_eDFT-unitary_formulation}

Let us consider a unitary operator $\hat{\mathcal{U}}^{\bxi}$
that connects the non-interacting eKS solutions to the exact physical ones $\{{\Psi}_\nu\}_{0\leq \nu<M}$, as follows, 
\be\label{eq:overline_wfs_from_unit_transf}
\ket{\overline{\Psi}_\nu}=\hat{\mathcal{U}}^{\bxi}\ket{\Phi^{\bxi}_\nu},\,\forall\nu\geq 0,
\ee
where the $M$ first (orthonormal) interacting wavefunctions $\{\overline{\Psi}_\nu\}_{0\leq \nu<M}$ are not necessarily the solutions to the Schr\"{o}dinger equation but they form an orthonormal basis of the true ensemble subspace: 
\be\label{eq:overline_and_true_spaces_match}
\sum_{0\leq \nu<M}\ket{\overline{\Psi}_\nu}\bra{\overline{\Psi}_\nu}=\sum_{0\leq \nu<M}\ket{\Psi_\nu}\bra{\Psi_\nu}.
\ee
Note that $\hat{\mathcal{U}}^{\bxi}$ can be written explicitly as follows, 
\be
\hat{\mathcal{U}}^{\bxi}\equiv\sum_{\mu\geq 0}\ket{\overline{\Psi}_\mu}\bra{\Phi^{\bxi}_\mu}.
\ee
A direct consequence of Eq.~(\ref{eq:overline_and_true_spaces_match}) is the strict decoupling (energy wise) between the unitary transformed eKS subspace and its orthogonal complement, \ie,
\be\label{eq:decoupling_unitary_PT}
\langle {\overline{\Psi}_\mu}\vert \hat{H}\vert {\overline{\Psi}_\nu}\rangle\underset{\mu\geq M}{\underset{0\leq \nu<M }{=}}0, 
\ee
or, equivalently,
\be\label{eq:decoupling_unitary_PT_in_terms_of_U}
\langle \Phi^{\bxi}_\mu\vert (\hat{\mathcal{U}}^{\bxi})^\dagger \hat{H}\hat{\mathcal{U}}^{\bxi}\vert \Phi^{\bxi}_\nu\rangle\underset{\mu\geq M}{\underset{0\leq \nu<M }{=}}0. 
\ee
As a result, we can focus on the targeted ensemble subspace or, equivalently, on the following {\it Hermitian} effective Hamiltonian, 
\be
\hat{\mathcal{H}}_{\rm eff}^{\bxi}=\hat{P}^{\bxi}(\hat{\mathcal{U}}^{\bxi})^\dagger \hat{H}\hat{\mathcal{U}}^{\bxi}\hat{P}^{\bxi},
\ee
which is restricted to the eKS subspace (see Eq.~(\ref{eq:projector_ortho_basis_KS_model_space})) and 
whose matrix elements read, according to Eq.~(\ref{eq:overline_wfs_from_unit_transf}),
\be
\langle \Phi^{\bxi}_\mu \vert \hat{\mathcal{H}}_{\rm eff}^{\bxi} \vert \Phi^{\bxi}_\nu \rangle {\underset{0\leq \mu,\nu<M }{=}} \langle \overline{\Psi}_\mu \vert\hat{H}\vert \overline{\Psi}_\nu\rangle. 
\ee
The above equation gives immediately the solutions to the diagonalization of the effective Hamiltonian matrix, \ie, 
\be
\sum_{0\leq \mu,\nu<M} \left(\mathcal{V}_{\mu\lambda}^{\bxi}\right)^*\mathcal{V}_{\nu\kappa}^{\bxi} 
\langle \Phi^{\bxi}_\mu \vert \hat{\mathcal{H}}_{\rm eff}^{\bxi} \vert \Phi^{\bxi}_\nu \rangle{\underset{0\leq \lambda,\kappa<M }{=}}\delta_{\lambda\kappa}E_\kappa,
\ee
where
\be
\mathcal{V}_{\mu\nu}^{\bxi}{\underset{0\leq \mu,\nu<M }{=}}
\langle \overline{\Psi}_\mu\vert \Psi_\nu\rangle, 
\ee
which translates as follows in terms of operators and wavefunctions,
\be\label{eff_hermitian_eigenvalue_prob}
\hat{\mathcal{H}}_{\rm eff}^{\bxi}\ket{\overline{\Phi}^{\bxi}_\kappa}=E_\kappa \ket{\overline{\Phi}^{\bxi}_\kappa},\, 0\leq \kappa<M,
\ee
with
\be
\ket{\overline{\Phi}^{\bxi}_\kappa}
=\hat{\mathcal{V}}^{\bxi}\ket{\Phi^{\bxi}_\kappa},
\ee
and
\be\label{eq:Vxi_op_full_expansion}
\begin{split}
\hat{\mathcal{V}}^{\bxi}&:=\sum_{0\leq \mu,\nu<M}\mathcal{V}_{\mu\nu}^{\bxi}\ket{\Phi^{\bxi}_\mu}\bra{\Phi^{\bxi}_\nu}
\\
&=\sum_{0\leq \mu,\nu<M}\langle \overline{\Psi}_\mu\vert \Psi_\nu\rangle\ket{\Phi^{\bxi}_\mu}\bra{\Phi^{\bxi}_\nu}
\end{split}
\ee
being unitary within the eKS space.\\

At this point it is important to realize that, in this unitary formulation, the effective Hamiltonian and, consequently, its diagonalizers are not uniquely defined, even though the corresponding eigenvalues (\ie, the true energies) are. The reason is that the targeted ensemble subspace admits an infinite number of orthonormal bases $\{\overline{\Psi}_\nu\}_{0\leq \nu<M}$. On the other hand, applying the transformation $\hat{\mathcal{V}}^{\bxi}$ to the ($M$ lowest-in-energy) eKS solutions and then applying the decoupling unitary transformation $\hat{\mathcal{U}}^{\bxi}$ uniquely defines the wave operator that, when applied to the latter solutions, delivers the exact eigenfunctions and energies, \ie,   
\be\label{eq:unique_wave_op_through_UV}
\begin{split}
\hat{\mathcal{U}}^{\bxi}\hat{\mathcal{V}}^{\bxi}
&=\sum_{0\leq \mu,\nu<M}\mathcal{V}_{\mu\nu}^{\bxi}\ket{\overline{\Psi}_\mu}\bra{\Phi^{\bxi}_\nu}
\\
&=\sum_{0\leq \nu<M}\ket{\Psi_\nu}\bra{\Phi^{\bxi}_\nu},
\end{split}
\ee
and
\be\label{eq:unitary_wave_op_applied}
\ket{\Psi_\nu}\underset{0\leq \nu<M}{=}\hat{\mathcal{U}}^{\bxi}\hat{\mathcal{V}}^{\bxi}\ket{\Phi^{\bxi}_\nu}=\hat{\mathcal{U}}^{\bxi}\ket{\overline{\Phi}^{\bxi}_\nu}.
\ee
In other words, the eigenfunctions $\{\overline{\Phi}^{\bxi}_\kappa\}_{0\leq \kappa<M}$ of the effective Hamiltonian are intermediate solutions that depend on the choice made for $\hat{\mathcal{U}}^{\bxi}$. Obviously, the decoupling constraint of Eq.~(\ref{eq:decoupling_unitary_PT}) allows for an infinite number of unitary transformations. For example, any additional rotation within the targeted physical ensemble subspace,
\be
\hat{\mathcal{R}}^{\bxi}=\sum_{0\leq \nu<M}\ket{\check{\Psi}_\nu}\bra{\overline{\Psi}_\nu}+\sum_{\nu\geq M}\ket{\overline{\Psi}_\nu}\bra{\overline{\Psi}_\nu},
\ee
where
\be
\sum_{0\leq \nu<M}\ket{\check{\Psi}_\nu}\bra{\check{\Psi}_\nu}=
\sum_{0\leq \nu<M}\ket{\overline{\Psi}_\nu}\bra{\overline{\Psi}_\nu},
\ee
is acceptable. In this case, the effective Hamiltonian and its matrix elements are modified as follows,
\be
\hat{\mathcal{H}}_{\rm eff}^{\bxi}\rightarrow \hat{P}^{\bxi}(\hat{\mathcal{U}}^{\bxi})^\dagger (\hat{\mathcal{R}}^{\bxi})^\dagger\hat{H}\hat{\mathcal{R}}^{\bxi}\hat{\mathcal{U}}^{\bxi}\hat{P}^{\bxi}
\ee
and
\be
\langle \Phi^{\bxi}_\mu \vert \hat{\mathcal{H}}_{\rm eff}^{\bxi} \vert \Phi^{\bxi}_\nu \rangle {\underset{0\leq \mu,\nu<M }{\rightarrow}} \langle \check{\Psi}_\mu \vert\hat{H}\vert \check{\Psi}_\nu\rangle, 
\ee
respectively. Therefore, according to Eq.~(\ref{eq:Vxi_op_full_expansion}), the transformation $\hat{\mathcal{V}}^{\bxi}$ that diagonalizes the effective Hamiltonian becomes 
\be
\hat{\mathcal{V}}^{\bxi}\rightarrow 
\sum_{0\leq \mu,\nu<M}
\langle \overline{\Psi}_\mu\vert (\hat{\mathcal{R}}^{\bxi})^\dagger \vert\Psi_\nu\rangle
\ket{\Phi^{\bxi}_\mu}\bra{\Phi^{\bxi}_\nu},
\ee
where the impact of the rotation on the diagonalizers can now be readily seen. However, as expected, the composition of the full unitary transformation $\hat{\mathcal{R}}^{\bxi}\hat{\mathcal{U}}^{\bxi}$ with the diagonalization (see Eq.~(\ref{eq:unique_wave_op_through_UV})) is invariant under the rotation $\hat{\mathcal{R}}^{\bxi}$:   
\be
\begin{split}
\hat{\mathcal{U}}^{\bxi}\hat{\mathcal{V}}^{\bxi}
&\rightarrow\hat{\mathcal{R}}^{\bxi}\sum_{0\leq \mu,\nu<M}
\langle \overline{\Psi}_\mu\vert (\hat{\mathcal{R}}^{\bxi})^\dagger \vert\Psi_\nu\rangle
\ket{\overline{\Psi}_\mu}\bra{\Phi^{\bxi}_\nu} 
\\
&\quad=\sum_{0\leq \nu<M}\ket{\Psi_\nu}\bra{\Phi^{\bxi}_\nu}
\\
&\quad=\hat{\mathcal{U}}^{\bxi}\hat{\mathcal{V}}^{\bxi}.
\end{split}
\ee
In summary, even though it is technically possible to work with orthonormal effective eKS solutions, they are not uniquely defined, unlike in the RS-eDFT introduced in Sec.~\ref{eq:ensemble_RS_theory}. In the latter case, the unicity comes from the projection of the exact solutions onto the eKS subspace. While the non-uniqueness prevents us from proposing alternative and clear definitions of the ensemble Hartree and exchange energies based on orthonormal effective eKS wavefunctions, for example, this is not a problem when it comes to perform practical perturbative calculations. Indeed, while the perturbation expansion of $\hat{\mathcal{U}}^{\bxi}$ would be deduced order by order from Eq.~(\ref{eq:decoupling_unitary_PT_in_terms_of_U}) and the ensemble density constraint simplified as follows (see Eq.~(\ref{eq:unitary_wave_op_applied})),
\be\label{eq:ens_dens_constraint_unitary_theory}
\begin{split}
&\sum_{0\leq \nu<M}\xi_\nu\langle \overline{\Phi}^{\bxi}_\nu \vert (\hat{\mathcal{U}}^{\bxi})^\dagger\hat{n}(\br)\hat{\mathcal{U}}^{\bxi} \vert \overline{\Phi}^{\bxi}_\nu\rangle
\\
&= \sum_{0\leq \nu<M}\xi_\nu n_{\Phi^{\bxi}_\nu}(\br),
\end{split}
\ee
which is the true benefit of a unitary formulation (when comparison is made with Eq.~(\ref{eq:ind_density_expression_with_normalization})), 
the targeted energy levels would be determined from the following exact expression (see Eq.~(\ref{eff_hermitian_eigenvalue_prob})),
\be
E_\nu=\langle \overline{\Phi}^{\bxi}_\nu \vert (\hat{\mathcal{U}}^{\bxi})^\dagger\hat{H}\hat{\mathcal{U}}^{\bxi}\vert \overline{\Phi}^{\bxi}_\nu\rangle. 
\ee
For completeness, the practical VV perturbative formulation of such a theory is briefly sketched in Appendix~\ref{appendix:VV-EDFT}.

\subsection{Quantum embedding of excited states through the eDFT formalism}\label{sec:ensemble_LPFET}

\subsubsection{Motivation}

Density matrix embedding theory (DMET)~\cite{knizia2012density,wouters2016practical,Verma26_Multireference_Embedding} has emerged over the last decade as a promising quantum embedding strategy for dealing, in particular, with the description of strong electron correlations. While DMET has been mostly (but not exclusively~\cite{tran2019using,ye2021accurate,mitra2021excited,chen2014intermediate,booth2015spectral,Verma2023_Optical,Lau2024_Optical,Ai2022_Efficient,Li2026_Towards-excitations}) developed for ground states, its extension to low-lying excited states through the ensemble formalism is an appealing strategy that has been successfully tested on prototypical systems in Ref.~\citenum{cernatic2024fragment}. In the latter seminal work, a single-shot (\ie, at the zeroth iteration of the self-consistency loop described for ground states in Sec.~\ref{sec:GS_DFET}) embedding of single orbitals has been used. A general (\ie, applicable to multiple-orbital embeddings) and formally exact formulation of the method has not been derived yet. In the following, we will highlight some of the key features of such a theory (its full derivation and implementation is left for future work), with a particular focus on the design of quantum baths (in which fragments of the system under study are embedded) for ensembles. Our starting point will be the formulation (in Sec.~\ref{sec:lattice_DFT}) of DFT in a basis of (arbitrarily chosen) orthonormal localized orbitals, from which a lattice-like density-functional embedding theory can be designed (see Sec.~\ref{sec:GS_DFET}). In Sec.~\ref{sec:gs_embedding_KS}, we give an intuitive introduction to quantum baths for pure ground states, thus allowing for a straightforward extension to ensembles in Sec.~\ref{sec:ensemble_embedding}. This last step is a comprehensive generalization to multiple-orbital fragments of the single-orbital ensemble embedding implemented in Ref.~\citenum{cernatic2024fragment} through successive unitary Householder transformations, which are applied to the ensemble one-electron reduced density matrix (1RDM). By using, ultimately, the ensemble density only (\ie, the localized orbitals ensemble occupation in this context) as basic variable, the resulting ensemble quantum embedding theory will benefit from the in-principle exactness of eDFT reviewed in Sec.~\ref{sec:review_section_NcEDFT}.

\subsubsection{Lattice formulation of DFT}\label{sec:lattice_DFT}

In DMET, the electronic structure problem is formulated in an orthonormal basis of {\it localized} orbitals $\left\{\chi_p({\bf r})\right\}$. The representation of operators or reduced density matrices in that basis will be referred to as lattice representation in the following, by analogy with the lattice models (such as the Hubbard model) that are popular in condensed matter physics. Such a choice of representation allows for a fragmentation of the system under study, each fragment being ultimately embedded into a so-called quantum bath, so that strong local electron correlations can be treated adequately. Note that the general formulation of the present quantum embedding theory does not depend on the orbital localization scheme. The connection between the real space representation used in Eqs.~(\ref{eq:kinetic_ener_op_SQ})--(\ref{eq:ext_pot_operator_sq}) and the lattice one reads as follows, in second quantization, 
\be
\hat{\Psi}_{\sigma}({\bf r})=\sum_p \chi_p({\bf r})\hat{c}_{p\sigma},
\ee
so that the physical {\it ab initio} Hamiltonian of Eq.~(\ref{eq:physical_Hamil}) can be equivalently rewritten in the lattice representation as (from now on we use real algebra, for simplicity),
\begin{equation}
    \begin{split}
\hat{H}&=\sum_\sigma\sum_{pq}h_{pq}\hat{c}_{p\sigma}^\dagger\hat{c}_{q\sigma}
\\
&\quad+\dfrac{1}{2}\sum_{\sigma,\sigma'}\sum_{pqrs}\langle pq\vert rs\rangle \hat{c}_{p\sigma}^\dagger\hat{c}_{q\sigma'}^\dagger\hat{c}_{s\sigma'}\hat{c}_{r\sigma},
\end{split}
\end{equation}
where the one- and two-electron integrals read (in real algebra)
\be
h_{pq}=\int d{\bf r}\,\chi_p({\bf r})\left(-\dfrac{\nabla^2_{\bf r}}{2}+v_{\rm ext}({\bf r})\right)\chi_q({\bf r})
\ee
and
\be
\langle pq\vert rs\rangle=\int d{\bf r}\int d{\bf r}'\,\dfrac{\chi_p({\bf r})\chi_q({\bf r}')\chi_r({\bf r})\chi_s({\bf r}')}{\vert {\bf r}-{\bf r}'\vert},
\ee
respectively.\\

In order to turn DMET (or, more precisely, its density embedding theory (DET) flavors~\cite{bulik2014density,bulik2014electron,fulde2017dealing,plat2020entanglement,mordovina2019self,sekaran2022local}) into a formally exact theory, in the sense of lattice DFT, the Hamiltonian can be re-decomposed as follows~\cite{makhlouf2025_local_potential},  
\be
\hat{H}=\hat{\mathcal{T}}+\hat{W}_{\rm ee}+\hat{\mathcal{V}}^{\rm ext},
\ee
where the off-diagonal contributions to the full {\it ab initio} one-electron Hamiltonian,     
\be
\hat{\mathcal{T}}:=\sum_\sigma\sum_{p\neq q}h_{pq}\hat{c}_{p\sigma}^\dagger\hat{c}_{q\sigma},
\ee
play the role of a hopping operator (the analog of the kinetic energy operator in a lattice model~\cite{DFT_ModelHamiltonians}) while the diagonal terms
\be
\hat{\mathcal{V}}^{\rm ext}:=\sum_{p} h_{pp} \hat{n}_p,
\ee
with
\be 
\hat{n}_p=\sum_\sigma \hat{c}_{p\sigma}^\dagger\hat{c}_{p\sigma},
\ee
play the role of the external local potential in the lattice. In this context, the KS Hamiltonian reads
\be
\hat{\mathcal{H}}^{\rm KS}=\hat{\mathcal{T}}+\hat{\mathcal{V}}^{\rm ext}+ \hat{\mathcal{V}}^{\rm Hxc},
\ee
where the Hxc potential operator
\be\label{eq:Hxc_pot_op_lattice}
\hat{\mathcal{V}}^{\rm Hxc}=\sum_{p}v_p^{\rm Hxc}\hat{n}_p,
\ee
which is {\it local} in the lattice representation, is such that the $N$-electron ground state of the lattice KS Hamiltonian  
\be\label{eq:lattice_KS_det_def}
\Phi_{\rm s}=\argmin_{\Psi\rightarrow N}\langle\Psi \vert\hat{\mathcal{H}}^{\rm KS}\vert\Psi\rangle
\ee
reproduces the exact physical (Full Configuration Interaction, in practical calculations where finite basis sets are used) localized orbitals ground-state occupation:
\be\label{eq:density_mapping_lattice_DFT}
\langle\Phi_{\rm s}\vert\hat{n}_p\vert\Phi_{\rm s}\rangle\overset{!}{=}\langle\Psi_0\vert\hat{n}_p\vert\Psi_0\rangle,\;\forall p. 
\ee
In the following, we refer to the above orbital occupations as the density profile (or just density), by analogy with regular real-space DFT (see Eqs.~(\ref{eq:dens_op_SQ}) and (\ref{eq:dens_Psi_from_dens_op})).
  
\subsubsection{KS density-functionalization of quantum embedding theory}\label{sec:GS_DFET}

The basic idea of quantum embedding theory is to evaluate the orbital occupation in the right-hand side of Eq.~(\ref{eq:density_mapping_lattice_DFT}) {\it locally}, \ie, by solving a drastically reduced-in-size (in terms of Hilbert space size) Schr\"{o}dinger equation for an embedded fragment $F$ (that contains the localized orbital of interest $\chi_p$) of the full system. To be more precise, what we refer to as ``fragment'' can be a single localized orbital or a collection of localized orbitals, \ie, $F\equiv \left\{\chi_p\right\}_{p\in F}$. Its dimension is arbitrarily chosen and depends on the system under study. Moreover, and this is the most crucial point, the recycling of conventional (pure-state) wave-function-based quantum chemistry methods in this context is only possible if the fragment $F$ (which is an open quantum system) is complemented by a so-called quantum bath $B$ that plays the role of an electronic reservoir. The resulting one-electron Hilbert space $C=F\oplus B$, which is referred to as embedding cluster, is the proper playground for evaluating exactly ({\it via} appropriate density-functional potential corrections to the true many-body Hamiltonian's projection onto the cluster~\cite{makhlouf2025_local_potential}) or approximately the local density profile 
\be\label{eq:mapping_local_dens_cluster}
\langle\Psi_0\vert\hat{n}_p\vert\Psi_0\rangle\overset{p\in F}{\equiv} \langle\Psi_0^{C}\vert\hat{n}_p\vert\Psi^{C}_0\rangle,
\ee
where $\Psi_0^{C}$ is the few-electron ground-state solution to the (reduced-in-size) cluster's Schr\"{o}dinger equation. As discussed in further detail in the following, it is convenient (practically but also formally) to design the bath from the lattice KS system, thus making the local density profile a functional of the (to-be-determined) local Hxc potential ${\bf v}^{\rm Hxc}=\left\{v_p^{\rm Hxc}\right\}$ over the full lattice (see Eq.~(\ref{eq:Hxc_pot_op_lattice})):
\be
\Psi_0^{C}\equiv \Psi_0^{C}\left({\bf v}^{\rm Hxc}\right). 
\ee
Since the lattice KS determinant $\Phi_{\rm s}\equiv \Phi_{\rm s}\left({\bf v}^{\rm Hxc}\right)$ introduced in Eq.~(\ref{eq:lattice_KS_det_def}) is itself a functional of the Hxc potential, the density constraint of Eq.~(\ref{eq:density_mapping_lattice_DFT}) provides the self-consistency loop from which ${\bf v}^{\rm Hxc}$ can be determined, by collecting the local density constraints associated with each embedded fragment (see Eq.~(\ref{eq:mapping_local_dens_cluster})), \ie, 
\be\label{eq:dens_mapp_constraint_LPFET}
\langle\hat{n}_p\rangle_{\Phi_{\rm s}\left({\bf v}^{\rm Hxc}\right)}\underset{p\in F}{\overset{!}{=}}\langle\hat{n}_p\rangle_{\Psi_0^{C}\left({\bf v}^{\rm Hxc}\right)},\;\forall F, 
\ee
where we used the shorthand notation $\langle \ldots\rangle_\Psi=\langle\Psi\vert \ldots\vert\Psi\rangle$. The recently derived LPFET~\cite{makhlouf2025_local_potential} (see also its generalized variant~\cite{makhlouf2026generalizedlocalpotentialfunctional}, in the sense of generalized KS theory) is a practical realization of the above equation. In the rest of the paper we will focus on the construction of the quantum bath and discuss its extension to ensembles.

\subsubsection{Ground-state density-functional quantum bath}\label{sec:gs_embedding_KS}

In practical quantum embedding calculations the bath construction usually relies on a (pure ground-state) full-size single-determinant reference wavefunction (the lattice KS determinant $\Phi_{\rm s}$, in the present context). In this special case, the idempotency of the corresponding 1RDM can be exploited to prove the exactness of the clusterization procedure (\ie, the perfect disentanglement of the cluster from its environment~\cite{sekaran2023unified}). Even though the clusterization is not exact anymore for correlated electrons~\cite{sekaran2021householder}, it can still be exactified density-wise, in the sense of DFT~\cite{makhlouf2025_local_potential}. Most importantly for our purpose, the idempotency of the 1RDM is lost as soon as we consider an ensemble state (rather than a pure ground state), even when dealing with non-interacting electrons. This is a consequence of the fractional occupation of the EKS orbitals that is controlled by the ensemble weights (see Eq.~(\ref{eq:ens_dens_frac_occ_EKS_orbs})). As a result, designing a quantum bath for an ensemble of pure ground and excited states is nontrivial~\cite{cernatic2024fragment}. In the following, we propose an alternative (but equivalent) approach to ground-state quantum baths whose ``ensemblization'' (according to the terminology of Ref.~\onlinecite{gould2026ensemblization}) can be deduced straightforwardly, thus paving the way toward an ensemble LPFET.\\

Let us start from the following expression of the $N$-electron lattice KS determinant (that we assume to be closed-shell, for simplicity), 
\be
\Phi_{\rm s}\equiv \vert\varphi_1^2\varphi_2^2\ldots \varphi^2_{\frac{N}{2}}\vert, 
\ee
where, according to Eq.~(\ref{eq:lattice_KS_det_def}), the occupied KS orbitals are the (lowest-in-energy) solutions to the one-electron KS equation,
\be
\hat{h}^{\rm KS}\vert \varphi_k\rangle=\varepsilon_k\vert \varphi_k\rangle,
\ee
$\hat{h}^{\rm KS}=\sum_{pq}(h_{pq}+\delta_{pq}v^{\rm Hxc}_p)\vert\chi_p\rangle\langle \chi_q\vert$ being the one-electron lattice KS Hamiltonian. For a given fragment $F$ of interest, we can project each localized orbital belonging to $F$ onto the subspace of the occupied KS orbitals in $\Phi_{\rm s}$:
\be\label{eq:projection_fragment_orb_into_occ}
\chi_p(\br)\underset{p\in F}{\rightarrow} \chi^{\rm occ}_p(\br)=\sum^{N/2}_{k=1}\langle \varphi_k\vert\chi_p\rangle\varphi_k(\br),
\ee
where $\langle \varphi_k\vert\chi_p\rangle=\int d\br'\varphi_k(\br')\chi_p(\br')$ describes the overlap of the $k$th occupied orbital with the fragment's orbital $\chi_p$. An orthonormal basis of the resulting one-electron subspace that we refer to as the subspace of {\it fragment-occupied orbitals} and whose dimension is (at most) that of the fragment is obtained as follows,  
\be\label{eq:final_orthonormal_frag-occ_orbitals}
\left\{\tilde{\varphi}_p(\br)=\sum_{q\in F}\left[{\bf S}^{-\frac{1}{2}}\right]_{pq}\chi^{\rm occ}_q(\br)
\right\}_{p\in F},
\ee
where
\be
{\bf S}\equiv\left\{\langle \chi^{\rm occ}_p\vert \chi^{\rm occ}_q\rangle\right\}_{p\in F,q\in F} 
\ee
is the overlap matrix between the projected fragment orbitals. In the following, we will assume (and this is usually the case~\cite{sekaran2023unified}) that the fragment orbitals remain linearly independent after projection onto the occupied orbitals subspace, which is equivalent to assuming that the eigenvalues of ${\bf S}$ are strictly positive since, for any column vector ${\bf X}\equiv \left\{X_p\right\}_{p\in F}$,
\be
\begin{split}
{\bf X}^\dagger{\bf S}{\bf X}
&=
\sum_{p\in F,q\in F}X_pS_{pq}X_q
\\
&\quad=
\left\langle \sum_{p\in F} X_p \chi^{\rm occ}_p \middle\vert \sum_{q\in F} X_q \chi^{\rm occ}_q\right\rangle>0.
\end{split}
\ee
The so-called {\it core orbitals} are then constructed as the orthonormal complement to the fragment-occupied orbitals within the occupied orbitals subspace:
\be\label{eq:decomp_occ_space_frag-occ_plus_core}
\left\{\varphi_k\right\}_{1\leq k\leq \frac{N}{2}}
=\left\{\tilde{\varphi}_p\right\}_{p\in F}
\oplus \left\{\tilde{\varphi}_q\right\}_{q\in {\rm core}}.
\ee
In practice, they can be computed by projecting out of the fragment-occupied orbitals subspace the original occupied lattice KS orbitals $\varphi_k$ (we could take the $\frac{N}{2}-N_{\rm orb}^{F}$ lowest-in-energy ones, for example, where $N_{\rm orb}^{F}$ denotes the total number of orbitals in the fragment $F$), and orthonormalizing, like in Eq.~(\ref{eq:final_orthonormal_frag-occ_orbitals}), \ie, 
\begin{subequations}
\begin{align}
&\left\{\varphi_k(\br)\right\}_{1\leq k\leq \frac{N}{2}-N_{\rm orb}^{F}}
\\\label{subeq:projecting_out_frag-occ}
&\rightarrow 
\left\{
\varphi_k(\br)-\sum_{p\in F}\langle \tilde{\varphi}_p\vert \varphi_k\rangle \tilde{\varphi}_p(\br)
\right\}_{1\leq k\leq \frac{N}{2}-N_{\rm orb}^{F}}
\\
&\overset{\rm orthon.}{\rightarrow}  \left\{\tilde{\varphi}_q\right\}_{q\in {\rm core}}. 
\end{align}
\end{subequations}

An important observation to make at this point is the fact that the core orbitals have {\it no overlap with the fragment}. This is a key motivation (in addition to the locality of the strong correlations we are interested in) for proceeding with the embedding of interacting electrons along the same lines. Indeed, by construction (see Eqs.~(\ref{eq:projection_fragment_orb_into_occ}) and (\ref{eq:decomp_occ_space_frag-occ_plus_core})), 
\be\label{eq:core-fragment_no_overlap}
\begin{split}
\langle \chi_p  \vert \tilde{\varphi}_q \rangle
&\underset{q\in {\rm core}}{\overset{p\in F}{=}}
\sum^{\frac{N}{2}}_{k=1}
\langle \chi_p \vert \varphi_k \rangle \langle \varphi_k \vert \tilde{\varphi}_q\rangle
\\
&\quad=\langle \chi^{\rm occ}_p \vert \tilde{\varphi}_q\rangle 
\\
&\quad=\sum_{r\in F}\left[{\bf S}^{\frac{1}{2}}\right]_{pr}\langle \tilde{\varphi}_r \vert \tilde{\varphi}_q\rangle 
\\
&\quad=0,
\end{split}
\ee
where we used the relation
\be
\sum_{p\in F}\left[{\bf S}^{\frac{1}{2}}\right]_{rp}\tilde{\varphi}_p(\br)
=\chi^{\rm occ}_r(\br)
\ee
that is deduced from Eq.~(\ref{eq:final_orthonormal_frag-occ_orbitals}).
Eq.~(\ref{eq:core-fragment_no_overlap}) and the fact that the occupied KS orbitals can be freely rotated within the full-size KS determinant, \ie,
\be
\Phi_{\rm s}\equiv \left\vert\prod_{q\in{\rm core}}\tilde{\varphi}_q^2\prod_{p\in F}\tilde{\varphi}_p^2\right\vert=\left\vert\Phi_{\rm core}\Phi^{C}\right\vert,
\ee
are the reasons why 1RDM elements involving at least one orbital in the fragment can be evaluated exactly (at the lattice KS level of calculation) from the reduced-in-size $2N_{\rm orb}^{F}$-electron Slater determinant $\Phi^{C}$ (the factor 2 coming from the doubly-occupied fragment-occupied orbitals) that is referred to as KS embedding cluster's wavefunction. We refer the reader to Refs.~\citenum{makhlouf2025_local_potential} and \citenum{makhlouf2026generalizedlocalpotentialfunctional} for more technical details. In connection with Eq.~(\ref{eq:core-fragment_no_overlap}), it is important to realize that the expected complete delocalization of the occupied KS orbitals $\left\{\varphi_k\right\}_{1\leq k\leq \frac{N}{2}}$ that diagonalize the KS Hamiltonian and, in particular, the fact that they have a non-zero overlap with the fragment's environment (\ie, the orthogonal complement to the fragment in the lattice) are central in the construction of the core orbitals. This becomes clear when rewriting the projection of Eq.~(\ref{subeq:projecting_out_frag-occ}) as follows, according to Eq.~(\ref{eq:decomp_occ_space_frag-occ_plus_core}), 
\be
\begin{split}
&\varphi_k(\br)-\sum_{p\in F}\langle \tilde{\varphi}_p\vert \varphi_k\rangle \tilde{\varphi}_p(\br)
\\
&=\sum_{q\in{\rm core}}\langle \tilde{\varphi}_q\vert \varphi_k\rangle \tilde{\varphi}_q(\br)
\\
&=\sum_{e\notin F}\langle\chi_e\vert \varphi_k \rangle
\left(\sum_{q\in{\rm core}}\langle \tilde{\varphi}_q\vert \chi_e \rangle \tilde{\varphi}_q(\br)\right),
\end{split}
\ee
where, for a given diagonalizing occupied KS orbital $\varphi_k$, the ``contraction'' coefficient for the environment's orbital $\chi_e$ projected onto the core subspace is $\langle\chi_e\vert \varphi_k \rangle$, as readily seen from the last line of the above equation. Note that the core electrons play a role (through the mean field they create within the cluster) once two-electron repulsions have been introduced into the embedding calculation~\cite{wouters2016practical,makhlouf2025_local_potential}.\\

The last and key step consists in defining the one-electron quantum bath subspace. In the present approach, it simply consists of the ``tails'' of the fragment-occupied orbitals beyond the fragment. Mathematically, a (non-orthonormal) basis of bath orbitals is obtained by projecting the fragment-occupied orbitals out of the fragment:   
\be\label{eq:bath_from_frag-occ_orbs}
\begin{split}
B\equiv\Bigg\{
&\chi_p^{\rm occ}({\br})
-\sum_{q\in F}\langle \chi_q\vert \chi_p^{\rm occ}\rangle \chi_q(\br)
\\
&=\tilde{\varphi}_p^{\rm bath}(\br)
\Bigg\}_{p\in F}
.
\end{split}
\ee
Thus, applying the variational principle within the cluster $C=F\oplus B$ ensures an exact reconstruction of the fragment-occupied orbitals and, therefore, an exact description of the fragment properties, at the lattice KS level of calculation.    
For completeness, let us point out that in the present pure ground-state theory, we recover from Eq.~(\ref{eq:bath_from_frag-occ_orbs}) the (to-be-orthonormalized) bath orbitals expression of Ref.~\citenum{sekaran2023unified}, 
\begin{subequations}
\begin{align}
\tilde{\varphi}_p^{\rm bath}(\br)&\overset{p\in F}{=}\sum_{q\notin F}\langle \chi_q\vert \chi_p^{\rm occ}\rangle \chi_q(\br)
\\
&=
\sum_{q\notin F}\sum^{N/2}_{k=1}\langle \varphi_k\vert\chi_p\rangle \langle \chi_q\vert \varphi_k\rangle \chi_q(\br)
\\
\label{subeq:bath_1RDM_func}
&=\sum_{q\notin F}\langle \Phi_{\rm s}\vert \hat{c}_{p\sigma}^\dagger \hat{c}_{q\sigma} \vert \Phi_{\rm s}\rangle \chi_q(\br)
,
\end{align}
\end{subequations}
which is a simple functional of the (idempotent) KS 1RDM.\\

In conclusion, the desired pure ground-state embedding cluster consists of the fragment and the bath (which contains the exact same number $N_{\rm orb}^F$ of orbitals as the fragment) and it is half-filled, as the electrons it contains are those that doubly occupy the fragment-occupied orbitals ($2N_{\rm orb}^F$ electrons in total).


\subsubsection{Extension to density-functional ensembles}\label{sec:ensemble_embedding}

In the light of the previous section, we can now revisit the single-orbital ($N_{\rm orb}^F=1$) ensemble embedding of Ref.~\citenum{cernatic2024fragment} in a more general and comprehensive way. We focus on TGOK ensembles, for simplicity, and consider a fragment of arbitrary size $N_{\rm orb}^F\geq 1$ as well as an arbitrary number $N_{\rm orb}^{\rm act}$ of active (\ie, fractionally occupied) eKS orbitals. We denote $N_{\rm e}^{\rm act}$ the number of (active) electrons that are distributed among the active orbitals in order to construct the KS ensemble.\\

As shown in Ref.~\citenum{cernatic2024fragment}, the trivial strategy that consists in replacing the ground-state 1RDM in Eq.~(\ref{subeq:bath_1RDM_func}) by the ensemble one does not achieve an exact ensemble embedding at the KS level of calculation, simply because the ensemble 1RDM is not idempotent, unlike the pure ground-state one. Instead, we should reconsider the construction of Sec.~\ref{sec:gs_embedding_KS} by focusing first on the inactive single determinant $\Phi_{\rm s}^{\rm in}$ that is common to all KS wavefunctions in the ensemble and where the inactive orbitals $\left\{\varphi_k\right\}_{k\in{\rm in}}$ (from which we never excite) are always doubly occupied. Projecting the fragment orbitals onto the inactive orbital subspace,
\be
\chi_p({\br})\overset{p\in F}{\rightarrow} \chi_p^{\rm in}({\br})=
\sum_{k\in {\rm in}}\langle \varphi_k\vert\chi_p\rangle\varphi_k(\br)
,
\ee
and keeping their ``tails'' (through a projection out of the fragment) defines what we refer to as the inactive bath: 
\be
B^{\rm in}\equiv\left\{
\chi_p^{\rm in}({\br})
-\sum_{q\in F}\langle \chi_q\vert \chi_p^{\rm in}\rangle \chi_q(\br)
\right\}_{p\in F}.
\ee
Then, in order to make sure that the active KS orbitals can be fully reconstructed from the ensemble embedding cluster, which guarantees, in particular, that contributions from the fragment to the eKS 1RDM are reproduced exactly, we need to complement the above inactive bath with the ``tails'' of the active orbitals. They define the active part of the bath: 
\be
B^{\rm act}\equiv\left\{
\varphi_k({\br})
-\sum_{q\in F}\langle \chi_q\vert \varphi_k\rangle \chi_q(\br)
\right\}_{k\in {\rm act}}
.
\ee
Note that, even though the active orbitals are orthogonal to the inactives, their projections onto the fragment can have a non-zero overlap, which means that $B^{\rm in}$ and $B^{\rm act}$ overlap, in principle, as readily seen from the simplified overlap expression below:   
\be
\begin{split}
&\left\langle \chi_p^{\rm in}
-\sum_{q\in F}\langle \chi_q\vert \chi_p^{\rm in}\rangle \chi_q\middle \vert \varphi_k
-\sum_{r\in F}\langle \chi_r\vert \varphi_k\rangle \chi_r
\right\rangle
\\
&\underset{k\in {\rm act}}{\overset{p\in F}{=}}
-\sum_{r\in F}\langle \chi_r\vert \varphi_k\rangle \langle \chi_p^{\rm in}\vert \chi_r\rangle
\\
&=-\sum_{r\in F}\sum_{k'\in {\rm in}}\langle \chi_p\vert \varphi_{k'}\rangle\langle \varphi_{k'}\vert \chi_r\rangle\langle \chi_r\vert \varphi_k\rangle
.
\end{split}
\ee
Therefore, the complete ensemble bath reads 
\be
B^{\rm ens}\equiv B^{\rm in}+B^{\rm act}, 
\ee
for which an orthonormal basis can be determined along the lines of Eq.~(\ref{eq:final_orthonormal_frag-occ_orbitals}).\\ 

In conclusion, the desired ensemble embedding cluster consists of the fragment and a larger ensemble bath that contains $N_{\rm orb}^F+N_{\rm orb}^{\rm act}$ bath orbitals in total. The total number of electrons within the cluster is $2N_{\rm orb}^F+N_{\rm e}^{\rm act}$, $2N_{\rm orb}^F$ electrons occupying the only $N_{\rm orb}^F$ inactive orbitals that have a non-zero overlap with the fragment and $N_{\rm e}^{\rm act}$ electrons that are distributed among the active orbitals. This is in perfect agreement with the numerical analysis of Ref.~\citenum{cernatic2024fragment}. As the ensemble bath enables the reconstruction of the complete set of active orbitals, it should allow for the description of delocalized excitation processes such as charge transfers, for example. In other words, if an interacting bath~\cite{makhlouf2025_local_potential} is ultimately employed, we expect the local treatment of electron correlations (which is central in the standard quantum embedding of pure states) to apply only to the inactive electrons. Combining this construction with the analog for ensembles of the local density mapping constraint of Eq.~(\ref{eq:dens_mapp_constraint_LPFET}) lays the foundations of an ensemble LPFET whose complete derivation and implementation are left for future work.

\section{Conclusions and perspectives}\label{sec:conclusions}

The $N$-centered (Nc) generalization of ensemble density functional theory (eDFT) is a quite recent and promising formalism describing neutral and charged electronic excitations within a unified, time-independent framework. Beyond the computation of excitation energies, the stationarity of the ensemble density-functional energy levels yields exact expressions for individual-state densities~\cite{Fromager2025indvElevel} and static linear response functions~\cite{dupuy2025_exact_static}, from which many properties of molecules and materials can be deduced. On one hand, eDFT can now be used to compute insightful reactivity descriptors of conceptual DFT~\cite{CDFTrev2003}, as in a recent work by some of the authors on the Nc eDFT of Fukui functions~\cite{dupuy2026fukui}. On the other hand, the above-mentioned developments show that the application of eDFT outside of static calculations and into excited-state molecular dynamics simulations~\cite{Filatov24_Unraveling}, for example, through the computation of excited states' forces acting on nuclei, has solid theoretical foundations.

As summarized in this work, eDFT in its Nc form also brings a clearer picture of subtle yet important features of DFT for charged and neutral excitations. Derivative discontinuities of the Hxc functional at integer electron numbers, a puzzling property of the fractional electron number
PPLB approach to charged electronic excitations~\cite{perdew1982density,mori2008localization,cohen2008fractional,cohen2008insights,stein2010fundamental,zheng2011improving,cohen2011challenges,kraisler2013piecewise,kraisler2014fundamental,perdew2017understanding}, is replaced by neatly-defined Hxc ensemble density-functional weight derivatives~\cite{senjean2018unified,PRA21_Hodgson_exact_Nc-eDFT_1D,Cernatic2022}. This allows for a systematic and straightforward exactification of Koopman's theorem without ever invoking the asymptotic behavior of the ensemble density. Hence, it should also be valid for extended systems. This holds not only for charged excitations, but also for neutral excitations~\cite{Cernatic2024_Neutral,cernatic2024extended_doubles} that are usually treated within time-dependent DFT (TDDFT).

For completeness, a clear connection between exact eDFT and the popular $\Delta$-SCF approach has been highlighted. It relies, among other well-identified approximations, on the expression of the ensemble Hxc density functional in terms of the individual-state KS densities. Use of individual KS wavefunctions for computation of ensemble Hx energies has been investigated previously, with a growing body of studies on the development and use of (orbital-dependent) hybrid functionals through different ensemble generalized KS extensions of the theory~\cite{filatov2015spin, Gould2020_Approximately,loos2020weightdependent,Gould2021_Ensemble_ugly,Cernatic2022}. Their performance is encouraging, motivating further work, notably to analyze and circumvent issues of ghost interactions and $v$-representability. The question of how to compute the correlation energy from individual densities also has to be addressed in order to further improve $\Delta$-SCF methods from the perspective of eDFT. To that end, identification of the missing state- and density-driven (SD and DD) correlation contributions~\cite{PRL19_Gould_DD_correlation,Fromager_2020,Cernatic2022} in the so-called ground-state individual-correlation (Gs-ic) approximation of eDFT~\cite{Cernatic2022,filatov2015spin} is the natural next step. A lot of progress has recently been made in the analysis and approximation~\cite{gould2025stationaryconditionsexcitedstates,Gould2025_PRL_tate-Specific,gould2026ensemblization} of SD/DD correlations, but there is no obvious nor unique definition of their energy contributions~\cite{Fromager_2020,Cernatic2022}, as also suggested by the Rayleigh--Schr\"{o}dinger reformulation of eDFT presented in this work (see below). Changing the reference KS states from which the missing correlation is to be recovered by adopting the Gs-ic ansatz from the onset~\cite{Cernatic2022} could bring precious insights and new developments in both $\Delta$-SCF and SD/DD ensemble correlation theory. Work is currently in progress in these directions.\\

In order to put the exact theory into a broader and more practical perspective, several strategies have been suggested and discussed in the present work. The first one consists in recycling regular (ground-state) DFAs by dressing them with a weight-dependent scaling function. We showed how the scaling function can be determined from exact properties of eDFT, focusing on the calculation of energy levels in the context of neutral excitations, as a proof of principle, while referring to Ref.~\citenum{dupuy2026fukui} for applications to charged excitations and, more precisely, to the Nc eDFT of Fukui functions. In the latter case, designing a scaling function for the Hxc potential directly, rather than for the Hxc functional, is a better strategy, as also illustrated in this work. As the reformulation of DFT as a potential-functional theory is now well-established~\cite{Perdew03OEP,Weitao04PFT,heaton2007optimized,sekaran2022local,makhlouf2025_local_potential,makhlouf2026generalizedlocalpotentialfunctional,gould2025stationaryconditionsexcitedstates}, it seems appealing to pursue the development of weight-dependent scaling functions, starting from the ensemble Hxc potential as basic ingredient. The Levy--Zahariev-shifted Hxc potential~\cite{kraisler2013piecewise,levy2014ground} may even be a better ingredient, as readily seen from the integral expression of Eq.~\eqref{eq:LZ_intrel_EV}. This is left for future work.

The second approach that we discussed is multi-reference ensemble density-functional perturbation theory, for the purpose of developing orbital-dependent ensemble Hxc DFAs. Interestingly, the in-principle exact Rayleigh--Schr\"{o}dinger formulation of the theory suggests alternative definitions of the ensemble Hartree, exchange, and correlation energies, separately. This relates to the fact that, in this formalism, the projections of the exact physical states onto the ensemble KS Hilbert space are the natural reference wavefunctions, from which the perturbation expansion of any targeted quantity, like the ensemble density, for example, is obtained. Unlike in  ensemble G\"{o}rling--Levy perturbation theory, the above-mentioned projections can actually be refined up to a given order of perturbation, which might be important for tackling nearly-degenerate situations. Setting the corrections to the ensemble density to zero, at each order of perturbation, is the key constraint from which the perturbation expansion of the ensemble Hxc potential (and, consequently, the perturbation operator itself) can be determined. For completeness, we considered a unitary formulation of the theory, in the spirit of Van Vleck perturbation theory~\cite{VanVleck29_On,SOKOLOV2024121}, and discussed its practical advantages and formal drawbacks. Both implementations should definitely be explored further in the future. 

Lastly, we revisited the quantum density matrix embedding theory (DMET) of excited states from the perspective of eDFT or, more precisely, lattice eDFT. By using the ensemble density as the basic variable of the theory, here the ensemble occupation of the localized orbitals from which fragments of the system under study can be defined and then embedded, our ensemble quantum embedding theory inherits the rigorous foundations and key results of eDFT. Particular attention has been paid to the construction of quantum baths for ensembles, thus providing a comprehensive and straightforward generalization to multiple-orbital fragments of the single-orbital ensemble embedding explored in Ref.~\onlinecite{cernatic2024fragment}. By construction, the approach can deal with non-local excitations, with charge transfers being the natural application. The full derivation and implementation of the theory, along the lines of the local potential functional embedding theory of ground states~\cite{makhlouf2025_local_potential,makhlouf2026generalizedlocalpotentialfunctional}, is left for future work.\\       

Even though they have not been mentioned up to this point, several other promising developments can be foreseen in the context of eDFT. The first one deals with the description of spin in DFT and, more specifically, with the density-functional derivative discontinuities induced by the variation of spin-up/down electron numbers~\cite{goshen2026manyelectronsystemsfractionalelectron,Goshen24_Ensemble,Hayman25_spin-migration,Burgess24_Tilted-Plane,Mori-Sanchez09_Discontinuous,Neil-Qiang-Su18_Describing}. Extending the Nc approach to spin-DFT is expected to shed a complementary light on the challenging modeling of such features. A second topic of interest is the reformulation of ensemble theories such as thermal DFT~\cite{smith2016exact,Mermin65_Thermal,Hilleke25_Fully,PRL11_Pittalis_exact_conds_thermalDFT} in terms of a pure quantum state~\cite{Harsha19_Thermofield,Harsha22_Thermal,Benavides-Riveros22_Excitations}. Such an approach is expected to be impactful in the context of quantum embedding theory~\cite{Sun20_Finite-temperature}, for example. More generally, its density functionalization would imply a redefinition of the fictitious KS system, whose pure ground state would reproduce some exact properties of the ensemble, like its density. 
A third line of research that seems promising deals with the calculation of electronic properties that are needed for non-adiabatic dynamics simulations. This is one of the major appeals a DFT of excited states can possess, and one that has definitively favored TDDFT over eDFT until now. While static linear response eDFT provides molecular forces of ground and excited states (through the evaluation of their individual densities) as well as their density-density linear response functions, in principle exactly, there remains to understand how to compute non-adiabatic couplings (NACs) or, more generally, how to predict transition dipole moments and oscillator strengths. While time-dependent linear response eDFT~\cite{daas2025ensembletimedependentdensityfunctional} is a natural route to NACs in terms of established TDDFT concepts, it is probably worth exploring further how much can be learned from the simpler static response theory, where the possibility to vary the ensemble weights might be exploited. Finally, all aforementioned approaches for excited-state calculations still rely on the Born--Oppenheimer (BO) separation between electrons and nuclei, which effectively breaks down in the vicinity of conical intersections. Some of the authors have proposed recently an exact beyond-BO DFT treating both electrons and nuclei on the same footing~\cite{Fromager_2024_Density}. It has subsequently been reformulated within the Exact Factorization picture~\cite{dupuy2026exactlyfactorizedmolecularkohnsham} to obtain disentangled (but coupled) electronic and nuclear KS equations including non-adiabatic effects from the onset (see also the related Refs.~\citenum{Requist16_Exact} and~\citenum{li2025bornoppenheimertimedependentdensityfunctional}). Development and testing of practical DFAs tailored to this framework, with connections to eDFT~\cite{Fromager_2024_Density} to be explored, is underway. In order to provide novel non-adiabatic dynamics schemes, the approach shall be extended to time-dependent processes~\cite{li2025bornoppenheimertimedependentdensityfunctional}. Work is currently in progress in this direction.

\section*{Data availability statement}
The data that supports the findings of this study are available from the corresponding author upon reasonable request.

\section*{Acknowledgements}
E.~F. is grateful to Gianluca Levi and Gaurav Harsha for stimulating and enlightening discussions. E.~F. would also like to thank Dr. Masahisa Tsuchiizu for hosting him at Nara Women's University in Japan. This work has benefited from support provided by the University of Strasbourg Institute for Advanced Study (USIAS) for a Fellowship, within the French national programme “Investment for the future” (IdEx-Unistra).
J\'{e}r\'{e}my Morere acknowledges mobility funding provided by GDR 2161, Th\'{e}MoSiA, funded by CNRS.

\appendix

\section{Uniform coordinate scaling relation extended to $N$-centered ensembles}\label{appendix:details_scaled_eDFA}

In the following, we consider Lieb's formulation (as a Legendre--Fenchel transform~\cite{Cernatic2024_Neutral}) of the $N$-centered ensemble Levy--Lieb functional introduced in Eq.~(\ref{eq:LL_nce_func_separate_def}),
\be\label{eqapp:Nce_LF_transf_for_scaling}
F^{\bxi}[n]=\max_v\left\{E^{\bxi}[v]-(v\vert n)\right\},
\ee
where $(v\vert n)=\int d\br\,v(\br)n(\br)$, and we evaluate the changes in the functional when the ($N$-centered ensemble) density $n$ undergoes the following uniform coordinate scaling,
\be
n(\br)\rightarrow n_\gamma(\br)=\gamma^3n(\gamma \br),\;\gamma>0.
\ee
As the scaling can be transferred from the density to the potential in the second term on the right-hand side of Eq.~(\ref{eqapp:Nce_LF_transf_for_scaling}), \ie, 
\be\label{eqapp:from_scaled_dens_to_scaled_pot}
\begin{split}
(v\vert n_\gamma)&=\gamma^3\int d\br\, v(\br)n(\gamma \br) 
\\
&=\int d\br\, v(\br/\gamma)\,n(\br)
\\
&=:(v_{{1}/{\gamma}}\vert n),
\end{split}
\ee
evaluating $F^{\bxi}[n_\gamma]$ requires in fact considering the $\mathcal{N}$-electron ($\mathcal{N}=N,N\pm1,N\pm2,\ldots$ being an integer) Schr\"{o}dinger equation, for the ground and the excited states, with the transformed potential $v_{{1}/{\gamma}}(\br)=v(\br/\gamma)$ in the Hamiltonian (see Eq.~(\ref{eq:first_quantized_physical_Hamil})):
\be\label{eqapp:SE_scaled_pot}
\begin{split}
&\left[-\dfrac{1}{2}\sum^{\mathcal{N}}_{i=1}\nabla_{\br_i}^2+\sum_{1\leq i<j}^{\mathcal{N}}\dfrac{1}{\vert {\bf r}_i-{\bf r}_j\vert}+\sum^{\mathcal{N}}_{i=1}v_{1/\gamma}({\bf r}_i)\right]\Psi_\nu
\\
&=E_\nu[v_{1/\gamma}]\Psi_\nu,\;\forall \nu,
\end{split}
\ee
where $\Psi_\nu\equiv \Psi_\nu(\br_1,\br_2,\ldots,\br_{\mathcal{N}})$. By introducing $\Psi^{\gamma}_\nu\equiv\Psi_\nu(\gamma\br_1,\gamma\br_2,\ldots,\gamma\br_{\mathcal{N}})$ or, equivalently, 
\be
\Psi_\nu(\br_1,\br_2,\ldots,\br_{\mathcal{N}})=\Psi^{\gamma}_\nu
\left(\br_1/\gamma,\br_2/\gamma,\ldots,\br_{\mathcal{N}}/\gamma\right),
\ee
where no normalization factor is introduced, as we focus exclusively on the eigenvalues of the Hamiltonian, Eq.~(\ref{eqapp:SE_scaled_pot}) can be rewritten as follows, 
\be
\begin{split}
&\left[-\dfrac{1}{2\gamma^2}\sum^{\mathcal{N}}_{i=1}\nabla_{\br_i}^2+\sum_{1\leq i<j}^{\mathcal{N}}\dfrac{1}{\gamma\vert {\bf r}_i-{\bf r}_j\vert}+\sum^{\mathcal{N}}_{i=1}v({\bf r}_i)\right]\Psi^\gamma_\nu
\\
&=E_\nu[v_{1/\gamma}]\Psi^\gamma_\nu,\;\forall \nu,
\end{split}
\ee
or, equivalently,
\be
\begin{split}
&\left[-\dfrac{1}{2}\sum^{\mathcal{N}}_{i=1}\nabla_{\br_i}^2+\sum_{1\leq i<j}^{\mathcal{N}}\dfrac{\gamma}{\vert {\bf r}_i-{\bf r}_j\vert}+\sum^{\mathcal{N}}_{i=1}\gamma^2v({\bf r}_i)\right]\Psi^\gamma_\nu
\\
&=\gamma^2E_\nu[v_{1/\gamma}]\Psi^\gamma_\nu,\;\forall \nu.
\end{split}
\ee
As a result, the energies $E^\gamma_\nu[\gamma^2 v]$ of the above Schr\"{o}dinger equation, where the electronic repulsion and the local potential $v$ are scaled by $\gamma$ and $\gamma^2$, respectively, can be obtained straightforwardly from those of the fully-interacting ($\gamma=1$) Eq.~(\ref{eqapp:SE_scaled_pot}), as follows,    
\be
E^\gamma_\nu[\gamma^2 v]=
\gamma^2E_\nu[v_{1/\gamma}]=:\gamma^2E^{\gamma=1}_\nu[v_{1/\gamma}].
\ee
As the relation holds for both ground and excited states, and for any electron number $\mathcal{N}$, it holds for all the states within the $N$-centered ensemble, and therefore, it applies to the $N$-centered ensemble energy: 
\be\label{eqapp:Nce_energy_scaling_relation_first}
\gamma^2E^{\bxi}[v_{1/\gamma}]=E^{\bxi,\gamma}[\gamma^2 v],\;\forall v,\;\forall \gamma.
\ee
In order to further simplify $F^{\bxi}[n_\gamma]$, we need to relate $E^{\bxi}[v]$ (first term on the right-hand side of Eq.~(\ref{eqapp:Nce_LF_transf_for_scaling})) to $v_{1/\gamma}$, according to Eq.~(\ref{eqapp:from_scaled_dens_to_scaled_pot}). This is achieved from Eq.~(\ref{eqapp:Nce_energy_scaling_relation_first}) and the substitution $v\rightarrow v_\gamma$, which gives 
\be
\gamma^2E^{\bxi}[v]=E^{\bxi,\gamma}[\gamma^2 v_\gamma],\;\forall \gamma,
\ee
thus leading, from the substitution $\gamma\rightarrow 1/\gamma$, to the desired relation
\be
E^{\bxi}[v]=\gamma^2 E^{\bxi,1/\gamma}\left[\dfrac{v_{1/\gamma}}{\gamma^2}\right].
\ee
We can now proceed with the following simplification,
\be
\begin{split}
F^{\bxi}[n_{\gamma}]&=\max_v\left\{\gamma^2 E^{\bxi,1/\gamma}\left[\dfrac{v_{1/\gamma}}{\gamma^2}\right]-(v_{1/\gamma}\vert n)\right\}
\\
&=\gamma^2\max_v\left\{E^{\bxi,1/\gamma}\left[\dfrac{v_{1/\gamma}}{\gamma^2}\right]-\left(\dfrac{v_{1/\gamma}}{\gamma^2}\middle\vert n\right)\right\},
\end{split}
\ee
where the one-to-one correspondence between $v$ and ${v_{1/\gamma}}/(\gamma^2)$, according to the relation 
\be
v(\br)=\gamma^2\left(\dfrac{v_{1/\gamma}(\gamma\br)}{\gamma^2}\right), 
\ee
allows to maximize with respect to ${v_{1/\gamma}}/(\gamma^2)$ instead. We therefore conclude that 
\be
\begin{split}
F^{\bxi}[n_{\gamma}]&=\gamma^2\max_v\left\{E^{\bxi,1/\gamma}[v]-(v\vert n)\right\}
\\
&=:\gamma^2 F^{\bxi,1/\gamma}[n],
\end{split}
\ee

or, equivalently, from the substitution $\gamma\rightarrow 1/\gamma$,  
\be
F^{\bxi,\gamma}[n]=\gamma^2 F^{\bxi}[n_{1/\gamma}],
\ee
which gives in the non-interacting case,
\be
T^{\bxi}_{\rm s}[n]=\gamma^2T^{\bxi}_{\rm s}[n_{1/\gamma}].
\ee
By taking the difference of the two above expressions we finally obtain the well-known scaling relation which is shown here to hold for $N$-centered ensembles too:
\be\label{eqapp:final_exact_constraint_scaling_Nc_ensemble}
E^{\bxi,\gamma}_{\rm Hxc}[n]=\gamma^2E^{\bxi}_{\rm Hxc}[n_{1/\gamma}],\;\forall \bxi.
\ee
We now introduce, for any interaction strength $\gamma$, the simple ansatz
\be
\begin{split}
E^{\bxi,\gamma}_{\rm Hxc}[n]
\approx \mathcal{E}^{\bxi,\gamma}_{\rm Hxc}[n]
&= s^{\gamma}(\bxi)E^{\bxi=0,\gamma}_{\rm Hxc}[n]
\\
&=s^{\gamma}(\bxi)E^{\gamma}_{\rm Hxc}[n],
\end{split}
\ee
where, at least for now, the scaling function $s^{\gamma}(\bxi)$ is in principle $\gamma$-dependent, while reducing to the scaling function introduced in Eq.~(\ref{eq:ehxc_scaling}) in the fully-interacting case, \ie,  
\be
s^{\gamma=1}(\bxi)=s(\bxi),
\ee
and
\be
\mathcal{E}^{\bxi}_{\rm Hxc}[n]=\mathcal{E}^{\bxi,\gamma=1}_{\rm Hxc}[n]=s(\bxi)E_{\rm Hxc}[n].
\ee
Applying the exact constraint of Eq.~(\ref{eqapp:final_exact_constraint_scaling_Nc_ensemble}) in the $\bxi=0$ ($N$-electron ground-state) limit leads to the relation   
\be
\begin{split}
\mathcal{E}^{\bxi,\gamma}_{\rm Hxc}[n]&=\gamma^2 s^{\gamma}(\bxi)
E_{\rm Hxc}[n_{1/\gamma}]
\\
&=\gamma^2 \left(\dfrac{s^{\gamma}(\bxi)}{s(\bxi)}\right)\mathcal{E}^{\bxi}_{\rm Hxc}[n_{1/\gamma}]. 
\end{split}
\ee
Consequently, in order to ensure that the approximate functional $\mathcal{E}^{\bxi,\gamma}_{\rm Hxc}[n]$ fulfills the exact constraint of Eq.~(\ref{eqapp:final_exact_constraint_scaling_Nc_ensemble}), for any weight, the scaling function must be $\gamma$-{\it independent}: 
\be
s^{\gamma}(\bxi)=s(\bxi),\;\forall \gamma>0.
\ee
In summary, the approximation underlying the density functional of Eq.~(\ref{eq:ehxc_scaling}), which is recovered in the fully-interacting $\gamma=1$ case, is
\be
E^{\bxi,\gamma}_{\rm Hxc}[n]
\approx \mathcal{E}^{\bxi,\gamma}_{\rm Hxc}[n]=s(\bxi)E^{\gamma}_{\rm Hxc}[n],\;\forall \gamma>0.
\ee

\section{VV ensemble density-functional perturbation theory}\label{appendix:VV-EDFT}
The unitary operator $\hat{\mathcal{U}}^{\bxi}$ introduced in section \ref{sec:mrpt_eDFT-unitary_formulation} is inspired by VV perturbation theory~\cite{VanVleck29_On,SOKOLOV2024121}. The goal of this approach is to uncouple the ensemble subspace and its orthogonal complement through a unitary transformation written as an exponential of an anti-Hermitian generator $\hat{\mathcal{S}}^{\bxi}$. It is assumed that this operator connects the eKS subspace block and its orthogonal complement only, thus leading to the explicit expression
\be
\hat{\mathcal{S}}^{\bxi} = 
\underset{\mu\geq M}{\underset{0\leq \nu<M }{\sum}}
\mathcal{S}_{\mu\nu}^{\bxi}
\left(
\ket{\Phi_\nu^{\bxi}}\bra{\Phi_\mu^{\bxi}} - \ket{\Phi_\mu^{\bxi}}\bra{\Phi_\nu^{\bxi}}
\right).
\ee
The Hamiltonian of the system is transformed by the unitary operator $\hat{\mathcal{U}}^{\bxi}=e^{-\hat{\mathcal{S}}^{\bxi}}$ into an effective Hamiltonian,

\be
\hat{\mathcal{H}}_{\rm eff}^{\bxi}=
e^{\hat{\mathcal{S}}^{\bxi}}
\hat{H}
e^{-\hat{\mathcal{S}}^{\bxi}}.
\ee
The effective Hamiltonian and the physical Hamiltonian share the same set of eigenvalues. 
We now focus on the coefficients of the generator $\hat{\mathcal{S}}^{\bxi}$ which are determined in such a way that the effective Hamiltonian is block-diagonal in the complete eKS basis, according to Eq.~(\ref{eq:decoupling_unitary_PT_in_terms_of_U}), \ie, 
\be\label{eq:cond-block-diagonal}
\bra{\Phi_\mu^{\bxi}} \hat{\mathcal{H}}_{\rm eff}^{\bxi} \ket{\Phi_\nu^{\bxi}}\underset{\mu\geq M}{\underset{0\leq \nu<M }{=}}0.
\ee
From the perturbation expansion of the Hamiltonian that we write in the general form (see Eqs.~(\ref{eq:splitting_true_Hamilt_into_KS_plus_pert}) and (\ref{eq:perturb_in_RSEDFPT}))
\be
\hat{H}\rightarrow \hat{H}(\alpha)=\hat{H}_{\rm KS}^{\bxi}+\alpha\hat{\mathcal{W}}^{\bxi(1)}+\alpha^2\hat{\mathcal{W}}^{\bxi(2)}+\ldots
\ee
and the expansion of the generator operator $\hat{\mathcal{S}}^{\bxi}$, 
\be
\hat{\mathcal{S}}^{\bxi} \rightarrow \hat{\mathcal{S}}^{\bxi}(\alpha) = \alpha\hat{\mathcal{S}}^{\bxi(1)} + \alpha^2\hat{\mathcal{S}}^{\bxi(2)} + \ldots,
\ee
or, equivalently,
\be
\mathcal{S}_{\mu\nu}^{\bxi} 
\rightarrow 
\mathcal{S}_{\mu\nu}^{\bxi}(\alpha) = \alpha\mathcal{S}_{\mu\nu}^{\bxi(1)} + \alpha^2\mathcal{S}_{\mu\nu}^{\bxi(2)} + \ldots,
\ee
the effective Hamiltonian can itself be expanded in perturbation through the following Baker--Campbell--Hausdorff expansion,   
\be\label{eqapp:BCH_exp_eff_Hamil}
\begin{split}
&
e^{\hat{\mathcal{S}}^{\bxi}(\alpha)}
\hat{H}(\alpha)
e^{-\hat{\mathcal{S}}^{\bxi}(\alpha)} = 
  \hat{H}(\alpha)
 + \left[\hat{\mathcal{S}}^{\bxi}(\alpha),\hat{H}(\alpha)\right]\\
&+ \dfrac{1}{2}\left[\hat{\mathcal{S}}^{\bxi}(\alpha),\left[\hat{\mathcal{S}}^{\bxi}(\alpha),\hat{H}(\alpha)\right]\right]
+ \cdots
\end{split}
\ee
Note that the ensemble density would be evaluated similarly by substituting in Eq.~(\ref{eqapp:BCH_exp_eff_Hamil}) the density operator $\hat{n}(\br)$ for $\hat{H}(\alpha)$, thus leading, ultimately, to the perturbation expansion of the ensemble Hxc potential, according to the ensemble density constraint of Eq.~(\ref{eq:ens_dens_constraint_unitary_theory}).\\

In a regular VV perturbation theory, where the perturbation contains first-order contributions only, the decoupling constraint of Eq.~(\ref{eq:cond-block-diagonal}) that we rewrite equivalently as follows,
\be
\left\langle\Phi_\nu^{\bxi}\middle\vert e^{\hat{\mathcal{S}}^{\bxi}(\alpha)}
\hat{H}(\alpha)
e^{-\hat{\mathcal{S}}^{\bxi}(\alpha)}\middle\vert \Phi_\mu^{\bxi}\right\rangle\underset{\mu\geq M}{\underset{0\leq \nu<M }{=}}0,\;\forall \alpha,
\ee
leads to the following expressions of the generator coefficients at first and second order, respectively,
\be
\mathcal{S}_{\mu\nu}^{\bxi(1)} = \dfrac{\bra{\Phi_\mu^{\bxi}}\hat{\mathcal{W}}^{\bxi(1)}\ket{\Phi_\nu^{\bxi}}}{\mathcal{E}^{\bxi}_{\nu}-\mathcal{E}^{\bxi}_{\mu}}
\ee
and
\be
\begin{split}
&\mathcal{S}_{\mu\nu}^{\bxi(2)} = 
\dfrac{1}{\mathcal{E}^{\bxi}_{\mu}-\mathcal{E}^{\bxi}_{\nu}}
\\
&\times
\left[\sum_{0\leq\lambda<M}
\dfrac{
\bra{\Phi_\mu^{\bxi}}    \hat{\mathcal{W}}^{\bxi(1)}\ket{\Phi_\lambda^{\bxi}}
\bra{\Phi_\lambda^{\bxi}}\hat{\mathcal{W}}^{\bxi(1)}\ket{\Phi_\nu^{\bxi}}}
{\mathcal{E}^{\bxi}_{\lambda}-\mathcal{E}^{\bxi}_{\mu}} \right.
\\
- &\left.\sum_{\kappa \geq M}
\dfrac{
\bra{\Phi_\mu^{\bxi}}   \hat{\mathcal{W}}^{\bxi(1)}\ket{\Phi_\kappa^{\bxi}}
\bra{\Phi_\kappa^{\bxi}}\hat{\mathcal{W}}^{\bxi(1)}\ket{\Phi_\nu^{\bxi}}}
{\mathcal{E}^{\bxi}_{\nu}-\mathcal{E}^{\bxi}_{\kappa}} \right]
.
\end{split}
\ee
In the present context of eDFT, the ensemble correlation potential will bring second- and higher-order contributions that can be incorporated straightforwardly into the perturbation expansions~\cite{srDFT_densitymatrixformulation}. As a final note, it may be necessary for practical purposes to refine the unperturbed Hamiltonian, which has been so far the bare non-interacting eKS Hamiltonian, in order to treat adequately nearly-degenerate situations, for example.


\clearpage 

\bibliography{biblio}

\newcommand{\Aa}[0]{Aa}
\begin{thebibliography}{160}%
\makeatletter
\providecommand \@ifxundefined [1]{%
 \@ifx{#1\undefined}
}%
\providecommand \@ifnum [1]{%
 \ifnum #1\expandafter \@firstoftwo
 \else \expandafter \@secondoftwo
 \fi
}%
\providecommand \@ifx [1]{%
 \ifx #1\expandafter \@firstoftwo
 \else \expandafter \@secondoftwo
 \fi
}%
\providecommand \natexlab [1]{#1}%
\providecommand \enquote  [1]{``#1''}%
\providecommand \bibnamefont  [1]{#1}%
\providecommand \bibfnamefont [1]{#1}%
\providecommand \citenamefont [1]{#1}%
\providecommand \href@noop [0]{\@secondoftwo}%
\providecommand \href [0]{\begingroup \@sanitize@url \@href}%
\providecommand \@href[1]{\@@startlink{#1}\@@href}%
\providecommand \@@href[1]{\endgroup#1\@@endlink}%
\providecommand \@sanitize@url [0]{\catcode `\\12\catcode `\$12\catcode
  `\&12\catcode `\#12\catcode `\^12\catcode `\_12\catcode `\%12\relax}%
\providecommand \@@startlink[1]{}%
\providecommand \@@endlink[0]{}%
\providecommand \url  [0]{\begingroup\@sanitize@url \@url }%
\providecommand \@url [1]{\endgroup\@href {#1}{\urlprefix }}%
\providecommand \urlprefix  [0]{URL }%
\providecommand \Eprint [0]{\href }%
\providecommand \doibase [0]{http://dx.doi.org/}%
\providecommand \selectlanguage [0]{\@gobble}%
\providecommand \bibinfo  [0]{\@secondoftwo}%
\providecommand \bibfield  [0]{\@secondoftwo}%
\providecommand \translation [1]{[#1]}%
\providecommand \BibitemOpen [0]{}%
\providecommand \bibitemStop [0]{}%
\providecommand \bibitemNoStop [0]{.\EOS\space}%
\providecommand \EOS [0]{\spacefactor3000\relax}%
\providecommand \BibitemShut  [1]{\csname bibitem#1\endcsname}%
\let\auto@bib@innerbib\@empty
\bibitem [{\citenamefont {Hohenberg}\ and\ \citenamefont
  {Kohn}(1964)}]{hktheo}%
  \BibitemOpen
  \bibfield  {author} {\bibinfo {author} {\bibfnamefont {P}~\bibnamefont
  {Hohenberg}}\ and\ \bibinfo {author} {\bibfnamefont {W}~\bibnamefont
  {Kohn}},\ }\bibfield  {title} {\enquote {\bibinfo {title} {Inhomogeneous
  electron gas},}\ }\href {https://doi.org/10.1103/PhysRev.136.B864} {\bibfield
   {journal} {\bibinfo  {journal} {Phys. Rev.}\ }\textbf {\bibinfo {volume}
  {136}},\ \bibinfo {pages} {B864} (\bibinfo {year} {1964})}\BibitemShut
  {NoStop}%
\bibitem [{\citenamefont {Kohn}\ and\ \citenamefont {Sham}(1965)}]{KS}%
  \BibitemOpen
  \bibfield  {author} {\bibinfo {author} {\bibfnamefont {W.}~\bibnamefont
  {Kohn}}\ and\ \bibinfo {author} {\bibfnamefont {L.J.}\ \bibnamefont {Sham}},\
  }\bibfield  {title} {\enquote {\bibinfo {title} {Self-consistent equations
  including exchange and correlation effects},}\ }\href {\doibase
  10.1103/PhysRev.140.A1133} {\bibfield  {journal} {\bibinfo  {journal} {Phys.
  Rev.}\ }\textbf {\bibinfo {volume} {140}},\ \bibinfo {pages} {A1133}
  (\bibinfo {year} {1965})}\BibitemShut {NoStop}%
\bibitem [{\citenamefont {Teale}\ \emph {et~al.}(2022)\citenamefont {Teale},
  \citenamefont {Helgaker}, \citenamefont {Savin}, \citenamefont {Adamo},
  \citenamefont {Aradi}, \citenamefont {Arbuznikov}, \citenamefont {Ayers},
  \citenamefont {Baerends}, \citenamefont {Barone}, \citenamefont
  {Calaminici},\ and\ \citenamefont {{et {\it al.}}}}]{Teale2022_DFT_exchange}%
  \BibitemOpen
  \bibfield  {author} {\bibinfo {author} {\bibfnamefont {Andrew~M.}\
  \bibnamefont {Teale}}, \bibinfo {author} {\bibfnamefont {Trygve}\
  \bibnamefont {Helgaker}}, \bibinfo {author} {\bibfnamefont {Andreas}\
  \bibnamefont {Savin}}, \bibinfo {author} {\bibfnamefont {Carlo}\ \bibnamefont
  {Adamo}}, \bibinfo {author} {\bibfnamefont {B{\'a}lint}\ \bibnamefont
  {Aradi}}, \bibinfo {author} {\bibfnamefont {Alexei~V.}\ \bibnamefont
  {Arbuznikov}}, \bibinfo {author} {\bibfnamefont {Paul~W.}\ \bibnamefont
  {Ayers}}, \bibinfo {author} {\bibfnamefont {Evert~Jan}\ \bibnamefont
  {Baerends}}, \bibinfo {author} {\bibfnamefont {Vincenzo}\ \bibnamefont
  {Barone}}, \bibinfo {author} {\bibfnamefont {Patrizia}\ \bibnamefont
  {Calaminici}}, \ and\ \bibinfo {author} {\bibnamefont {{et {\it al.}}}},\
  }\bibfield  {title} {\enquote {\bibinfo {title} {Dft exchange: sharing
  perspectives on the workhorse of quantum chemistry and materials science},}\
  }\href {\doibase 10.1039/D2CP02827A} {\bibfield  {journal} {\bibinfo
  {journal} {Phys. Chem. Chem. Phys.}\ }\textbf {\bibinfo {volume} {24}},\
  \bibinfo {pages} {28700--28781} (\bibinfo {year} {2022})}\BibitemShut
  {NoStop}%
\bibitem [{\citenamefont {Perdew}\ \emph {et~al.}(1996)\citenamefont {Perdew},
  \citenamefont {Burke},\ and\ \citenamefont
  {Ernzerhof}}]{perdew1996generalized}%
  \BibitemOpen
  \bibfield  {author} {\bibinfo {author} {\bibfnamefont {John~P}\ \bibnamefont
  {Perdew}}, \bibinfo {author} {\bibfnamefont {Kieron}\ \bibnamefont {Burke}},
  \ and\ \bibinfo {author} {\bibfnamefont {Matthias}\ \bibnamefont
  {Ernzerhof}},\ }\bibfield  {title} {\enquote {\bibinfo {title} {Generalized
  gradient approximation made simple},}\ }\href@noop {} {\bibfield  {journal}
  {\bibinfo  {journal} {Phys. Rev. Lett.}\ }\textbf {\bibinfo {volume} {77}},\
  \bibinfo {pages} {3865} (\bibinfo {year} {1996})}\BibitemShut {NoStop}%
\bibitem [{\citenamefont {Burke}(2012)}]{Burke2012perspectiveDFA}%
  \BibitemOpen
  \bibfield  {author} {\bibinfo {author} {\bibfnamefont {Kieron}\ \bibnamefont
  {Burke}},\ }\bibfield  {title} {\enquote {\bibinfo {title} {Perspective on
  density functional theory},}\ }\href {\doibase 10.1063/1.4704546} {\bibfield
  {journal} {\bibinfo  {journal} {The Journal of Chemical Physics}\ }\textbf
  {\bibinfo {volume} {136}},\ \bibinfo {pages} {150901} (\bibinfo {year}
  {2012})},\ \Eprint
  {http://arxiv.org/abs/https://pubs.aip.org/aip/jcp/article-pdf/doi/10.1063/1.4704546/19858608/150901\_1\_1.4704546.pdf}
  {https://pubs.aip.org/aip/jcp/article-pdf/doi/10.1063/1.4704546/19858608/150901\_1\_1.4704546.pdf}
  \BibitemShut {NoStop}%
\bibitem [{\citenamefont {Goerigk}\ \emph {et~al.}(2017)\citenamefont
  {Goerigk}, \citenamefont {Hansen}, \citenamefont {Bauer}, \citenamefont
  {Ehrlich}, \citenamefont {Najibi},\ and\ \citenamefont
  {Grimme}}]{Goerigk2017dftzoo}%
  \BibitemOpen
  \bibfield  {author} {\bibinfo {author} {\bibfnamefont {Lars}\ \bibnamefont
  {Goerigk}}, \bibinfo {author} {\bibfnamefont {Andreas}\ \bibnamefont
  {Hansen}}, \bibinfo {author} {\bibfnamefont {Christoph}\ \bibnamefont
  {Bauer}}, \bibinfo {author} {\bibfnamefont {Stephan}\ \bibnamefont
  {Ehrlich}}, \bibinfo {author} {\bibfnamefont {Asim}\ \bibnamefont {Najibi}},
  \ and\ \bibinfo {author} {\bibfnamefont {Stefan}\ \bibnamefont {Grimme}},\
  }\bibfield  {title} {\enquote {\bibinfo {title} {A look at the density
  functional theory zoo with the advanced gmtkn55 database for general main
  group thermochemistry{,} kinetics and noncovalent interactions},}\ }\href
  {\doibase 10.1039/C7CP04913G} {\bibfield  {journal} {\bibinfo  {journal}
  {Phys. Chem. Chem. Phys.}\ }\textbf {\bibinfo {volume} {19}},\ \bibinfo
  {pages} {32184--32215} (\bibinfo {year} {2017})}\BibitemShut {NoStop}%
\bibitem [{\citenamefont {Tanaka}\ \emph {et~al.}(2024)\citenamefont {Tanaka},
  \citenamefont {Saito}, \citenamefont {Murata}, \citenamefont {Nakata},\ and\
  \citenamefont {Miyazaki}}]{Tanaka2024largescaleDFT}%
  \BibitemOpen
  \bibfield  {author} {\bibinfo {author} {\bibfnamefont {Atsushi}\ \bibnamefont
  {Tanaka}}, \bibinfo {author} {\bibfnamefont {Atsuki}\ \bibnamefont {Saito}},
  \bibinfo {author} {\bibfnamefont {Takashi}\ \bibnamefont {Murata}}, \bibinfo
  {author} {\bibfnamefont {Ayako}\ \bibnamefont {Nakata}}, \ and\ \bibinfo
  {author} {\bibfnamefont {Tsuyoshi}\ \bibnamefont {Miyazaki}},\ }\bibfield
  {title} {\enquote {\bibinfo {title} {Large-scale dft calculations of
  multi-component glass systems (sio2)0.70(al2o3)0.13(xo)0.17 (x = mg, ca, sr,
  ba) : Accuracy of classical force fields},}\ }\href {\doibase
  https://doi.org/10.1016/j.jnoncrysol.2023.122714} {\bibfield  {journal}
  {\bibinfo  {journal} {Journal of Non-Crystalline Solids}\ }\textbf {\bibinfo
  {volume} {625}},\ \bibinfo {pages} {122714} (\bibinfo {year}
  {2024})}\BibitemShut {NoStop}%
\bibitem [{\citenamefont {Runge}\ and\ \citenamefont
  {Gross}(1984)}]{runge1984density}%
  \BibitemOpen
  \bibfield  {author} {\bibinfo {author} {\bibfnamefont {Erich}\ \bibnamefont
  {Runge}}\ and\ \bibinfo {author} {\bibfnamefont {Eberhard~KU}\ \bibnamefont
  {Gross}},\ }\bibfield  {title} {\enquote {\bibinfo {title}
  {Density-functional theory for time-dependent systems},}\ }\href
  {https://doi.org/10.1103/PhysRevLett.52.997} {\bibfield  {journal} {\bibinfo
  {journal} {Phys. Rev. Lett.}\ }\textbf {\bibinfo {volume} {52}},\ \bibinfo
  {pages} {997} (\bibinfo {year} {1984})}\BibitemShut {NoStop}%
\bibitem [{\citenamefont {Casida}(1995)}]{casida1995timedependent}%
  \BibitemOpen
  \bibfield  {author} {\bibinfo {author} {\bibfnamefont {Mark~E.}\ \bibnamefont
  {Casida}},\ }\bibfield  {title} {\enquote {\bibinfo {title} {Time-dependent
  density functional response theory for molecules},}\ }in\ \href@noop {}
  {\emph {\bibinfo {booktitle} {Recent Advances in Density Functional
  Methods}}},\ Vol.~\bibinfo {volume} {1},\ \bibinfo {editor} {edited by\
  \bibinfo {editor} {\bibfnamefont {D.~P.}\ \bibnamefont {Chong}}}\ (\bibinfo
  {publisher} {World Scientific},\ \bibinfo {year} {1995})\ pp.\ \bibinfo
  {pages} {155--192}\BibitemShut {NoStop}%
\bibitem [{\citenamefont {Casida}\ and\ \citenamefont
  {Huix-Rotllant}(2012)}]{Casida_tddft_review_2012}%
  \BibitemOpen
  \bibfield  {author} {\bibinfo {author} {\bibfnamefont {M.E.}\ \bibnamefont
  {Casida}}\ and\ \bibinfo {author} {\bibfnamefont {M.}~\bibnamefont
  {Huix-Rotllant}},\ }\bibfield  {title} {\enquote {\bibinfo {title} {Progress
  in time-dependent density-functional theory},}\ }\href
  {https://doi.org/10.1146/annurev-physchem-032511-143803} {\bibfield
  {journal} {\bibinfo  {journal} {Annu. Rev. Phys. Chem.}\ }\textbf {\bibinfo
  {volume} {63}},\ \bibinfo {pages} {287} (\bibinfo {year} {2012})}\BibitemShut
  {NoStop}%
\bibitem [{\citenamefont {Elliott}\ \emph {et~al.}(2011)\citenamefont
  {Elliott}, \citenamefont {Goldson}, \citenamefont {Canahui},\ and\
  \citenamefont {Maitra}}]{elliott2011perspectives}%
  \BibitemOpen
  \bibfield  {author} {\bibinfo {author} {\bibfnamefont {Peter}\ \bibnamefont
  {Elliott}}, \bibinfo {author} {\bibfnamefont {Sharma}\ \bibnamefont
  {Goldson}}, \bibinfo {author} {\bibfnamefont {Chris}\ \bibnamefont
  {Canahui}}, \ and\ \bibinfo {author} {\bibfnamefont {Neepa~T.}\ \bibnamefont
  {Maitra}},\ }\bibfield  {title} {\enquote {\bibinfo {title} {Perspectives on
  double-excitations in tddft},}\ }\href {\doibase
  https://doi.org/10.1016/j.chemphys.2011.03.020} {\bibfield  {journal}
  {\bibinfo  {journal} {Chem. Phys.}\ }\textbf {\bibinfo {volume} {391}},\
  \bibinfo {pages} {110--119} (\bibinfo {year} {2011})},\ \bibinfo {note} {open
  problems and new solutions in time dependent density functional
  theory}\BibitemShut {NoStop}%
\bibitem [{\citenamefont {Lacombe}\ and\ \citenamefont
  {Maitra}(2023)}]{Lacombe2023_Non-adiabatic}%
  \BibitemOpen
  \bibfield  {author} {\bibinfo {author} {\bibfnamefont {L.}~\bibnamefont
  {Lacombe}}\ and\ \bibinfo {author} {\bibfnamefont {N.T.}\ \bibnamefont
  {Maitra}},\ }\bibfield  {title} {\enquote {\bibinfo {title} {Non-adiabatic
  approximations in time-dependent density functional theory: progress and
  prospects},}\ }\href {https://doi.org/10.1038/s41524-023-01061-0} {\bibfield
  {journal} {\bibinfo  {journal} {npj Comput Mater}\ }\textbf {\bibinfo
  {volume} {9}},\ \bibinfo {pages} {124} (\bibinfo {year} {2023})}\BibitemShut
  {NoStop}%
\bibitem [{\citenamefont {Maitra}\ \emph
  {et~al.}(2004{\natexlab{a}})\citenamefont {Maitra}, \citenamefont {Zhang},
  \citenamefont {Cave},\ and\ \citenamefont {Burke}}]{Maitra_DTDDFT04}%
  \BibitemOpen
  \bibfield  {author} {\bibinfo {author} {\bibfnamefont {Neepa~T.}\
  \bibnamefont {Maitra}}, \bibinfo {author} {\bibfnamefont {Fan}\ \bibnamefont
  {Zhang}}, \bibinfo {author} {\bibfnamefont {Robert~J.}\ \bibnamefont {Cave}},
  \ and\ \bibinfo {author} {\bibfnamefont {Kieron}\ \bibnamefont {Burke}},\
  }\bibfield  {title} {\enquote {\bibinfo {title} {Double excitations within
  time-dependent density functional theory linear response},}\ }\href {\doibase
  10.1063/1.1651060} {\bibfield  {journal} {\bibinfo  {journal} {The Journal of
  Chemical Physics}\ }\textbf {\bibinfo {volume} {120}},\ \bibinfo {pages}
  {5932--5937} (\bibinfo {year} {2004}{\natexlab{a}})}\BibitemShut {NoStop}%
\bibitem [{\citenamefont {Dar}\ and\ \citenamefont
  {Maitra}(2023)}]{Dar2023_TDDFT_DE}%
  \BibitemOpen
  \bibfield  {author} {\bibinfo {author} {\bibfnamefont {Davood~B.}\
  \bibnamefont {Dar}}\ and\ \bibinfo {author} {\bibfnamefont {Neepa~T.}\
  \bibnamefont {Maitra}},\ }\bibfield  {title} {\enquote {\bibinfo {title}
  {Oscillator strengths and excited-state couplings for double excitations in
  time-dependent density functional theory},}\ }\href {\doibase
  10.1063/5.0176705} {\bibfield  {journal} {\bibinfo  {journal} {The Journal of
  Chemical Physics}\ }\textbf {\bibinfo {volume} {159}},\ \bibinfo {pages}
  {211104} (\bibinfo {year} {2023})},\ \Eprint
  {http://arxiv.org/abs/https://pubs.aip.org/aip/jcp/article-pdf/doi/10.1063/5.0176705/18233586/211104\_1\_5.0176705.pdf}
  {https://pubs.aip.org/aip/jcp/article-pdf/doi/10.1063/5.0176705/18233586/211104\_1\_5.0176705.pdf}
  \BibitemShut {NoStop}%
\bibitem [{\citenamefont {Dar}\ and\ \citenamefont
  {Maitra}(2025)}]{Dar2025_DTDDFT_CI}%
  \BibitemOpen
  \bibfield  {author} {\bibinfo {author} {\bibfnamefont {Davood~B.}\
  \bibnamefont {Dar}}\ and\ \bibinfo {author} {\bibfnamefont {Neepa~T.}\
  \bibnamefont {Maitra}},\ }\bibfield  {title} {\enquote {\bibinfo {title}
  {Capturing the elusive curve-crossing in low-lying states of butadiene with
  dressed tddft},}\ }\href {\doibase 10.1021/acs.jpclett.4c03167} {\bibfield
  {journal} {\bibinfo  {journal} {The Journal of Physical Chemistry Letters}\
  }\textbf {\bibinfo {volume} {16}},\ \bibinfo {pages} {703--709} (\bibinfo
  {year} {2025})},\ \Eprint
  {http://arxiv.org/abs/https://doi.org/10.1021/acs.jpclett.4c03167}
  {https://doi.org/10.1021/acs.jpclett.4c03167} \BibitemShut {NoStop}%
\bibitem [{\citenamefont {Baranova}\ and\ \citenamefont
  {Maitra}(2025)}]{baranova2025_excited-state_densities}%
  \BibitemOpen
  \bibfield  {author} {\bibinfo {author} {\bibfnamefont {Anna}\ \bibnamefont
  {Baranova}}\ and\ \bibinfo {author} {\bibfnamefont {Neepa~T.}\ \bibnamefont
  {Maitra}},\ }\bibfield  {title} {\enquote {\bibinfo {title} {Excited-state
  densities from time-dependent density functional response theory},}\ }\href
  {\doibase 10.1021/acs.jctc.5c00909} {\bibfield  {journal} {\bibinfo
  {journal} {Journal of Chemical Theory and Computation}\ }\textbf {\bibinfo
  {volume} {21}},\ \bibinfo {pages} {10437--10451} (\bibinfo {year}
  {2025})}\BibitemShut {NoStop}%
\bibitem [{\citenamefont {Matsika}(2021)}]{Matsika2021elstructCI}%
  \BibitemOpen
  \bibfield  {author} {\bibinfo {author} {\bibfnamefont {Spiridoula}\
  \bibnamefont {Matsika}},\ }\bibfield  {title} {\enquote {\bibinfo {title}
  {Electronic structure methods for the description of nonadiabatic effects and
  conical intersections},}\ }\href {\doibase 10.1021/acs.chemrev.1c00074}
  {\bibfield  {journal} {\bibinfo  {journal} {Chemical Reviews}\ }\textbf
  {\bibinfo {volume} {121}},\ \bibinfo {pages} {9407--9449} (\bibinfo {year}
  {2021})},\ \Eprint
  {http://arxiv.org/abs/https://doi.org/10.1021/acs.chemrev.1c00074}
  {https://doi.org/10.1021/acs.chemrev.1c00074} \BibitemShut {NoStop}%
\bibitem [{\citenamefont {Maitra}(2017)}]{maitra2017charge}%
  \BibitemOpen
  \bibfield  {author} {\bibinfo {author} {\bibfnamefont {Neepa~T}\ \bibnamefont
  {Maitra}},\ }\bibfield  {title} {\enquote {\bibinfo {title} {Charge transfer
  in time-dependent density functional theory},}\ }\href@noop {} {\bibfield
  {journal} {\bibinfo  {journal} {J. Phys. Condens. Matter}\ }\textbf {\bibinfo
  {volume} {29}},\ \bibinfo {pages} {423001} (\bibinfo {year}
  {2017})}\BibitemShut {NoStop}%
\bibitem [{\citenamefont {Perdew}\ \emph {et~al.}(1982)\citenamefont {Perdew},
  \citenamefont {Parr}, \citenamefont {Levy},\ and\ \citenamefont
  {Balduz~Jr}}]{perdew1982density}%
  \BibitemOpen
  \bibfield  {author} {\bibinfo {author} {\bibfnamefont {John~P}\ \bibnamefont
  {Perdew}}, \bibinfo {author} {\bibfnamefont {Robert~G}\ \bibnamefont {Parr}},
  \bibinfo {author} {\bibfnamefont {Mel}\ \bibnamefont {Levy}}, \ and\ \bibinfo
  {author} {\bibfnamefont {Jose~L}\ \bibnamefont {Balduz~Jr}},\ }\bibfield
  {title} {\enquote {\bibinfo {title} {Density-functional theory for fractional
  particle number: derivative discontinuities of the energy},}\ }\href
  {https://doi.org/10.1103/PhysRevLett.49.1691} {\bibfield  {journal} {\bibinfo
   {journal} {Phys. Rev. Lett.}\ }\textbf {\bibinfo {volume} {49}},\ \bibinfo
  {pages} {1691} (\bibinfo {year} {1982})}\BibitemShut {NoStop}%
\bibitem [{\citenamefont {Mori-S{\'a}nchez}\ \emph {et~al.}(2008)\citenamefont
  {Mori-S{\'a}nchez}, \citenamefont {Cohen},\ and\ \citenamefont
  {Yang}}]{mori2008localization}%
  \BibitemOpen
  \bibfield  {author} {\bibinfo {author} {\bibfnamefont {Paula}\ \bibnamefont
  {Mori-S{\'a}nchez}}, \bibinfo {author} {\bibfnamefont {Aron~J}\ \bibnamefont
  {Cohen}}, \ and\ \bibinfo {author} {\bibfnamefont {Weitao}\ \bibnamefont
  {Yang}},\ }\bibfield  {title} {\enquote {\bibinfo {title} {Localization and
  delocalization errors in density functional theory and implications for
  band-gap prediction},}\ }\href@noop {} {\bibfield  {journal} {\bibinfo
  {journal} {Phys. Rev. Lett.}\ }\textbf {\bibinfo {volume} {100}},\ \bibinfo
  {pages} {146401} (\bibinfo {year} {2008})}\BibitemShut {NoStop}%
\bibitem [{\citenamefont {Cohen}\ \emph
  {et~al.}(2008{\natexlab{a}})\citenamefont {Cohen}, \citenamefont
  {Mori-S{\'a}nchez},\ and\ \citenamefont {Yang}}]{cohen2008fractional}%
  \BibitemOpen
  \bibfield  {author} {\bibinfo {author} {\bibfnamefont {Aron~J}\ \bibnamefont
  {Cohen}}, \bibinfo {author} {\bibfnamefont {Paula}\ \bibnamefont
  {Mori-S{\'a}nchez}}, \ and\ \bibinfo {author} {\bibfnamefont {Weitao}\
  \bibnamefont {Yang}},\ }\bibfield  {title} {\enquote {\bibinfo {title}
  {Fractional charge perspective on the band gap in density-functional
  theory},}\ }\href {https://doi.org/10.1103/PhysRevB.77.115123} {\bibfield
  {journal} {\bibinfo  {journal} {Phys. Rev. B}\ }\textbf {\bibinfo {volume}
  {77}},\ \bibinfo {pages} {115123} (\bibinfo {year}
  {2008}{\natexlab{a}})}\BibitemShut {NoStop}%
\bibitem [{\citenamefont {Cohen}\ \emph
  {et~al.}(2008{\natexlab{b}})\citenamefont {Cohen}, \citenamefont
  {Mori-S{\'a}nchez},\ and\ \citenamefont {Yang}}]{cohen2008insights}%
  \BibitemOpen
  \bibfield  {author} {\bibinfo {author} {\bibfnamefont {Aron~J}\ \bibnamefont
  {Cohen}}, \bibinfo {author} {\bibfnamefont {Paula}\ \bibnamefont
  {Mori-S{\'a}nchez}}, \ and\ \bibinfo {author} {\bibfnamefont {Weitao}\
  \bibnamefont {Yang}},\ }\bibfield  {title} {\enquote {\bibinfo {title}
  {Insights into current limitations of density functional theory},}\
  }\href@noop {} {\bibfield  {journal} {\bibinfo  {journal} {Science}\ }\textbf
  {\bibinfo {volume} {321}},\ \bibinfo {pages} {792--794} (\bibinfo {year}
  {2008}{\natexlab{b}})}\BibitemShut {NoStop}%
\bibitem [{\citenamefont {Stein}\ \emph {et~al.}(2010)\citenamefont {Stein},
  \citenamefont {Eisenberg}, \citenamefont {Kronik},\ and\ \citenamefont
  {Baer}}]{stein2010fundamental}%
  \BibitemOpen
  \bibfield  {author} {\bibinfo {author} {\bibfnamefont {Tamar}\ \bibnamefont
  {Stein}}, \bibinfo {author} {\bibfnamefont {Helen}\ \bibnamefont
  {Eisenberg}}, \bibinfo {author} {\bibfnamefont {Leeor}\ \bibnamefont
  {Kronik}}, \ and\ \bibinfo {author} {\bibfnamefont {Roi}\ \bibnamefont
  {Baer}},\ }\bibfield  {title} {\enquote {\bibinfo {title} {Fundamental gaps
  in finite systems from eigenvalues of a generalized kohn-sham method},}\
  }\href@noop {} {\bibfield  {journal} {\bibinfo  {journal} {Phys. Rev. Lett.}\
  }\textbf {\bibinfo {volume} {105}},\ \bibinfo {pages} {266802} (\bibinfo
  {year} {2010})}\BibitemShut {NoStop}%
\bibitem [{\citenamefont {Zheng}\ \emph {et~al.}(2011)\citenamefont {Zheng},
  \citenamefont {Cohen}, \citenamefont {Mori-S{\'a}nchez}, \citenamefont {Hu},\
  and\ \citenamefont {Yang}}]{zheng2011improving}%
  \BibitemOpen
  \bibfield  {author} {\bibinfo {author} {\bibfnamefont {Xiao}\ \bibnamefont
  {Zheng}}, \bibinfo {author} {\bibfnamefont {Aron~J}\ \bibnamefont {Cohen}},
  \bibinfo {author} {\bibfnamefont {Paula}\ \bibnamefont {Mori-S{\'a}nchez}},
  \bibinfo {author} {\bibfnamefont {Xiangqian}\ \bibnamefont {Hu}}, \ and\
  \bibinfo {author} {\bibfnamefont {Weitao}\ \bibnamefont {Yang}},\ }\bibfield
  {title} {\enquote {\bibinfo {title} {Improving band gap prediction in density
  functional theory from molecules to solids},}\ }\href
  {https://doi.org/10.1103/PhysRevLett.107.026403} {\bibfield  {journal}
  {\bibinfo  {journal} {Phys. Rev. Lett.}\ }\textbf {\bibinfo {volume} {107}},\
  \bibinfo {pages} {026403} (\bibinfo {year} {2011})}\BibitemShut {NoStop}%
\bibitem [{\citenamefont {Cohen}\ \emph {et~al.}(2011)\citenamefont {Cohen},
  \citenamefont {Mori-S{\'a}nchez},\ and\ \citenamefont
  {Yang}}]{cohen2011challenges}%
  \BibitemOpen
  \bibfield  {author} {\bibinfo {author} {\bibfnamefont {Aron~J}\ \bibnamefont
  {Cohen}}, \bibinfo {author} {\bibfnamefont {Paula}\ \bibnamefont
  {Mori-S{\'a}nchez}}, \ and\ \bibinfo {author} {\bibfnamefont {Weitao}\
  \bibnamefont {Yang}},\ }\bibfield  {title} {\enquote {\bibinfo {title}
  {Challenges for density functional theory},}\ }\href {\doibase
  10.1021/cr200107z} {\bibfield  {journal} {\bibinfo  {journal} {Chem. Rev.}\
  }\textbf {\bibinfo {volume} {112}},\ \bibinfo {pages} {289--320} (\bibinfo
  {year} {2011})}\BibitemShut {NoStop}%
\bibitem [{\citenamefont {Kraisler}\ and\ \citenamefont
  {Kronik}(2013)}]{kraisler2013piecewise}%
  \BibitemOpen
  \bibfield  {author} {\bibinfo {author} {\bibfnamefont {Eli}\ \bibnamefont
  {Kraisler}}\ and\ \bibinfo {author} {\bibfnamefont {Leeor}\ \bibnamefont
  {Kronik}},\ }\bibfield  {title} {\enquote {\bibinfo {title} {Piecewise
  linearity of approximate density functionals revisited: implications for
  frontier orbital energies},}\ }\href
  {https://doi.org/10.1103/PhysRevLett.110.126403} {\bibfield  {journal}
  {\bibinfo  {journal} {Phys. Rev. Lett.}\ }\textbf {\bibinfo {volume} {110}},\
  \bibinfo {pages} {126403} (\bibinfo {year} {2013})}\BibitemShut {NoStop}%
\bibitem [{\citenamefont {Kraisler}\ and\ \citenamefont
  {Kronik}(2014)}]{kraisler2014fundamental}%
  \BibitemOpen
  \bibfield  {author} {\bibinfo {author} {\bibfnamefont {Eli}\ \bibnamefont
  {Kraisler}}\ and\ \bibinfo {author} {\bibfnamefont {Leeor}\ \bibnamefont
  {Kronik}},\ }\bibfield  {title} {\enquote {\bibinfo {title} {Fundamental gaps
  with approximate density functionals: The derivative discontinuity revealed
  from ensemble considerations},}\ }\href@noop {} {\bibfield  {journal}
  {\bibinfo  {journal} {J. Chem. Phys.}\ }\textbf {\bibinfo {volume} {140}},\
  \bibinfo {pages} {18A540} (\bibinfo {year} {2014})}\BibitemShut {NoStop}%
\bibitem [{\citenamefont {Perdew}\ \emph {et~al.}(2017)\citenamefont {Perdew},
  \citenamefont {Yang}, \citenamefont {Burke}, \citenamefont {Yang},
  \citenamefont {Gross}, \citenamefont {Scheffler}, \citenamefont {Scuseria},
  \citenamefont {Henderson}, \citenamefont {Zhang}, \citenamefont {Ruzsinszky}
  \emph {et~al.}}]{perdew2017understanding}%
  \BibitemOpen
  \bibfield  {author} {\bibinfo {author} {\bibfnamefont {John~P}\ \bibnamefont
  {Perdew}}, \bibinfo {author} {\bibfnamefont {Weitao}\ \bibnamefont {Yang}},
  \bibinfo {author} {\bibfnamefont {Kieron}\ \bibnamefont {Burke}}, \bibinfo
  {author} {\bibfnamefont {Zenghui}\ \bibnamefont {Yang}}, \bibinfo {author}
  {\bibfnamefont {Eberhard K~U}\ \bibnamefont {Gross}}, \bibinfo {author}
  {\bibfnamefont {Matthias}\ \bibnamefont {Scheffler}}, \bibinfo {author}
  {\bibfnamefont {Gustavo~E}\ \bibnamefont {Scuseria}}, \bibinfo {author}
  {\bibfnamefont {Thomas~M}\ \bibnamefont {Henderson}}, \bibinfo {author}
  {\bibfnamefont {Igor~Ying}\ \bibnamefont {Zhang}}, \bibinfo {author}
  {\bibfnamefont {Adrienn}\ \bibnamefont {Ruzsinszky}},  \emph {et~al.},\
  }\bibfield  {title} {\enquote {\bibinfo {title} {Understanding band gaps of
  solids in generalized kohn--sham theory},}\ }\href@noop {} {\bibfield
  {journal} {\bibinfo  {journal} {Proc. Natl. Acad. Sci.}\ }\textbf {\bibinfo
  {volume} {114}},\ \bibinfo {pages} {2801--2806} (\bibinfo {year}
  {2017})}\BibitemShut {NoStop}%
\bibitem [{\citenamefont {Hellgren}\ and\ \citenamefont
  {Gross}(2012)}]{Hellgren12_Effect}%
  \BibitemOpen
  \bibfield  {author} {\bibinfo {author} {\bibfnamefont {Maria}\ \bibnamefont
  {Hellgren}}\ and\ \bibinfo {author} {\bibfnamefont {E.~K.~U.}\ \bibnamefont
  {Gross}},\ }\bibfield  {title} {\enquote {\bibinfo {title} {{Effect of
  discontinuities in Kohn-Sham-based chemical reactivity theory}},}\ }\href
  {\doibase 10.1063/1.3694103} {\bibfield  {journal} {\bibinfo  {journal} {The
  Journal of Chemical Physics}\ }\textbf {\bibinfo {volume} {136}},\ \bibinfo
  {pages} {114102} (\bibinfo {year} {2012})}\BibitemShut {NoStop}%
\bibitem [{\citenamefont {Hodgson}\ \emph {et~al.}(2017)\citenamefont
  {Hodgson}, \citenamefont {Kraisler}, \citenamefont {Schild},\ and\
  \citenamefont {Gross}}]{hodgson2017interatomic}%
  \BibitemOpen
  \bibfield  {author} {\bibinfo {author} {\bibfnamefont {Matthew~JP}\
  \bibnamefont {Hodgson}}, \bibinfo {author} {\bibfnamefont {Eli}\ \bibnamefont
  {Kraisler}}, \bibinfo {author} {\bibfnamefont {Axel}\ \bibnamefont {Schild}},
  \ and\ \bibinfo {author} {\bibfnamefont {Eberhard~KU}\ \bibnamefont
  {Gross}},\ }\bibfield  {title} {\enquote {\bibinfo {title} {How interatomic
  steps in the exact kohn--sham potential relate to derivative discontinuities
  of the energy},}\ }\href {https://doi.org/10.1021/acs.jpclett.7b02615}
  {\bibfield  {journal} {\bibinfo  {journal} {J. Phys. Chem. Lett.}\ }\textbf
  {\bibinfo {volume} {8}},\ \bibinfo {pages} {5974--5980} (\bibinfo {year}
  {2017})}\BibitemShut {NoStop}%
\bibitem [{\citenamefont {Kraisler}\ \emph {et~al.}(2021)\citenamefont
  {Kraisler}, \citenamefont {Hodgson},\ and\ \citenamefont
  {Gross}}]{Kraisler21_From}%
  \BibitemOpen
  \bibfield  {author} {\bibinfo {author} {\bibfnamefont {Eli}\ \bibnamefont
  {Kraisler}}, \bibinfo {author} {\bibfnamefont {M.~J.~P.}\ \bibnamefont
  {Hodgson}}, \ and\ \bibinfo {author} {\bibfnamefont {E.~K.~U.}\ \bibnamefont
  {Gross}},\ }\bibfield  {title} {\enquote {\bibinfo {title} {From kohn–sham
  to many-electron energies via step structures in the exchange-correlation
  potential},}\ }\href {\doibase 10.1021/acs.jctc.0c01093} {\bibfield
  {journal} {\bibinfo  {journal} {Journal of Chemical Theory and Computation}\
  }\textbf {\bibinfo {volume} {17}},\ \bibinfo {pages} {1390--1407} (\bibinfo
  {year} {2021})}\BibitemShut {NoStop}%
\bibitem [{\citenamefont {Cernatic}\ \emph {et~al.}(2022)\citenamefont
  {Cernatic}, \citenamefont {Senjean}, \citenamefont {Robert},\ and\
  \citenamefont {Fromager}}]{Cernatic2022}%
  \BibitemOpen
  \bibfield  {author} {\bibinfo {author} {\bibfnamefont {F.}~\bibnamefont
  {Cernatic}}, \bibinfo {author} {\bibfnamefont {B.}~\bibnamefont {Senjean}},
  \bibinfo {author} {\bibfnamefont {V.}~\bibnamefont {Robert}}, \ and\ \bibinfo
  {author} {\bibfnamefont {E.}~\bibnamefont {Fromager}},\ }\bibfield  {title}
  {\enquote {\bibinfo {title} {Ensemble density functional theory of neutral
  and charged excitations},}\ }\href
  {https://doi.org/10.1007/s41061-021-00359-1} {\bibfield  {journal} {\bibinfo
  {journal} {Top Curr Chem (Z)}\ }\textbf {\bibinfo {volume} {380}},\ \bibinfo
  {pages} {4} (\bibinfo {year} {2022})}\BibitemShut {NoStop}%
\bibitem [{\citenamefont {Gould}\ \emph {et~al.}(2026)\citenamefont {Gould},
  \citenamefont {Kronik},\ and\ \citenamefont
  {Pittalis}}]{gould2026ensemblization}%
  \BibitemOpen
  \bibfield  {author} {\bibinfo {author} {\bibfnamefont {Tim}\ \bibnamefont
  {Gould}}, \bibinfo {author} {\bibfnamefont {Leeor}\ \bibnamefont {Kronik}}, \
  and\ \bibinfo {author} {\bibfnamefont {Stefano}\ \bibnamefont {Pittalis}},\
  }\bibfield  {title} {\enquote {\bibinfo {title} {“ensemblization” of
  density functional theory},}\ }\href {\doibase 10.1063/5.0274509} {\bibfield
  {journal} {\bibinfo  {journal} {The Journal of Chemical Physics}\ }\textbf
  {\bibinfo {volume} {164}},\ \bibinfo {pages} {040901} (\bibinfo {year}
  {2026})}\BibitemShut {NoStop}%
\bibitem [{\citenamefont {Theophilou}(1979)}]{JPC79_Theophilou_equi-ensembles}%
  \BibitemOpen
  \bibfield  {author} {\bibinfo {author} {\bibfnamefont {A~K}\ \bibnamefont
  {Theophilou}},\ }\bibfield  {title} {\enquote {\bibinfo {title} {The energy
  density functional formalism for excited states},}\ }\href
  {https://iopscience.iop.org/article/10.1088/0022-3719/12/24/013} {\bibfield
  {journal} {\bibinfo  {journal} {J. Phys. C: Solid State Phys.}\ }\textbf
  {\bibinfo {volume} {12}},\ \bibinfo {pages} {5419} (\bibinfo {year}
  {1979})}\BibitemShut {NoStop}%
\bibitem [{\citenamefont {Theophilou}(1987)}]{theophilou_book}%
  \BibitemOpen
  \bibfield  {author} {\bibinfo {author} {\bibfnamefont {A~K}\ \bibnamefont
  {Theophilou}},\ }\enquote {\bibinfo {title} {The single particle density in
  physics and chemistry},}\ \ (\bibinfo  {publisher} {Academic Press},\
  \bibinfo {year} {1987})\ pp.\ \bibinfo {pages} {210--212}\BibitemShut
  {NoStop}%
\bibitem [{\citenamefont {Gross}\ \emph
  {et~al.}(1988{\natexlab{a}})\citenamefont {Gross}, \citenamefont {Oliveira},\
  and\ \citenamefont {Kohn}}]{gross1988rayleigh}%
  \BibitemOpen
  \bibfield  {author} {\bibinfo {author} {\bibfnamefont {E.~K.~U.}\
  \bibnamefont {Gross}}, \bibinfo {author} {\bibfnamefont {L.~N.}\ \bibnamefont
  {Oliveira}}, \ and\ \bibinfo {author} {\bibfnamefont {W.}~\bibnamefont
  {Kohn}},\ }\bibfield  {title} {\enquote {\bibinfo {title} {Rayleigh-ritz
  variational principle for ensembles of fractionally occupied states},}\
  }\href {\doibase 10.1103/PhysRevA.37.2805} {\bibfield  {journal} {\bibinfo
  {journal} {Phys. Rev. A}\ }\textbf {\bibinfo {volume} {37}},\ \bibinfo
  {pages} {2805} (\bibinfo {year} {1988}{\natexlab{a}})}\BibitemShut {NoStop}%
\bibitem [{\citenamefont {Gross}\ \emph
  {et~al.}(1988{\natexlab{b}})\citenamefont {Gross}, \citenamefont {Oliveira},\
  and\ \citenamefont {Kohn}}]{gross1988density}%
  \BibitemOpen
  \bibfield  {author} {\bibinfo {author} {\bibfnamefont {E.~K.~U.}\
  \bibnamefont {Gross}}, \bibinfo {author} {\bibfnamefont {L.~N.}\ \bibnamefont
  {Oliveira}}, \ and\ \bibinfo {author} {\bibfnamefont {W.}~\bibnamefont
  {Kohn}},\ }\bibfield  {title} {\enquote {\bibinfo {title} {Density-functional
  theory for ensembles of fractionally occupied states. i. basic formalism},}\
  }\href {\doibase 10.1103/PhysRevA.37.2809} {\bibfield  {journal} {\bibinfo
  {journal} {Phys. Rev. A}\ }\textbf {\bibinfo {volume} {37}},\ \bibinfo
  {pages} {2809} (\bibinfo {year} {1988}{\natexlab{b}})}\BibitemShut {NoStop}%
\bibitem [{\citenamefont {Sagredo}\ and\ \citenamefont
  {Burke}(2018)}]{sagredo2018can}%
  \BibitemOpen
  \bibfield  {author} {\bibinfo {author} {\bibfnamefont {Francisca}\
  \bibnamefont {Sagredo}}\ and\ \bibinfo {author} {\bibfnamefont {Kieron}\
  \bibnamefont {Burke}},\ }\bibfield  {title} {\enquote {\bibinfo {title}
  {Accurate double excitations from ensemble density functional
  calculations},}\ }\href {\doibase 10.1063/1.5043411} {\bibfield  {journal}
  {\bibinfo  {journal} {J. Chem. Phys.}\ }\textbf {\bibinfo {volume} {149}},\
  \bibinfo {pages} {134103} (\bibinfo {year} {2018})}\BibitemShut {NoStop}%
\bibitem [{\citenamefont {Marut}\ \emph {et~al.}(2020)\citenamefont {Marut},
  \citenamefont {Senjean}, \citenamefont {Fromager},\ and\ \citenamefont
  {Loos}}]{marut2020weight}%
  \BibitemOpen
  \bibfield  {author} {\bibinfo {author} {\bibfnamefont {Clotilde}\
  \bibnamefont {Marut}}, \bibinfo {author} {\bibfnamefont {Bruno}\ \bibnamefont
  {Senjean}}, \bibinfo {author} {\bibfnamefont {Emmanuel}\ \bibnamefont
  {Fromager}}, \ and\ \bibinfo {author} {\bibfnamefont {Pierre-Fran{\c c}ois}\
  \bibnamefont {Loos}},\ }\bibfield  {title} {\enquote {\bibinfo {title}
  {Weight dependence of local exchange--correlation functionals in ensemble
  density-functional theory: double excitations in two-electron systems},}\
  }\href {\doibase 10.1039/D0FD00059K} {\bibfield  {journal} {\bibinfo
  {journal} {Faraday Discuss.}\ }\textbf {\bibinfo {volume} {224}},\ \bibinfo
  {pages} {402--423} (\bibinfo {year} {2020})}\BibitemShut {NoStop}%
\bibitem [{\citenamefont {Gould}\ \emph {et~al.}(2021)\citenamefont {Gould},
  \citenamefont {Kronik},\ and\ \citenamefont {Pittalis}}]{Gould2021_Double}%
  \BibitemOpen
  \bibfield  {author} {\bibinfo {author} {\bibfnamefont {Tim}\ \bibnamefont
  {Gould}}, \bibinfo {author} {\bibfnamefont {Leeor}\ \bibnamefont {Kronik}}, \
  and\ \bibinfo {author} {\bibfnamefont {Stefano}\ \bibnamefont {Pittalis}},\
  }\bibfield  {title} {\enquote {\bibinfo {title} {Double excitations in
  molecules from ensemble density functionals: Theory and approximations},}\
  }\href {\doibase 10.1103/PhysRevA.104.022803} {\bibfield  {journal} {\bibinfo
   {journal} {Phys. Rev. A}\ }\textbf {\bibinfo {volume} {104}},\ \bibinfo
  {pages} {022803} (\bibinfo {year} {2021})}\BibitemShut {NoStop}%
\bibitem [{\citenamefont {Ullrich}\ and\ \citenamefont
  {Kohn}(2001)}]{Ullrich2001EDFTdegen}%
  \BibitemOpen
  \bibfield  {author} {\bibinfo {author} {\bibfnamefont {C.~A.}\ \bibnamefont
  {Ullrich}}\ and\ \bibinfo {author} {\bibfnamefont {W.}~\bibnamefont {Kohn}},\
  }\bibfield  {title} {\enquote {\bibinfo {title} {Kohn-sham theory for
  ground-state ensembles},}\ }\href {\doibase 10.1103/PhysRevLett.87.093001}
  {\bibfield  {journal} {\bibinfo  {journal} {Phys. Rev. Lett.}\ }\textbf
  {\bibinfo {volume} {87}},\ \bibinfo {pages} {093001} (\bibinfo {year}
  {2001})}\BibitemShut {NoStop}%
\bibitem [{\citenamefont {Deur}\ \emph {et~al.}(2018)\citenamefont {Deur},
  \citenamefont {Mazouin}, \citenamefont {Senjean},\ and\ \citenamefont
  {Fromager}}]{deur2018exploring}%
  \BibitemOpen
  \bibfield  {author} {\bibinfo {author} {\bibfnamefont {Killian}\ \bibnamefont
  {Deur}}, \bibinfo {author} {\bibfnamefont {Laurent}\ \bibnamefont {Mazouin}},
  \bibinfo {author} {\bibfnamefont {Bruno}\ \bibnamefont {Senjean}}, \ and\
  \bibinfo {author} {\bibfnamefont {Emmanuel}\ \bibnamefont {Fromager}},\
  }\bibfield  {title} {\enquote {\bibinfo {title} {Exploring weight-dependent
  density-functional approximations for ensembles in the hubbard dimer},}\
  }\href {https://doi.org/10.1140/epjb/e2018-90124-7} {\bibfield  {journal}
  {\bibinfo  {journal} {Eur. Phys. J. B}\ }\textbf {\bibinfo {volume} {91}},\
  \bibinfo {pages} {162} (\bibinfo {year} {2018})}\BibitemShut {NoStop}%
\bibitem [{\citenamefont {Deur}\ and\ \citenamefont
  {Fromager}(2019)}]{deur2019ground}%
  \BibitemOpen
  \bibfield  {author} {\bibinfo {author} {\bibfnamefont {Killian}\ \bibnamefont
  {Deur}}\ and\ \bibinfo {author} {\bibfnamefont {Emmanuel}\ \bibnamefont
  {Fromager}},\ }\bibfield  {title} {\enquote {\bibinfo {title} {Ground and
  excited energy levels can be extracted exactly from a single ensemble
  density-functional theory calculation},}\ }\href
  {https://doi.org/10.1063/1.5084312} {\bibfield  {journal} {\bibinfo
  {journal} {J. Chem. Phys.}\ }\textbf {\bibinfo {volume} {150}},\ \bibinfo
  {pages} {094106} (\bibinfo {year} {2019})}\BibitemShut {NoStop}%
\bibitem [{\citenamefont {Fromager}(2020)}]{Fromager_2020}%
  \BibitemOpen
  \bibfield  {author} {\bibinfo {author} {\bibfnamefont {Emmanuel}\
  \bibnamefont {Fromager}},\ }\bibfield  {title} {\enquote {\bibinfo {title}
  {Individual correlations in ensemble density functional theory: State- and
  density-driven decompositions without additional kohn-sham systems},}\ }\href
  {\doibase 10.1103/PhysRevLett.124.243001} {\bibfield  {journal} {\bibinfo
  {journal} {Phys. Rev. Lett.}\ }\textbf {\bibinfo {volume} {124}},\ \bibinfo
  {pages} {243001} (\bibinfo {year} {2020})}\BibitemShut {NoStop}%
\bibitem [{\citenamefont {Scott}\ \emph {et~al.}(2024)\citenamefont {Scott},
  \citenamefont {Kozlowski}, \citenamefont {Crisostomo}, \citenamefont
  {Pribram-Jones},\ and\ \citenamefont {Burke}}]{scott2024exact}%
  \BibitemOpen
  \bibfield  {author} {\bibinfo {author} {\bibfnamefont {Thais~R.}\
  \bibnamefont {Scott}}, \bibinfo {author} {\bibfnamefont {John}\ \bibnamefont
  {Kozlowski}}, \bibinfo {author} {\bibfnamefont {Steven}\ \bibnamefont
  {Crisostomo}}, \bibinfo {author} {\bibfnamefont {Aurora}\ \bibnamefont
  {Pribram-Jones}}, \ and\ \bibinfo {author} {\bibfnamefont {Kieron}\
  \bibnamefont {Burke}},\ }\bibfield  {title} {\enquote {\bibinfo {title}
  {Exact conditions for ensemble density functional theory},}\ }\href {\doibase
  10.1103/PhysRevB.109.195120} {\bibfield  {journal} {\bibinfo  {journal}
  {Phys. Rev. B}\ }\textbf {\bibinfo {volume} {109}},\ \bibinfo {pages}
  {195120} (\bibinfo {year} {2024})}\BibitemShut {NoStop}%
\bibitem [{\citenamefont {Giarrusso}\ and\ \citenamefont
  {Loos}(2023)}]{Giarrusso2023_Exact}%
  \BibitemOpen
  \bibfield  {author} {\bibinfo {author} {\bibfnamefont {Sara}\ \bibnamefont
  {Giarrusso}}\ and\ \bibinfo {author} {\bibfnamefont {Pierre-Fran{\c c}ois}\
  \bibnamefont {Loos}},\ }\bibfield  {title} {\enquote {\bibinfo {title} {Exact
  excited-state functionals of the asymmetric hubbard dimer},}\ }\href
  {\doibase 10.1021/acs.jpclett.3c02052} {\bibfield  {journal} {\bibinfo
  {journal} {J. Phys. Chem. Lett.}\ }\textbf {\bibinfo {volume} {14}},\
  \bibinfo {pages} {8780--8786} (\bibinfo {year} {2023})}\BibitemShut {NoStop}%
\bibitem [{\citenamefont {Loos}\ and\ \citenamefont
  {Giarrusso}(2025)}]{Giarrusso_excDFT25}%
  \BibitemOpen
  \bibfield  {author} {\bibinfo {author} {\bibfnamefont {Pierre-François}\
  \bibnamefont {Loos}}\ and\ \bibinfo {author} {\bibfnamefont {Sara}\
  \bibnamefont {Giarrusso}},\ }\bibfield  {title} {\enquote {\bibinfo {title}
  {Excited-state-specific kohn–sham formalism for the asymmetric hubbard
  dimer},}\ }\href {\doibase 10.1063/5.0255324} {\bibfield  {journal} {\bibinfo
   {journal} {The Journal of Chemical Physics}\ }\textbf {\bibinfo {volume}
  {162}},\ \bibinfo {pages} {144104} (\bibinfo {year} {2025})}\BibitemShut
  {NoStop}%
\bibitem [{\citenamefont {Loos}(2025)}]{Loos2025excstUEG}%
  \BibitemOpen
  \bibfield  {author} {\bibinfo {author} {\bibfnamefont {Pierre-François}\
  \bibnamefont {Loos}},\ }\bibfield  {title} {\enquote {\bibinfo {title}
  {Excited states of the uniform electron gas},}\ }\href {\doibase
  10.1063/5.0263799} {\bibfield  {journal} {\bibinfo  {journal} {The Journal of
  Chemical Physics}\ }\textbf {\bibinfo {volume} {162}},\ \bibinfo {pages}
  {204105} (\bibinfo {year} {2025})},\ \Eprint
  {http://arxiv.org/abs/https://pubs.aip.org/aip/jcp/article-pdf/doi/10.1063/5.0263799/20528988/204105\_1\_5.0263799.pdf}
  {https://pubs.aip.org/aip/jcp/article-pdf/doi/10.1063/5.0263799/20528988/204105\_1\_5.0263799.pdf}
  \BibitemShut {NoStop}%
\bibitem [{\citenamefont {Gould}\ and\ \citenamefont
  {Pittalis}(2017)}]{gould2017hartree}%
  \BibitemOpen
  \bibfield  {author} {\bibinfo {author} {\bibfnamefont {Tim}\ \bibnamefont
  {Gould}}\ and\ \bibinfo {author} {\bibfnamefont {Stefano}\ \bibnamefont
  {Pittalis}},\ }\bibfield  {title} {\enquote {\bibinfo {title} {Hartree and
  exchange in ensemble density functional theory: Avoiding the nonuniqueness
  disaster},}\ }\href {https://doi.org/10.1103/PhysRevLett.119.243001}
  {\bibfield  {journal} {\bibinfo  {journal} {Phys. Rev. Lett.}\ }\textbf
  {\bibinfo {volume} {119}},\ \bibinfo {pages} {243001} (\bibinfo {year}
  {2017})}\BibitemShut {NoStop}%
\bibitem [{\citenamefont {Gould}\ \emph {et~al.}(2019)\citenamefont {Gould},
  \citenamefont {Pittalis}, \citenamefont {Toulouse}, \citenamefont
  {Kraisler},\ and\ \citenamefont {Kronik}}]{gould2019asymptotic}%
  \BibitemOpen
  \bibfield  {author} {\bibinfo {author} {\bibfnamefont {Tim}\ \bibnamefont
  {Gould}}, \bibinfo {author} {\bibfnamefont {Stefano}\ \bibnamefont
  {Pittalis}}, \bibinfo {author} {\bibfnamefont {Julien}\ \bibnamefont
  {Toulouse}}, \bibinfo {author} {\bibfnamefont {Eli}\ \bibnamefont
  {Kraisler}}, \ and\ \bibinfo {author} {\bibfnamefont {Leeor}\ \bibnamefont
  {Kronik}},\ }\bibfield  {title} {\enquote {\bibinfo {title} {Asymptotic
  behavior of the hartree-exchange and correlation potentials in ensemble
  density functional theory},}\ }\href {https://doi.org/10.1039/C9CP03633D}
  {\bibfield  {journal} {\bibinfo  {journal} {Phys. Chem. Chem. Phys.}\
  }\textbf {\bibinfo {volume} {21}},\ \bibinfo {pages} {19805--19815} (\bibinfo
  {year} {2019})}\BibitemShut {NoStop}%
\bibitem [{\citenamefont {Gould}\ \emph {et~al.}(2020)\citenamefont {Gould},
  \citenamefont {Stefanucci},\ and\ \citenamefont
  {Pittalis}}]{PRL20_Gould_Hartree_def_from_ACDF_th}%
  \BibitemOpen
  \bibfield  {author} {\bibinfo {author} {\bibfnamefont {Tim}\ \bibnamefont
  {Gould}}, \bibinfo {author} {\bibfnamefont {Gianluca}\ \bibnamefont
  {Stefanucci}}, \ and\ \bibinfo {author} {\bibfnamefont {Stefano}\
  \bibnamefont {Pittalis}},\ }\bibfield  {title} {\enquote {\bibinfo {title}
  {Ensemble density functional theory: Insight from the fluctuation-dissipation
  theorem},}\ }\href {\doibase 10.1103/PhysRevLett.125.233001} {\bibfield
  {journal} {\bibinfo  {journal} {Phys. Rev. Lett.}\ }\textbf {\bibinfo
  {volume} {125}},\ \bibinfo {pages} {233001} (\bibinfo {year}
  {2020})}\BibitemShut {NoStop}%
\bibitem [{\citenamefont {Gould}\ and\ \citenamefont
  {Pittalis}(2019)}]{PRL19_Gould_DD_correlation}%
  \BibitemOpen
  \bibfield  {author} {\bibinfo {author} {\bibfnamefont {Tim}\ \bibnamefont
  {Gould}}\ and\ \bibinfo {author} {\bibfnamefont {Stefano}\ \bibnamefont
  {Pittalis}},\ }\bibfield  {title} {\enquote {\bibinfo {title} {Density-driven
  correlations in many-electron ensembles: Theory and application for excited
  states},}\ }\href {\doibase 10.1103/PhysRevLett.123.016401} {\bibfield
  {journal} {\bibinfo  {journal} {Phys. Rev. Lett.}\ }\textbf {\bibinfo
  {volume} {123}},\ \bibinfo {pages} {016401} (\bibinfo {year}
  {2019})}\BibitemShut {NoStop}%
\bibitem [{\citenamefont
  {Gould}(2025)}]{gould2025stationaryconditionsexcitedstates}%
  \BibitemOpen
  \bibfield  {author} {\bibinfo {author} {\bibfnamefont {Tim}\ \bibnamefont
  {Gould}},\ }\bibfield  {title} {\enquote {\bibinfo {title} {Variational
  principles in ensemble and excited-state density- and potential-functional
  theories},}\ }\href {\doibase 10.1103/PhysRevA.111.032806} {\bibfield
  {journal} {\bibinfo  {journal} {Phys. Rev. A}\ }\textbf {\bibinfo {volume}
  {111}},\ \bibinfo {pages} {032806} (\bibinfo {year} {2025})}\BibitemShut
  {NoStop}%
\bibitem [{\citenamefont {Yang}\ \emph {et~al.}(2014)\citenamefont {Yang},
  \citenamefont {Trail}, \citenamefont {Pribram-Jones}, \citenamefont {Burke},
  \citenamefont {Needs},\ and\ \citenamefont {Ullrich}}]{yang2014exact}%
  \BibitemOpen
  \bibfield  {author} {\bibinfo {author} {\bibfnamefont {Zeng-hui}\
  \bibnamefont {Yang}}, \bibinfo {author} {\bibfnamefont {John~R}\ \bibnamefont
  {Trail}}, \bibinfo {author} {\bibfnamefont {Aurora}\ \bibnamefont
  {Pribram-Jones}}, \bibinfo {author} {\bibfnamefont {Kieron}\ \bibnamefont
  {Burke}}, \bibinfo {author} {\bibfnamefont {Richard~J}\ \bibnamefont
  {Needs}}, \ and\ \bibinfo {author} {\bibfnamefont {Carsten~A}\ \bibnamefont
  {Ullrich}},\ }\bibfield  {title} {\enquote {\bibinfo {title} {Exact and
  approximate kohn-sham potentials in ensemble density-functional theory},}\
  }\href {https://doi.org/10.1103/PhysRevA.90.042501} {\bibfield  {journal}
  {\bibinfo  {journal} {Phys. Rev. A}\ }\textbf {\bibinfo {volume} {90}},\
  \bibinfo {pages} {042501} (\bibinfo {year} {2014})}\BibitemShut {NoStop}%
\bibitem [{\citenamefont {Gould}\ \emph {et~al.}(2018)\citenamefont {Gould},
  \citenamefont {Kronik},\ and\ \citenamefont {Pittalis}}]{gould2018charge}%
  \BibitemOpen
  \bibfield  {author} {\bibinfo {author} {\bibfnamefont {Tim}\ \bibnamefont
  {Gould}}, \bibinfo {author} {\bibfnamefont {Leeor}\ \bibnamefont {Kronik}}, \
  and\ \bibinfo {author} {\bibfnamefont {Stefano}\ \bibnamefont {Pittalis}},\
  }\bibfield  {title} {\enquote {\bibinfo {title} {Charge transfer excitations
  from exact and approximate ensemble kohn-sham theory},}\ }\href
  {https://doi.org/10.1063/1.5022832} {\bibfield  {journal} {\bibinfo
  {journal} {J. Chem. Phys.}\ }\textbf {\bibinfo {volume} {148}},\ \bibinfo
  {pages} {174101} (\bibinfo {year} {2018})}\BibitemShut {NoStop}%
\bibitem [{\citenamefont {Loos}\ and\ \citenamefont
  {Fromager}(2020)}]{loos2020weightdependent}%
  \BibitemOpen
  \bibfield  {author} {\bibinfo {author} {\bibfnamefont {Pierre-Fran{\c c}ois}\
  \bibnamefont {Loos}}\ and\ \bibinfo {author} {\bibfnamefont {Emmanuel}\
  \bibnamefont {Fromager}},\ }\bibfield  {title} {\enquote {\bibinfo {title} {A
  weight-dependent local correlation density-functional approximation for
  ensembles},}\ }\href {\doibase 10.1063/5.0007388} {\bibfield  {journal}
  {\bibinfo  {journal} {J. Chem. Phys.}\ }\textbf {\bibinfo {volume} {152}},\
  \bibinfo {pages} {214101} (\bibinfo {year} {2020})}\BibitemShut {NoStop}%
\bibitem [{\citenamefont {Yang}(2021)}]{Yang2021_Second}%
  \BibitemOpen
  \bibfield  {author} {\bibinfo {author} {\bibfnamefont {Zeng-hui}\
  \bibnamefont {Yang}},\ }\bibfield  {title} {\enquote {\bibinfo {title}
  {Second-order perturbative correlation energy functional in the ensemble
  density-functional theory},}\ }\href {\doibase 10.1103/PhysRevA.104.052806}
  {\bibfield  {journal} {\bibinfo  {journal} {Phys. Rev. A}\ }\textbf {\bibinfo
  {volume} {104}},\ \bibinfo {pages} {052806} (\bibinfo {year}
  {2021})}\BibitemShut {NoStop}%
\bibitem [{\citenamefont {Gould}(2020)}]{Gould2020_Approximately}%
  \BibitemOpen
  \bibfield  {author} {\bibinfo {author} {\bibfnamefont {Tim}\ \bibnamefont
  {Gould}},\ }\bibfield  {title} {\enquote {\bibinfo {title} {Approximately
  self-consistent ensemble density functional theory: Toward inclusion of all
  correlations},}\ }\href {\doibase 10.1021/acs.jpclett.0c02894} {\bibfield
  {journal} {\bibinfo  {journal} {J. Phys. Chem. Lett.}\ }\textbf {\bibinfo
  {volume} {11}},\ \bibinfo {pages} {9907--9912} (\bibinfo {year}
  {2020})}\BibitemShut {NoStop}%
\bibitem [{\citenamefont {Gould}\ \emph {et~al.}(2023)\citenamefont {Gould},
  \citenamefont {Kooi}, \citenamefont {Gori-Giorgi},\ and\ \citenamefont
  {Pittalis}}]{Gould2023_Electronic}%
  \BibitemOpen
  \bibfield  {author} {\bibinfo {author} {\bibfnamefont {Tim}\ \bibnamefont
  {Gould}}, \bibinfo {author} {\bibfnamefont {Derk~P.}\ \bibnamefont {Kooi}},
  \bibinfo {author} {\bibfnamefont {Paola}\ \bibnamefont {Gori-Giorgi}}, \ and\
  \bibinfo {author} {\bibfnamefont {Stefano}\ \bibnamefont {Pittalis}},\
  }\bibfield  {title} {\enquote {\bibinfo {title} {Electronic excited states in
  extreme limits via ensemble density functionals},}\ }\href {\doibase
  10.1103/PhysRevLett.130.106401} {\bibfield  {journal} {\bibinfo  {journal}
  {Phys. Rev. Lett.}\ }\textbf {\bibinfo {volume} {130}},\ \bibinfo {pages}
  {106401} (\bibinfo {year} {2023})}\BibitemShut {NoStop}%
\bibitem [{\citenamefont {Gould}\ and\ \citenamefont
  {Pittalis}(2024)}]{gould2024local}%
  \BibitemOpen
  \bibfield  {author} {\bibinfo {author} {\bibfnamefont {Tim}\ \bibnamefont
  {Gould}}\ and\ \bibinfo {author} {\bibfnamefont {Stefano}\ \bibnamefont
  {Pittalis}},\ }\bibfield  {title} {\enquote {\bibinfo {title} {Local density
  approximation for excited states},}\ }\href {\doibase
  10.1103/PhysRevX.14.041045} {\bibfield  {journal} {\bibinfo  {journal} {Phys.
  Rev. X}\ }\textbf {\bibinfo {volume} {14}},\ \bibinfo {pages} {041045}
  (\bibinfo {year} {2024})}\BibitemShut {NoStop}%
\bibitem [{\citenamefont {Gould}\ \emph {et~al.}(2025)\citenamefont {Gould},
  \citenamefont {Dale}, \citenamefont {Kronik},\ and\ \citenamefont
  {Pittalis}}]{Gould2025_PRL_tate-Specific}%
  \BibitemOpen
  \bibfield  {author} {\bibinfo {author} {\bibfnamefont {Tim}\ \bibnamefont
  {Gould}}, \bibinfo {author} {\bibfnamefont {Stephen~G.}\ \bibnamefont
  {Dale}}, \bibinfo {author} {\bibfnamefont {Leeor}\ \bibnamefont {Kronik}}, \
  and\ \bibinfo {author} {\bibfnamefont {Stefano}\ \bibnamefont {Pittalis}},\
  }\bibfield  {title} {\enquote {\bibinfo {title} {State-specific density
  functionals for excited states via a density-driven correlation model},}\
  }\href {\doibase 10.1103/PhysRevLett.134.228001} {\bibfield  {journal}
  {\bibinfo  {journal} {Phys. Rev. Lett.}\ }\textbf {\bibinfo {volume} {134}},\
  \bibinfo {pages} {228001} (\bibinfo {year} {2025})}\BibitemShut {NoStop}%
\bibitem [{\citenamefont {Filatov}(2015)}]{filatov2015spin}%
  \BibitemOpen
  \bibfield  {author} {\bibinfo {author} {\bibfnamefont {Michael}\ \bibnamefont
  {Filatov}},\ }\bibfield  {title} {\enquote {\bibinfo {title} {Spin-restricted
  ensemble-referenced kohn--sham method: basic principles and application to
  strongly correlated ground and excited states of molecules},}\ }\href
  {\doibase 10.1002/wcms.1209} {\bibfield  {journal} {\bibinfo  {journal}
  {WIREs Comput. Mol. Sci.}\ }\textbf {\bibinfo {volume} {5}},\ \bibinfo
  {pages} {146} (\bibinfo {year} {2015})}\BibitemShut {NoStop}%
\bibitem [{\citenamefont {Filatov}\ \emph {et~al.}(2021)\citenamefont
  {Filatov}, \citenamefont {Lee},\ and\ \citenamefont {Choi}}]{Filatov21SSR}%
  \BibitemOpen
  \bibfield  {author} {\bibinfo {author} {\bibfnamefont {Michael}\ \bibnamefont
  {Filatov}}, \bibinfo {author} {\bibfnamefont {Seunghoon}\ \bibnamefont
  {Lee}}, \ and\ \bibinfo {author} {\bibfnamefont {Cheol~Ho}\ \bibnamefont
  {Choi}},\ }\bibfield  {title} {\enquote {\bibinfo {title} {Description of
  sudden polarization in the excited electronic states with an ensemble density
  functional theory method},}\ }\href {\doibase 10.1021/acs.jctc.1c00479}
  {\bibfield  {journal} {\bibinfo  {journal} {Journal of Chemical Theory and
  Computation}\ }\textbf {\bibinfo {volume} {17}},\ \bibinfo {pages}
  {5123--5139} (\bibinfo {year} {2021})}\BibitemShut {NoStop}%
\bibitem [{\citenamefont {Gould}\ and\ \citenamefont
  {Kronik}(2021)}]{Gould2021_Ensemble_ugly}%
  \BibitemOpen
  \bibfield  {author} {\bibinfo {author} {\bibfnamefont {Tim}\ \bibnamefont
  {Gould}}\ and\ \bibinfo {author} {\bibfnamefont {Leeor}\ \bibnamefont
  {Kronik}},\ }\bibfield  {title} {\enquote {\bibinfo {title} {{Ensemble
  generalized Kohn--Sham theory: The good, the bad, and the ugly}},}\ }\href
  {\doibase 10.1063/5.0040447} {\bibfield  {journal} {\bibinfo  {journal} {J.
  Chem. Phys.}\ }\textbf {\bibinfo {volume} {154}},\ \bibinfo {pages} {094125}
  (\bibinfo {year} {2021})}\BibitemShut {NoStop}%
\bibitem [{\citenamefont {Fromager}(2025)}]{Fromager2025indvElevel}%
  \BibitemOpen
  \bibfield  {author} {\bibinfo {author} {\bibfnamefont {Emmanuel}\
  \bibnamefont {Fromager}},\ }\bibfield  {title} {\enquote {\bibinfo {title}
  {Ensemble density functional theory of ground and excited energy levels},}\
  }\href {\doibase 10.1021/acs.jpca.4c06744} {\bibfield  {journal} {\bibinfo
  {journal} {The Journal of Physical Chemistry A}\ }\textbf {\bibinfo {volume}
  {129}},\ \bibinfo {pages} {1143--1155} (\bibinfo {year} {2025})}\BibitemShut
  {NoStop}%
\bibitem [{\citenamefont {Dupuy}\ and\ \citenamefont
  {Fromager}(2025)}]{dupuy2025_exact_static}%
  \BibitemOpen
  \bibfield  {author} {\bibinfo {author} {\bibfnamefont {Lucien}\ \bibnamefont
  {Dupuy}}\ and\ \bibinfo {author} {\bibfnamefont {Emmanuel}\ \bibnamefont
  {Fromager}},\ }\bibfield  {title} {\enquote {\bibinfo {title} {Exact static
  linear response of excited states from ensemble density functional theory},}\
  }\href {\doibase 10.1021/acs.jpca.5c04552} {\bibfield  {journal} {\bibinfo
  {journal} {The Journal of Physical Chemistry A}\ }\textbf {\bibinfo {volume}
  {129}},\ \bibinfo {pages} {9095--9109} (\bibinfo {year} {2025})}\BibitemShut
  {NoStop}%
\bibitem [{\citenamefont {Daas}\ \emph {et~al.}(2025)\citenamefont {Daas},
  \citenamefont {Crisostomo},\ and\ \citenamefont
  {Burke}}]{daas2025ensembletimedependentdensityfunctional}%
  \BibitemOpen
  \bibfield  {author} {\bibinfo {author} {\bibfnamefont {Kimberly~J.}\
  \bibnamefont {Daas}}, \bibinfo {author} {\bibfnamefont {Steven}\ \bibnamefont
  {Crisostomo}}, \ and\ \bibinfo {author} {\bibfnamefont {Kieron}\ \bibnamefont
  {Burke}},\ }\href {https://arxiv.org/abs/2507.19464} {\enquote {\bibinfo
  {title} {Ensemble time-dependent density functional theory},}\ } (\bibinfo
  {year} {2025}),\ \Eprint {http://arxiv.org/abs/2507.19464} {arXiv:2507.19464
  [physics.chem-ph]} \BibitemShut {NoStop}%
\bibitem [{\citenamefont {Senjean}\ and\ \citenamefont
  {Fromager}(2018)}]{senjean2018unified}%
  \BibitemOpen
  \bibfield  {author} {\bibinfo {author} {\bibfnamefont {Bruno}\ \bibnamefont
  {Senjean}}\ and\ \bibinfo {author} {\bibfnamefont {Emmanuel}\ \bibnamefont
  {Fromager}},\ }\bibfield  {title} {\enquote {\bibinfo {title} {Unified
  formulation of fundamental and optical gap problems in density-functional
  theory for ensembles},}\ }\href {https://doi.org/10.1103/PhysRevA.98.022513}
  {\bibfield  {journal} {\bibinfo  {journal} {Phys. Rev. A}\ }\textbf {\bibinfo
  {volume} {98}},\ \bibinfo {pages} {022513} (\bibinfo {year}
  {2018})}\BibitemShut {NoStop}%
\bibitem [{\citenamefont {Cernatic}\ \emph
  {et~al.}(2024{\natexlab{a}})\citenamefont {Cernatic}, \citenamefont {Loos},
  \citenamefont {Senjean},\ and\ \citenamefont
  {Fromager}}]{Cernatic2024_Neutral}%
  \BibitemOpen
  \bibfield  {author} {\bibinfo {author} {\bibfnamefont {Filip}\ \bibnamefont
  {Cernatic}}, \bibinfo {author} {\bibfnamefont {Pierre-Fran\mbox{\c{c}}ois}\
  \bibnamefont {Loos}}, \bibinfo {author} {\bibfnamefont {Bruno}\ \bibnamefont
  {Senjean}}, \ and\ \bibinfo {author} {\bibfnamefont {Emmanuel}\ \bibnamefont
  {Fromager}},\ }\bibfield  {title} {\enquote {\bibinfo {title} {Neutral
  electronic excitations and derivative discontinuities: An extended
  $n$-centered ensemble density functional theory perspective},}\ }\href
  {\doibase 10.1103/PhysRevB.109.235113} {\bibfield  {journal} {\bibinfo
  {journal} {Phys. Rev. B}\ }\textbf {\bibinfo {volume} {109}},\ \bibinfo
  {pages} {235113} (\bibinfo {year} {2024}{\natexlab{a}})}\BibitemShut
  {NoStop}%
\bibitem [{\citenamefont {Senjean}\ and\ \citenamefont
  {Fromager}(2020)}]{Senjean_2020}%
  \BibitemOpen
  \bibfield  {author} {\bibinfo {author} {\bibfnamefont {Bruno}\ \bibnamefont
  {Senjean}}\ and\ \bibinfo {author} {\bibfnamefont {Emmanuel}\ \bibnamefont
  {Fromager}},\ }\bibfield  {title} {\enquote {\bibinfo {title} {N-centered
  ensemble density-functional theory for open systems},}\ }\href {\doibase
  10.1002/qua.26190} {\bibfield  {journal} {\bibinfo  {journal} {Int. J.
  Quantum Chem.}\ }\textbf {\bibinfo {volume} {120}},\ \bibinfo {pages}
  {e26190} (\bibinfo {year} {2020})}\BibitemShut {NoStop}%
\bibitem [{\citenamefont {Hodgson}\ \emph {et~al.}(2021)\citenamefont
  {Hodgson}, \citenamefont {Wetherell},\ and\ \citenamefont
  {Fromager}}]{PRA21_Hodgson_exact_Nc-eDFT_1D}%
  \BibitemOpen
  \bibfield  {author} {\bibinfo {author} {\bibfnamefont {M.~J.~P.}\
  \bibnamefont {Hodgson}}, \bibinfo {author} {\bibfnamefont {J.}~\bibnamefont
  {Wetherell}}, \ and\ \bibinfo {author} {\bibfnamefont {Emmanuel}\
  \bibnamefont {Fromager}},\ }\bibfield  {title} {\enquote {\bibinfo {title}
  {Exact exchange-correlation potentials for calculating the fundamental gap
  with a fixed number of electrons},}\ }\href {\doibase
  10.1103/PhysRevA.103.012806} {\bibfield  {journal} {\bibinfo  {journal}
  {Phys. Rev. A}\ }\textbf {\bibinfo {volume} {103}},\ \bibinfo {pages}
  {012806} (\bibinfo {year} {2021})}\BibitemShut {NoStop}%
\bibitem [{\citenamefont {Pribram-Jones}\ \emph {et~al.}(2014)\citenamefont
  {Pribram-Jones}, \citenamefont {hui Yang}, \citenamefont {R.Trail},
  \citenamefont {Burke}, \citenamefont {J.Needs},\ and\ \citenamefont
  {A.Ullrich}}]{pribramjones2014excitations}%
  \BibitemOpen
  \bibfield  {author} {\bibinfo {author} {\bibfnamefont {Aurora}\ \bibnamefont
  {Pribram-Jones}}, \bibinfo {author} {\bibfnamefont {Zeng}\ \bibnamefont {hui
  Yang}}, \bibinfo {author} {\bibfnamefont {John}\ \bibnamefont {R.Trail}},
  \bibinfo {author} {\bibfnamefont {Kieron}\ \bibnamefont {Burke}}, \bibinfo
  {author} {\bibfnamefont {Richard}\ \bibnamefont {J.Needs}}, \ and\ \bibinfo
  {author} {\bibfnamefont {Carsten}\ \bibnamefont {A.Ullrich}},\ }\bibfield
  {title} {\enquote {\bibinfo {title} {Excitations and benchmark ensemble
  density functional theory for two electrons},}\ }\href {\doibase
  10.1063/1.4872255} {\bibfield  {journal} {\bibinfo  {journal} {J. Chem.
  Phys.}\ }\textbf {\bibinfo {volume} {140}},\ \bibinfo {pages} {18A541}
  (\bibinfo {year} {2014})}\BibitemShut {NoStop}%
\bibitem [{\citenamefont {Yang}\ \emph {et~al.}(2017)\citenamefont {Yang},
  \citenamefont {Pribram-Jones}, \citenamefont {Burke},\ and\ \citenamefont
  {Ullrich}}]{yang2017direct}%
  \BibitemOpen
  \bibfield  {author} {\bibinfo {author} {\bibfnamefont {Zeng-hui}\
  \bibnamefont {Yang}}, \bibinfo {author} {\bibfnamefont {Aurora}\ \bibnamefont
  {Pribram-Jones}}, \bibinfo {author} {\bibfnamefont {Kieron}\ \bibnamefont
  {Burke}}, \ and\ \bibinfo {author} {\bibfnamefont {Carsten~A}\ \bibnamefont
  {Ullrich}},\ }\bibfield  {title} {\enquote {\bibinfo {title} {Direct
  extraction of excitation energies from ensemble density-functional theory},}\
  }\href {https://doi.org/10.1103/PhysRevLett.119.033003} {\bibfield  {journal}
  {\bibinfo  {journal} {Phys. Rev. Lett.}\ }\textbf {\bibinfo {volume} {119}},\
  \bibinfo {pages} {033003} (\bibinfo {year} {2017})}\BibitemShut {NoStop}%
\bibitem [{\citenamefont {Lindgren}\ and\ \citenamefont
  {Morrison}(1986)}]{Lindgren1986}%
  \BibitemOpen
  \bibfield  {author} {\bibinfo {author} {\bibfnamefont {Ingvar}\ \bibnamefont
  {Lindgren}}\ and\ \bibinfo {author} {\bibfnamefont {John}\ \bibnamefont
  {Morrison}},\ }\enquote {\bibinfo {title} {Perturbation theory},}\ in\ \href
  {\doibase 10.1007/978-3-642-61640-2_9} {\emph {\bibinfo {booktitle} {Atomic
  Many-Body Theory}}}\ (\bibinfo  {publisher} {Springer Berlin Heidelberg},\
  \bibinfo {address} {Berlin, Heidelberg},\ \bibinfo {year} {1986})\ pp.\
  \bibinfo {pages} {184--211}\BibitemShut {NoStop}%
\bibitem [{\citenamefont {Van~Vleck}(1929)}]{VanVleck29_On}%
  \BibitemOpen
  \bibfield  {author} {\bibinfo {author} {\bibfnamefont {J.~H.}\ \bibnamefont
  {Van~Vleck}},\ }\bibfield  {title} {\enquote {\bibinfo {title} {On
  $\ensuremath{\sigma}$-type doubling and electron spin in the spectra of
  diatomic molecules},}\ }\href {\doibase 10.1103/PhysRev.33.467} {\bibfield
  {journal} {\bibinfo  {journal} {Phys. Rev.}\ }\textbf {\bibinfo {volume}
  {33}},\ \bibinfo {pages} {467--506} (\bibinfo {year} {1929})}\BibitemShut
  {NoStop}%
\bibitem [{\citenamefont {Sokolov}(2024)}]{SOKOLOV2024121}%
  \BibitemOpen
  \bibfield  {author} {\bibinfo {author} {\bibfnamefont {Alexander~Yu.}\
  \bibnamefont {Sokolov}},\ }\bibfield  {title} {\enquote {\bibinfo {title}
  {Chapter four - multireference perturbation theories based on the dyall
  hamiltonian},}\ }in\ \href {\doibase
  https://doi.org/10.1016/bs.aiq.2024.04.004} {\emph {\bibinfo {booktitle}
  {Novel Treatments of Strong Correlations}}},\ \bibinfo {series} {Advances in
  Quantum Chemistry}, Vol.~\bibinfo {volume} {90},\ \bibinfo {editor} {edited
  by\ \bibinfo {editor} {\bibfnamefont {Ramon A.~Miranda}\ \bibnamefont
  {Quintana}}\ and\ \bibinfo {editor} {\bibfnamefont {John~F.}\ \bibnamefont
  {Stanton}}}\ (\bibinfo  {publisher} {Academic Press},\ \bibinfo {year}
  {2024})\ pp.\ \bibinfo {pages} {121--155}\BibitemShut {NoStop}%
\bibitem [{\citenamefont {Makhlouf}\ \emph {et~al.}(2025)\citenamefont
  {Makhlouf}, \citenamefont {Senjean},\ and\ \citenamefont
  {Fromager}}]{makhlouf2025_local_potential}%
  \BibitemOpen
  \bibfield  {author} {\bibinfo {author} {\bibfnamefont {Wafa}\ \bibnamefont
  {Makhlouf}}, \bibinfo {author} {\bibfnamefont {Bruno}\ \bibnamefont
  {Senjean}}, \ and\ \bibinfo {author} {\bibfnamefont {Emmanuel}\ \bibnamefont
  {Fromager}},\ }\bibfield  {title} {\enquote {\bibinfo {title} {Local
  potential functional embedding theory of molecular systems: Localized
  orbital-based embedding from an exact density-functional perspective},}\
  }\href {\doibase 10.1021/acs.jctc.5c01256} {\bibfield  {journal} {\bibinfo
  {journal} {Journal of Chemical Theory and Computation}\ }\textbf {\bibinfo
  {volume} {21}},\ \bibinfo {pages} {10293--10314} (\bibinfo {year}
  {2025})}\BibitemShut {NoStop}%
\bibitem [{\citenamefont {Makhlouf}\ \emph {et~al.}(2026)\citenamefont
  {Makhlouf}, \citenamefont {Senjean},\ and\ \citenamefont
  {Fromager}}]{makhlouf2026generalizedlocalpotentialfunctional}%
  \BibitemOpen
  \bibfield  {author} {\bibinfo {author} {\bibfnamefont {Wafa}\ \bibnamefont
  {Makhlouf}}, \bibinfo {author} {\bibfnamefont {Bruno}\ \bibnamefont
  {Senjean}}, \ and\ \bibinfo {author} {\bibfnamefont {Emmanuel}\ \bibnamefont
  {Fromager}},\ }\href {https://arxiv.org/abs/2602.15624} {\enquote {\bibinfo
  {title} {Generalized local potential functional embedding theory of localized
  orbitals},}\ } (\bibinfo {year} {2026}),\ \Eprint
  {http://arxiv.org/abs/2602.15624} {arXiv:2602.15624 [cond-mat.str-el]}
  \BibitemShut {NoStop}%
\bibitem [{\citenamefont {Knizia}\ and\ \citenamefont
  {Chan}(2012)}]{knizia2012density}%
  \BibitemOpen
  \bibfield  {author} {\bibinfo {author} {\bibfnamefont {Gerald}\ \bibnamefont
  {Knizia}}\ and\ \bibinfo {author} {\bibfnamefont {Garnet Kin-Lic}\
  \bibnamefont {Chan}},\ }\bibfield  {title} {\enquote {\bibinfo {title}
  {Density matrix embedding: A simple alternative to dynamical mean-field
  theory},}\ }\href {\doibase 10.1103/PhysRevLett.109.186404} {\bibfield
  {journal} {\bibinfo  {journal} {Phys. Rev. Lett.}\ }\textbf {\bibinfo
  {volume} {109}},\ \bibinfo {pages} {186404} (\bibinfo {year}
  {2012})}\BibitemShut {NoStop}%
\bibitem [{\citenamefont {Cernatic}\ and\ \citenamefont
  {Fromager}(2024)}]{cernatic2024extended_doubles}%
  \BibitemOpen
  \bibfield  {author} {\bibinfo {author} {\bibfnamefont {Filip}\ \bibnamefont
  {Cernatic}}\ and\ \bibinfo {author} {\bibfnamefont {Emmanuel}\ \bibnamefont
  {Fromager}},\ }\bibfield  {title} {\enquote {\bibinfo {title} {Extended
  n-centered ensemble density functional theory of double electronic
  excitations},}\ }\href {\doibase https://doi.org/10.1002/jcc.27387}
  {\bibfield  {journal} {\bibinfo  {journal} {Journal of Computational
  Chemistry}\ }\textbf {\bibinfo {volume} {45}},\ \bibinfo {pages} {1945--1962}
  (\bibinfo {year} {2024})}\BibitemShut {NoStop}%
\bibitem [{\citenamefont {González}\ and\ \citenamefont
  {Ayers}(2026)}]{Gonzalez26_Counterexamples}%
  \BibitemOpen
  \bibfield  {author} {\bibinfo {author} {\bibfnamefont {Marco~Martínez}\
  \bibnamefont {González}}\ and\ \bibinfo {author} {\bibfnamefont {Paul~W.}\
  \bibnamefont {Ayers}},\ }\bibfield  {title} {\enquote {\bibinfo {title}
  {Counterexamples to the convexity of the energy for coulomb particles},}\
  }\href {\doibase 10.1021/acs.jpclett.5c03648} {\bibfield  {journal} {\bibinfo
   {journal} {The Journal of Physical Chemistry Letters}\ }\textbf {\bibinfo
  {volume} {17}},\ \bibinfo {pages} {2524--2529} (\bibinfo {year}
  {2026})}\BibitemShut {NoStop}%
\bibitem [{\citenamefont {Di~Marino}\ \emph {et~al.}(2024)\citenamefont
  {Di~Marino}, \citenamefont {Lewin},\ and\ \citenamefont
  {Nenna}}]{Di-Marino24_Ground}%
  \BibitemOpen
  \bibfield  {author} {\bibinfo {author} {\bibfnamefont {Simone}\ \bibnamefont
  {Di~Marino}}, \bibinfo {author} {\bibfnamefont {Mathieu}\ \bibnamefont
  {Lewin}}, \ and\ \bibinfo {author} {\bibfnamefont {Luca}\ \bibnamefont
  {Nenna}},\ }\bibfield  {title} {\enquote {\bibinfo {title} {Ground state
  energy is not always convex in the number of electrons},}\ }\href {\doibase
  10.1021/acs.jpca.4c06345} {\bibfield  {journal} {\bibinfo  {journal} {The
  Journal of Physical Chemistry A}\ }\textbf {\bibinfo {volume} {128}},\
  \bibinfo {pages} {10697--10706} (\bibinfo {year} {2024})}\BibitemShut
  {NoStop}%
\bibitem [{\citenamefont {Levy}(1979)}]{levy1979universal}%
  \BibitemOpen
  \bibfield  {author} {\bibinfo {author} {\bibfnamefont {Mel}\ \bibnamefont
  {Levy}},\ }\bibfield  {title} {\enquote {\bibinfo {title} {Universal
  variational functionals of electron densities, first-order density matrices,
  and natural spin-orbitals and solution of the v-representability problem},}\
  }\href {https://doi.org/10.1073/pnas.76.12.6062} {\bibfield  {journal}
  {\bibinfo  {journal} {Proc. Natl. Acad. Sci.}\ }\textbf {\bibinfo {volume}
  {76}},\ \bibinfo {pages} {6062--6065} (\bibinfo {year} {1979})}\BibitemShut
  {NoStop}%
\bibitem [{\citenamefont {Lieb}(1983)}]{LFTransform-Lieb}%
  \BibitemOpen
  \bibfield  {author} {\bibinfo {author} {\bibfnamefont {Elliott~H}\
  \bibnamefont {Lieb}},\ }\bibfield  {title} {\enquote {\bibinfo {title}
  {Density functionals for coulomb systems},}\ }\href
  {https://doi.org/10.1002/qua.560240302} {\bibfield  {journal} {\bibinfo
  {journal} {Int. J. Quantum Chem.}\ }\textbf {\bibinfo {volume} {24}},\
  \bibinfo {pages} {243} (\bibinfo {year} {1983})}\BibitemShut {NoStop}%
\bibitem [{\citenamefont {Senjean}\ \emph {et~al.}(2015)\citenamefont
  {Senjean}, \citenamefont {Knecht}, \citenamefont {Jensen},\ and\
  \citenamefont {Fromager}}]{senjean2015linear}%
  \BibitemOpen
  \bibfield  {author} {\bibinfo {author} {\bibfnamefont {Bruno}\ \bibnamefont
  {Senjean}}, \bibinfo {author} {\bibfnamefont {Stefan}\ \bibnamefont
  {Knecht}}, \bibinfo {author} {\bibfnamefont {Hans J{\o}rgen~{\Aa}}\
  \bibnamefont {Jensen}}, \ and\ \bibinfo {author} {\bibfnamefont {Emmanuel}\
  \bibnamefont {Fromager}},\ }\bibfield  {title} {\enquote {\bibinfo {title}
  {Linear interpolation method in ensemble kohn-sham and range-separated
  density-functional approximations for excited states},}\ }\href
  {https://doi.org/10.1103/PhysRevA.92.012518} {\bibfield  {journal} {\bibinfo
  {journal} {Phys. Rev. A}\ }\textbf {\bibinfo {volume} {92}},\ \bibinfo
  {pages} {012518} (\bibinfo {year} {2015})}\BibitemShut {NoStop}%
\bibitem [{\citenamefont {Levy}\ and\ \citenamefont
  {Zahariev}(2014)}]{levy2014ground}%
  \BibitemOpen
  \bibfield  {author} {\bibinfo {author} {\bibfnamefont {Mel}\ \bibnamefont
  {Levy}}\ and\ \bibinfo {author} {\bibfnamefont {Federico}\ \bibnamefont
  {Zahariev}},\ }\bibfield  {title} {\enquote {\bibinfo {title} {Ground-state
  energy as a simple sum of orbital energies in kohn-sham theory: A shift in
  perspective through a shift in potential},}\ }\href
  {https://doi.org/10.1103/PhysRevLett.113.113002} {\bibfield  {journal}
  {\bibinfo  {journal} {Phys. Rev. Lett.}\ }\textbf {\bibinfo {volume} {113}},\
  \bibinfo {pages} {113002} (\bibinfo {year} {2014})}\BibitemShut {NoStop}%
\bibitem [{\citenamefont {Maitra}\ \emph
  {et~al.}(2004{\natexlab{b}})\citenamefont {Maitra}, \citenamefont {Zhang},
  \citenamefont {Cave},\ and\ \citenamefont {Burke}}]{maitra2004double}%
  \BibitemOpen
  \bibfield  {author} {\bibinfo {author} {\bibfnamefont {Neepa~T}\ \bibnamefont
  {Maitra}}, \bibinfo {author} {\bibfnamefont {Fan}\ \bibnamefont {Zhang}},
  \bibinfo {author} {\bibfnamefont {Robert~J}\ \bibnamefont {Cave}}, \ and\
  \bibinfo {author} {\bibfnamefont {Kieron}\ \bibnamefont {Burke}},\ }\bibfield
   {title} {\enquote {\bibinfo {title} {Double excitations within
  time-dependent density functional theory linear response},}\ }\href
  {https://doi.org/10.1063/1.1651060} {\bibfield  {journal} {\bibinfo
  {journal} {J. Chem. Phys.}\ }\textbf {\bibinfo {volume} {120}},\ \bibinfo
  {pages} {5932} (\bibinfo {year} {2004}{\natexlab{b}})}\BibitemShut {NoStop}%
\bibitem [{\citenamefont {Cave}\ \emph {et~al.}(2004)\citenamefont {Cave},
  \citenamefont {Zhang}, \citenamefont {Maitra},\ and\ \citenamefont
  {Burke}}]{cave2004dressed}%
  \BibitemOpen
  \bibfield  {author} {\bibinfo {author} {\bibfnamefont {Robert~J.}\
  \bibnamefont {Cave}}, \bibinfo {author} {\bibfnamefont {Fan}\ \bibnamefont
  {Zhang}}, \bibinfo {author} {\bibfnamefont {Neepa~T.}\ \bibnamefont
  {Maitra}}, \ and\ \bibinfo {author} {\bibfnamefont {Kieron}\ \bibnamefont
  {Burke}},\ }\bibfield  {title} {\enquote {\bibinfo {title} {A dressed tddft
  treatment of the 21ag states of butadiene and hexatriene},}\ }\href {\doibase
  https://doi.org/10.1016/j.cplett.2004.03.051} {\bibfield  {journal} {\bibinfo
   {journal} {Chem. Phys. Lett.}\ }\textbf {\bibinfo {volume} {389}},\ \bibinfo
  {pages} {39--42} (\bibinfo {year} {2004})}\BibitemShut {NoStop}%
\bibitem [{\citenamefont {Huix-Rotllant}\ \emph {et~al.}(2011)\citenamefont
  {Huix-Rotllant}, \citenamefont {Ipatov}, \citenamefont {Rubio},\ and\
  \citenamefont {Casida}}]{Huix-Rotllant2011_Assessment}%
  \BibitemOpen
  \bibfield  {author} {\bibinfo {author} {\bibfnamefont {Miquel}\ \bibnamefont
  {Huix-Rotllant}}, \bibinfo {author} {\bibfnamefont {Andrei}\ \bibnamefont
  {Ipatov}}, \bibinfo {author} {\bibfnamefont {Angel}\ \bibnamefont {Rubio}}, \
  and\ \bibinfo {author} {\bibfnamefont {Mark~E.}\ \bibnamefont {Casida}},\
  }\bibfield  {title} {\enquote {\bibinfo {title} {Assessment of dressed
  time-dependent density-functional theory for the low-lying valence states of
  28 organic chromophores},}\ }\href {\doibase
  https://doi.org/10.1016/j.chemphys.2011.03.019} {\bibfield  {journal}
  {\bibinfo  {journal} {Chemical Physics}\ }\textbf {\bibinfo {volume} {391}},\
  \bibinfo {pages} {120--129} (\bibinfo {year} {2011})},\ \bibinfo {note} {open
  problems and new solutions in time dependent density functional
  theory}\BibitemShut {NoStop}%
\bibitem [{\citenamefont {Maitra}(2022)}]{maitra2022double}%
  \BibitemOpen
  \bibfield  {author} {\bibinfo {author} {\bibfnamefont {Neepa~T.}\
  \bibnamefont {Maitra}},\ }\bibfield  {title} {\enquote {\bibinfo {title}
  {Double and charge-transfer excitations in time-dependent density functional
  theory},}\ }\href {\doibase 10.1146/annurev-physchem-082720-124933}
  {\bibfield  {journal} {\bibinfo  {journal} {Annu. Rev. Phys. Chem.}\ }\textbf
  {\bibinfo {volume} {73}},\ \bibinfo {pages} {117--140} (\bibinfo {year}
  {2022})}\BibitemShut {NoStop}%
\bibitem [{\citenamefont {Perdew}\ and\ \citenamefont
  {Levy}(1983)}]{perdew1983physical}%
  \BibitemOpen
  \bibfield  {author} {\bibinfo {author} {\bibfnamefont {John~P}\ \bibnamefont
  {Perdew}}\ and\ \bibinfo {author} {\bibfnamefont {Mel}\ \bibnamefont
  {Levy}},\ }\bibfield  {title} {\enquote {\bibinfo {title} {Physical content
  of the exact kohn-sham orbital energies: band gaps and derivative
  discontinuities},}\ }\href {https://doi.org/10.1103/PhysRevLett.51.1884}
  {\bibfield  {journal} {\bibinfo  {journal} {Phys. Rev. Lett.}\ }\textbf
  {\bibinfo {volume} {51}},\ \bibinfo {pages} {1884} (\bibinfo {year}
  {1983})}\BibitemShut {NoStop}%
\bibitem [{\citenamefont {Levy}(1995)}]{levy1995excitation}%
  \BibitemOpen
  \bibfield  {author} {\bibinfo {author} {\bibfnamefont {Mel}\ \bibnamefont
  {Levy}},\ }\bibfield  {title} {\enquote {\bibinfo {title} {Excitation
  energies from density-functional orbital energies},}\ }\href
  {https://doi.org/10.1103/PhysRevA.52.R4313} {\bibfield  {journal} {\bibinfo
  {journal} {Phys. Rev. A}\ }\textbf {\bibinfo {volume} {52}},\ \bibinfo
  {pages} {R4313} (\bibinfo {year} {1995})}\BibitemShut {NoStop}%
\bibitem [{\citenamefont {Gould}\ \emph {et~al.}(2022)\citenamefont {Gould},
  \citenamefont {Hashimi}, \citenamefont {Kronik},\ and\ \citenamefont
  {Dale}}]{gould2022single}%
  \BibitemOpen
  \bibfield  {author} {\bibinfo {author} {\bibfnamefont {Tim}\ \bibnamefont
  {Gould}}, \bibinfo {author} {\bibfnamefont {Zahed}\ \bibnamefont {Hashimi}},
  \bibinfo {author} {\bibfnamefont {Leeor}\ \bibnamefont {Kronik}}, \ and\
  \bibinfo {author} {\bibfnamefont {Stephen~G.}\ \bibnamefont {Dale}},\
  }\bibfield  {title} {\enquote {\bibinfo {title} {Single excitation energies
  obtained from the ensemble "homo-lumo gap": Exact results and
  approximations},}\ }\href {https://doi.org/10.1021/acs.jpclett.2c00042}
  {\bibfield  {journal} {\bibinfo  {journal} {J. Phys. Chem. Lett.}\ }\textbf
  {\bibinfo {volume} {13}},\ \bibinfo {pages} {2452--2458} (\bibinfo {year}
  {2022})}\BibitemShut {NoStop}%
\bibitem [{\citenamefont {Janak}(1978)}]{janak1978proof}%
  \BibitemOpen
  \bibfield  {author} {\bibinfo {author} {\bibfnamefont {J~F}\ \bibnamefont
  {Janak}},\ }\bibfield  {title} {\enquote {\bibinfo {title} {Proof that
  $\partial e / \partial n_i = \varepsilon$ in density-functional theory},}\
  }\href {https://doi.org/10.1103/PhysRevB.18.7165} {\bibfield  {journal}
  {\bibinfo  {journal} {Phys. Rev. B}\ }\textbf {\bibinfo {volume} {18}},\
  \bibinfo {pages} {7165} (\bibinfo {year} {1978})}\BibitemShut {NoStop}%
\bibitem [{\citenamefont {Dupuy}\ and\ \citenamefont
  {Fromager}(2026)}]{dupuy2026fukui}%
  \BibitemOpen
  \bibfield  {author} {\bibinfo {author} {\bibfnamefont {Lucien}\ \bibnamefont
  {Dupuy}}\ and\ \bibinfo {author} {\bibfnamefont {Emmanuel}\ \bibnamefont
  {Fromager}},\ }\href {https://arxiv.org/abs/2601.02985} {\enquote {\bibinfo
  {title} {Charged excitations made neutral: N-centered ensemble density
  functional theory of fukui functions},}\ } (\bibinfo {year} {2026}),\ \Eprint
  {http://arxiv.org/abs/2601.02985} {arXiv:2601.02985 [physics.chem-ph]}
  \BibitemShut {NoStop}%
\bibitem [{\citenamefont {Heid}\ \emph {et~al.}(2018)\citenamefont {Heid},
  \citenamefont {Hunt},\ and\ \citenamefont {Schröder}}]{C7CP08549D}%
  \BibitemOpen
  \bibfield  {author} {\bibinfo {author} {\bibfnamefont {Esther}\ \bibnamefont
  {Heid}}, \bibinfo {author} {\bibfnamefont {Patricia~A.}\ \bibnamefont
  {Hunt}}, \ and\ \bibinfo {author} {\bibfnamefont {Christian}\ \bibnamefont
  {Schröder}},\ }\bibfield  {title} {\enquote {\bibinfo {title} {Evaluating
  excited state atomic polarizabilities of chromophores},}\ }\href {\doibase
  10.1039/C7CP08549D} {\bibfield  {journal} {\bibinfo  {journal} {Phys. Chem.
  Chem. Phys.}\ }\textbf {\bibinfo {volume} {20}},\ \bibinfo {pages}
  {8554--8563} (\bibinfo {year} {2018})}\BibitemShut {NoStop}%
\bibitem [{\citenamefont {Levi}\ \emph
  {et~al.}(2020{\natexlab{a}})\citenamefont {Levi}, \citenamefont {Ivanov},\
  and\ \citenamefont {J\'{o}nsson}}]{Levi20_Variational}%
  \BibitemOpen
  \bibfield  {author} {\bibinfo {author} {\bibfnamefont {Gianluca}\
  \bibnamefont {Levi}}, \bibinfo {author} {\bibfnamefont {Aleksei~V.}\
  \bibnamefont {Ivanov}}, \ and\ \bibinfo {author} {\bibfnamefont {Hannes}\
  \bibnamefont {J\'{o}nsson}},\ }\bibfield  {title} {\enquote {\bibinfo {title}
  {Variational calculations of excited states via direct optimization of the
  orbitals in dft},}\ }\href {\doibase 10.1039/D0FD00064G} {\bibfield
  {journal} {\bibinfo  {journal} {Faraday Discuss.}\ }\textbf {\bibinfo
  {volume} {224}},\ \bibinfo {pages} {448--466} (\bibinfo {year}
  {2020}{\natexlab{a}})}\BibitemShut {NoStop}%
\bibitem [{\citenamefont {Levi}\ \emph
  {et~al.}(2020{\natexlab{b}})\citenamefont {Levi}, \citenamefont {Ivanov},\
  and\ \citenamefont {Jónsson}}]{Levi_DOMOM_2020}%
  \BibitemOpen
  \bibfield  {author} {\bibinfo {author} {\bibfnamefont {Gianluca}\
  \bibnamefont {Levi}}, \bibinfo {author} {\bibfnamefont {Aleksei~V.}\
  \bibnamefont {Ivanov}}, \ and\ \bibinfo {author} {\bibfnamefont {Hannes}\
  \bibnamefont {Jónsson}},\ }\bibfield  {title} {\enquote {\bibinfo {title}
  {Variational density functional calculations of excited states via direct
  optimization},}\ }\href {\doibase 10.1021/acs.jctc.0c00597} {\bibfield
  {journal} {\bibinfo  {journal} {Journal of Chemical Theory and Computation}\
  }\textbf {\bibinfo {volume} {16}},\ \bibinfo {pages} {6968--6982} (\bibinfo
  {year} {2020}{\natexlab{b}})},\ \Eprint
  {http://arxiv.org/abs/https://doi.org/10.1021/acs.jctc.0c00597}
  {https://doi.org/10.1021/acs.jctc.0c00597} \BibitemShut {NoStop}%
\bibitem [{\citenamefont {Ivanov}\ \emph {et~al.}(2021)\citenamefont {Ivanov},
  \citenamefont {Levi}, \citenamefont {J\'{o}nsson},\ and\ \citenamefont
  {J\'{o}nsson}}]{Ivanov21_Method}%
  \BibitemOpen
  \bibfield  {author} {\bibinfo {author} {\bibfnamefont {Aleksei~V.}\
  \bibnamefont {Ivanov}}, \bibinfo {author} {\bibfnamefont {Gianluca}\
  \bibnamefont {Levi}}, \bibinfo {author} {\bibfnamefont {Elvar~\"{O}.}\
  \bibnamefont {J\'{o}nsson}}, \ and\ \bibinfo {author} {\bibfnamefont
  {Hannes}\ \bibnamefont {J\'{o}nsson}},\ }\bibfield  {title} {\enquote
  {\bibinfo {title} {Method for calculating excited electronic states using
  density functionals and direct orbital optimization with real space grid or
  plane-wave basis set},}\ }\href {\doibase 10.1021/acs.jctc.1c00157}
  {\bibfield  {journal} {\bibinfo  {journal} {Journal of Chemical Theory and
  Computation}\ }\textbf {\bibinfo {volume} {17}},\ \bibinfo {pages}
  {5034--5049} (\bibinfo {year} {2021})}\BibitemShut {NoStop}%
\bibitem [{\citenamefont {Hait}\ and\ \citenamefont
  {Head-Gordon}(2021)}]{Hait21_Orbital}%
  \BibitemOpen
  \bibfield  {author} {\bibinfo {author} {\bibfnamefont {Diptarka}\
  \bibnamefont {Hait}}\ and\ \bibinfo {author} {\bibfnamefont {Martin}\
  \bibnamefont {Head-Gordon}},\ }\bibfield  {title} {\enquote {\bibinfo {title}
  {Orbital optimized density functional theory for electronic excited
  states},}\ }\href {\doibase 10.1021/acs.jpclett.1c00744} {\bibfield
  {journal} {\bibinfo  {journal} {The Journal of Physical Chemistry Letters}\
  }\textbf {\bibinfo {volume} {12}},\ \bibinfo {pages} {4517--4529} (\bibinfo
  {year} {2021})}\BibitemShut {NoStop}%
\bibitem [{\citenamefont {Schmerwitz}\ \emph {et~al.}(2022)\citenamefont
  {Schmerwitz}, \citenamefont {Ivanov}, \citenamefont {J\'{o}nsson},
  \citenamefont {J\'{o}nsson},\ and\ \citenamefont
  {Levi}}]{Schmerwitz22_Variational}%
  \BibitemOpen
  \bibfield  {author} {\bibinfo {author} {\bibfnamefont {Yorick L.~A.}\
  \bibnamefont {Schmerwitz}}, \bibinfo {author} {\bibfnamefont {Aleksei~V.}\
  \bibnamefont {Ivanov}}, \bibinfo {author} {\bibfnamefont {Elvar~\"{O}.}\
  \bibnamefont {J\'{o}nsson}}, \bibinfo {author} {\bibfnamefont {Hannes}\
  \bibnamefont {J\'{o}nsson}}, \ and\ \bibinfo {author} {\bibfnamefont
  {Gianluca}\ \bibnamefont {Levi}},\ }\bibfield  {title} {\enquote {\bibinfo
  {title} {Variational density functional calculations of excited states:
  Conical intersection and avoided crossing in ethylene bond twisting},}\
  }\href {\doibase 10.1021/acs.jpclett.2c00741} {\bibfield  {journal} {\bibinfo
   {journal} {The Journal of Physical Chemistry Letters}\ }\textbf {\bibinfo
  {volume} {13}},\ \bibinfo {pages} {3990--3999} (\bibinfo {year}
  {2022})}\BibitemShut {NoStop}%
\bibitem [{\citenamefont {Schmerwitz}\ \emph {et~al.}(0)\citenamefont
  {Schmerwitz}, \citenamefont {Selenius},\ and\ \citenamefont
  {Levi}}]{Schmerwitz26_Freeze-and-Release}%
  \BibitemOpen
  \bibfield  {author} {\bibinfo {author} {\bibfnamefont {Yorick L.~A.}\
  \bibnamefont {Schmerwitz}}, \bibinfo {author} {\bibfnamefont {Elli}\
  \bibnamefont {Selenius}}, \ and\ \bibinfo {author} {\bibfnamefont {Gianluca}\
  \bibnamefont {Levi}},\ }\bibfield  {title} {\enquote {\bibinfo {title}
  {Freeze-and-release direct optimization method for variational calculations
  of excited electronic states},}\ }\href {\doibase 10.1021/acs.jctc.5c01974}
  {\bibfield  {journal} {\bibinfo  {journal} {Journal of Chemical Theory and
  Computation}\ }\textbf {\bibinfo {volume} {0}},\ \bibinfo {pages} {null}
  (\bibinfo {year} {0})}\BibitemShut {NoStop}%
\bibitem [{\citenamefont {Sigurdarson}\ \emph {et~al.}(2023)\citenamefont
  {Sigurdarson}, \citenamefont {Schmerwitz}, \citenamefont {Tveiten},
  \citenamefont {Levi},\ and\ \citenamefont
  {Jónsson}}]{OODFTRydberg_Levi2023}%
  \BibitemOpen
  \bibfield  {author} {\bibinfo {author} {\bibfnamefont {Alec~E.}\ \bibnamefont
  {Sigurdarson}}, \bibinfo {author} {\bibfnamefont {Yorick L.~A.}\ \bibnamefont
  {Schmerwitz}}, \bibinfo {author} {\bibfnamefont {Dagrún K.~V.}\ \bibnamefont
  {Tveiten}}, \bibinfo {author} {\bibfnamefont {Gianluca}\ \bibnamefont
  {Levi}}, \ and\ \bibinfo {author} {\bibfnamefont {Hannes}\ \bibnamefont
  {Jónsson}},\ }\bibfield  {title} {\enquote {\bibinfo {title}
  {Orbital-optimized density functional calculations of molecular rydberg
  excited states with real space grid representation and self-interaction
  correction},}\ }\href {\doibase 10.1063/5.0179271} {\bibfield  {journal}
  {\bibinfo  {journal} {The Journal of Chemical Physics}\ }\textbf {\bibinfo
  {volume} {159}},\ \bibinfo {pages} {214109} (\bibinfo {year} {2023})},\
  \Eprint
  {http://arxiv.org/abs/https://pubs.aip.org/aip/jcp/article-pdf/doi/10.1063/5.0179271/18236027/214109\_1\_5.0179271.pdf}
  {https://pubs.aip.org/aip/jcp/article-pdf/doi/10.1063/5.0179271/18236027/214109\_1\_5.0179271.pdf}
  \BibitemShut {NoStop}%
\bibitem [{\citenamefont {Selenius}\ \emph {et~al.}(2024)\citenamefont
  {Selenius}, \citenamefont {Sigurdarson}, \citenamefont {Schmerwitz},\ and\
  \citenamefont {Levi}}]{Levi_OODFTvsTDDFT2024}%
  \BibitemOpen
  \bibfield  {author} {\bibinfo {author} {\bibfnamefont {Elli}\ \bibnamefont
  {Selenius}}, \bibinfo {author} {\bibfnamefont {Alec~Elías}\ \bibnamefont
  {Sigurdarson}}, \bibinfo {author} {\bibfnamefont {Yorick L.~A.}\ \bibnamefont
  {Schmerwitz}}, \ and\ \bibinfo {author} {\bibfnamefont {Gianluca}\
  \bibnamefont {Levi}},\ }\bibfield  {title} {\enquote {\bibinfo {title}
  {Orbital-optimized versus time-dependent density functional calculations of
  intramolecular charge transfer excited states},}\ }\href {\doibase
  10.1021/acs.jctc.3c01319} {\bibfield  {journal} {\bibinfo  {journal} {Journal
  of Chemical Theory and Computation}\ }\textbf {\bibinfo {volume} {20}},\
  \bibinfo {pages} {3809--3822} (\bibinfo {year} {2024})},\ \Eprint
  {http://arxiv.org/abs/https://doi.org/10.1021/acs.jctc.3c01319}
  {https://doi.org/10.1021/acs.jctc.3c01319} \BibitemShut {NoStop}%
\bibitem [{\citenamefont {G\"orling}(1996)}]{Gorling95dftexc}%
  \BibitemOpen
  \bibfield  {author} {\bibinfo {author} {\bibfnamefont {Andreas}\ \bibnamefont
  {G\"orling}},\ }\bibfield  {title} {\enquote {\bibinfo {title}
  {Density-functional theory for excited states},}\ }\href {\doibase
  10.1103/PhysRevA.54.3912} {\bibfield  {journal} {\bibinfo  {journal} {Phys.
  Rev. A}\ }\textbf {\bibinfo {volume} {54}},\ \bibinfo {pages} {3912--3915}
  (\bibinfo {year} {1996})}\BibitemShut {NoStop}%
\bibitem [{\citenamefont {Yang}\ and\ \citenamefont
  {Ayers}(2024)}]{yang2024foundationdeltascfapproachdensity}%
  \BibitemOpen
  \bibfield  {author} {\bibinfo {author} {\bibfnamefont {Weitao}\ \bibnamefont
  {Yang}}\ and\ \bibinfo {author} {\bibfnamefont {Paul~W.}\ \bibnamefont
  {Ayers}},\ }\href {https://arxiv.org/abs/2403.04604} {\enquote {\bibinfo
  {title} {Foundation for the {$\Delta$}scf approach in density functional
  theory},}\ } (\bibinfo {year} {2024}),\ \Eprint
  {http://arxiv.org/abs/2403.04604} {arXiv:2403.04604 [physics.chem-ph]}
  \BibitemShut {NoStop}%
\bibitem [{\citenamefont {Sinyavskiy}\ \emph {et~al.}(2025)\citenamefont
  {Sinyavskiy}, \citenamefont {Mališ},\ and\ \citenamefont
  {Luber}}]{luber25DSCF}%
  \BibitemOpen
  \bibfield  {author} {\bibinfo {author} {\bibfnamefont {Andrey}\ \bibnamefont
  {Sinyavskiy}}, \bibinfo {author} {\bibfnamefont {Momir}\ \bibnamefont
  {Mališ}}, \ and\ \bibinfo {author} {\bibfnamefont {Sandra}\ \bibnamefont
  {Luber}},\ }\bibfield  {title} {\enquote {\bibinfo {title} {Bridging the gap
  between variational and perturbational dft-based methods for calculating
  excited states},}\ }\href {\doibase 10.1021/acs.jctc.5c00724} {\bibfield
  {journal} {\bibinfo  {journal} {Journal of Chemical Theory and Computation}\
  }\textbf {\bibinfo {volume} {21}},\ \bibinfo {pages} {7430--7449} (\bibinfo
  {year} {2025})}\BibitemShut {NoStop}%
\bibitem [{\citenamefont {Gidopoulos}\ \emph {et~al.}(2002)\citenamefont
  {Gidopoulos}, \citenamefont {Papaconstantinou},\ and\ \citenamefont
  {Gross}}]{ensemble_ghost_interaction}%
  \BibitemOpen
  \bibfield  {author} {\bibinfo {author} {\bibfnamefont {N~I}\ \bibnamefont
  {Gidopoulos}}, \bibinfo {author} {\bibfnamefont {P~G}\ \bibnamefont
  {Papaconstantinou}}, \ and\ \bibinfo {author} {\bibfnamefont {E~K~U}\
  \bibnamefont {Gross}},\ }\bibfield  {title} {\enquote {\bibinfo {title}
  {Spurious interactions, and their correction, in the ensemble-kohn-sham
  scheme for excited states},}\ }\href
  {https://doi.org/10.1103/PhysRevLett.88.033003} {\bibfield  {journal}
  {\bibinfo  {journal} {Phys. Rev. Lett.}\ }\textbf {\bibinfo {volume} {88}},\
  \bibinfo {pages} {033003} (\bibinfo {year} {2002})}\BibitemShut {NoStop}%
\bibitem [{\citenamefont {Nagy}(1995)}]{Nagy_ensAC}%
  \BibitemOpen
  \bibfield  {author} {\bibinfo {author} {\bibfnamefont {\'{A}}\ \bibnamefont
  {Nagy}},\ }\bibfield  {title} {\enquote {\bibinfo {title} {Coordinate scaling
  and adiabatic connection formula for ensembles of fractionally occupied
  excited states},}\ }\href {\doibase https://doi.org/10.1002/qua.560560406}
  {\bibfield  {journal} {\bibinfo  {journal} {Int. J. Quantum Chem.}\ }\textbf
  {\bibinfo {volume} {56}},\ \bibinfo {pages} {225--228} (\bibinfo {year}
  {1995})}\BibitemShut {NoStop}%
\bibitem [{\citenamefont {Carrascal}\ \emph {et~al.}(2015)\citenamefont
  {Carrascal}, \citenamefont {Ferrer}, \citenamefont {Smith},\ and\
  \citenamefont {Burke}}]{carrascal2015hubbard}%
  \BibitemOpen
  \bibfield  {author} {\bibinfo {author} {\bibfnamefont {D~J}\ \bibnamefont
  {Carrascal}}, \bibinfo {author} {\bibfnamefont {J}~\bibnamefont {Ferrer}},
  \bibinfo {author} {\bibfnamefont {J~C}\ \bibnamefont {Smith}}, \ and\
  \bibinfo {author} {\bibfnamefont {K}~\bibnamefont {Burke}},\ }\bibfield
  {title} {\enquote {\bibinfo {title} {The hubbard dimer: a density functional
  case study of a many-body problem},}\ }\href
  {http://stacks.iop.org/0953-8984/27/i=39/a=393001} {\bibfield  {journal}
  {\bibinfo  {journal} {J. Phys. Condens. Matter}\ }\textbf {\bibinfo {volume}
  {27}},\ \bibinfo {pages} {393001} (\bibinfo {year} {2015})}\BibitemShut
  {NoStop}%
\bibitem [{\citenamefont {Deur}\ \emph {et~al.}(2017)\citenamefont {Deur},
  \citenamefont {Mazouin},\ and\ \citenamefont {Fromager}}]{deur2017exact}%
  \BibitemOpen
  \bibfield  {author} {\bibinfo {author} {\bibfnamefont {Killian}\ \bibnamefont
  {Deur}}, \bibinfo {author} {\bibfnamefont {Laurent}\ \bibnamefont {Mazouin}},
  \ and\ \bibinfo {author} {\bibfnamefont {Emmanuel}\ \bibnamefont
  {Fromager}},\ }\bibfield  {title} {\enquote {\bibinfo {title} {Exact ensemble
  density functional theory for excited states in a model system: Investigating
  the weight dependence of the correlation energy},}\ }\href
  {https://doi.org/10.1103/PhysRevB.95.035120} {\bibfield  {journal} {\bibinfo
  {journal} {Phys. Rev. B}\ }\textbf {\bibinfo {volume} {95}},\ \bibinfo
  {pages} {035120} (\bibinfo {year} {2017})}\BibitemShut {NoStop}%
\bibitem [{\citenamefont {Trushin}\ \emph {et~al.}(2025)\citenamefont
  {Trushin}, \citenamefont {Fauser}, \citenamefont {M\"olkner}, \citenamefont
  {Erhard},\ and\ \citenamefont {G\"orling}}]{Trushin25_Accurate}%
  \BibitemOpen
  \bibfield  {author} {\bibinfo {author} {\bibfnamefont {Egor}\ \bibnamefont
  {Trushin}}, \bibinfo {author} {\bibfnamefont {Steffen}\ \bibnamefont
  {Fauser}}, \bibinfo {author} {\bibfnamefont {Andreas}\ \bibnamefont
  {M\"olkner}}, \bibinfo {author} {\bibfnamefont {Jannis}\ \bibnamefont
  {Erhard}}, \ and\ \bibinfo {author} {\bibfnamefont {Andreas}\ \bibnamefont
  {G\"orling}},\ }\bibfield  {title} {\enquote {\bibinfo {title} {Accurate
  correlation potentials from the self-consistent random phase
  approximation},}\ }\href {\doibase 10.1103/PhysRevLett.134.016402} {\bibfield
   {journal} {\bibinfo  {journal} {Phys. Rev. Lett.}\ }\textbf {\bibinfo
  {volume} {134}},\ \bibinfo {pages} {016402} (\bibinfo {year}
  {2025})}\BibitemShut {NoStop}%
\bibitem [{\citenamefont {Toulouse}\ \emph {et~al.}(2005)\citenamefont
  {Toulouse}, \citenamefont {Gori-Giorgi},\ and\ \citenamefont
  {Savin}}]{TousrXmd}%
  \BibitemOpen
  \bibfield  {author} {\bibinfo {author} {\bibfnamefont {J.}~\bibnamefont
  {Toulouse}}, \bibinfo {author} {\bibfnamefont {P.}~\bibnamefont
  {Gori-Giorgi}}, \ and\ \bibinfo {author} {\bibfnamefont {A.}~\bibnamefont
  {Savin}},\ }\bibfield  {title} {\enquote {\bibinfo {title} {A short-range
  correlation energy density functional with multi-determinantal reference},}\
  }\href {https://doi.org/10.1007/s00214-005-0688-2} {\bibfield  {journal}
  {\bibinfo  {journal} {Theor. Chem. Acc.}\ }\textbf {\bibinfo {volume}
  {114}},\ \bibinfo {pages} {305} (\bibinfo {year} {2005})}\BibitemShut
  {NoStop}%
\bibitem [{\citenamefont {Gori-Giorgi}\ and\ \citenamefont
  {Savin}(2009)}]{gori2009range}%
  \BibitemOpen
  \bibfield  {author} {\bibinfo {author} {\bibfnamefont {Paola}\ \bibnamefont
  {Gori-Giorgi}}\ and\ \bibinfo {author} {\bibfnamefont {Andreas}\ \bibnamefont
  {Savin}},\ }\bibfield  {title} {\enquote {\bibinfo {title} {Range separation
  combined with the overhauser model: Application to the h2 molecule along the
  dissociation curve},}\ }\href {\doibase https://doi.org/10.1002/qua.22034}
  {\bibfield  {journal} {\bibinfo  {journal} {International Journal of Quantum
  Chemistry}\ }\textbf {\bibinfo {volume} {109}},\ \bibinfo {pages}
  {1950--1961} (\bibinfo {year} {2009})}\BibitemShut {NoStop}%
\bibitem [{\citenamefont {Alam}\ \emph {et~al.}(2016)\citenamefont {Alam},
  \citenamefont {Knecht},\ and\ \citenamefont {Fromager}}]{alam2016ghost}%
  \BibitemOpen
  \bibfield  {author} {\bibinfo {author} {\bibfnamefont {Md~Mehboob}\
  \bibnamefont {Alam}}, \bibinfo {author} {\bibfnamefont {Stefan}\ \bibnamefont
  {Knecht}}, \ and\ \bibinfo {author} {\bibfnamefont {Emmanuel}\ \bibnamefont
  {Fromager}},\ }\bibfield  {title} {\enquote {\bibinfo {title}
  {Ghost-interaction correction in ensemble density-functional theory for
  excited states with and without range separation},}\ }\href
  {https://link.aps.org/doi/10.1103/PhysRevA.94.012511} {\bibfield  {journal}
  {\bibinfo  {journal} {Phys. Rev. A}\ }\textbf {\bibinfo {volume} {94}},\
  \bibinfo {pages} {012511} (\bibinfo {year} {2016})}\BibitemShut {NoStop}%
\bibitem [{\citenamefont {Fromager}\ and\ \citenamefont
  {Jensen}(2008)}]{srDFT_densitymatrixformulation}%
  \BibitemOpen
  \bibfield  {author} {\bibinfo {author} {\bibfnamefont {Emmanuel}\
  \bibnamefont {Fromager}}\ and\ \bibinfo {author} {\bibfnamefont {Hans
  J{\o}rgen~{\Aa}}\ \bibnamefont {Jensen}},\ }\bibfield  {title} {\enquote
  {\bibinfo {title} {Self-consistent many-body perturbation theory in
  range-separated density-functional theory: A one-electron
  reduced-density-matrix-based formulation},}\ }\href {\doibase
  10.1103/PhysRevA.78.022504} {\bibfield  {journal} {\bibinfo  {journal} {Phys.
  Rev. A}\ }\textbf {\bibinfo {volume} {78}},\ \bibinfo {pages} {022504}
  (\bibinfo {year} {2008})}\BibitemShut {NoStop}%
\bibitem [{\citenamefont {Wouters}\ \emph {et~al.}(2016)\citenamefont
  {Wouters}, \citenamefont {Jim{\'e}nez-Hoyos}, \citenamefont {Sun},\ and\
  \citenamefont {Chan}}]{wouters2016practical}%
  \BibitemOpen
  \bibfield  {author} {\bibinfo {author} {\bibfnamefont {Sebastian}\
  \bibnamefont {Wouters}}, \bibinfo {author} {\bibfnamefont {Carlos~A}\
  \bibnamefont {Jim{\'e}nez-Hoyos}}, \bibinfo {author} {\bibfnamefont {Qiming}\
  \bibnamefont {Sun}}, \ and\ \bibinfo {author} {\bibfnamefont {Garnet K-L}\
  \bibnamefont {Chan}},\ }\bibfield  {title} {\enquote {\bibinfo {title} {A
  practical guide to density matrix embedding theory in quantum chemistry},}\
  }\href {\doibase 10.1021/acs.jctc.6b00316} {\bibfield  {journal} {\bibinfo
  {journal} {J. Chem. Theory Comput.}\ }\textbf {\bibinfo {volume} {12}},\
  \bibinfo {pages} {2706--2719} (\bibinfo {year} {2016})}\BibitemShut {NoStop}%
\bibitem [{\citenamefont {Verma}\ \emph {et~al.}(2026)\citenamefont {Verma},
  \citenamefont {Mitra}, \citenamefont {Wang}, \citenamefont {D’Cunha},
  \citenamefont {Jangid}, \citenamefont {Hennefarth}, \citenamefont {Agarawal},
  \citenamefont {Otis}, \citenamefont {Haldar}, \citenamefont {Hermes},\ and\
  \citenamefont {Gagliardi}}]{Verma26_Multireference_Embedding}%
  \BibitemOpen
  \bibfield  {author} {\bibinfo {author} {\bibfnamefont {Shreya}\ \bibnamefont
  {Verma}}, \bibinfo {author} {\bibfnamefont {Abhishek}\ \bibnamefont {Mitra}},
  \bibinfo {author} {\bibfnamefont {Qiaohong}\ \bibnamefont {Wang}}, \bibinfo
  {author} {\bibfnamefont {Ruhee}\ \bibnamefont {D’Cunha}}, \bibinfo {author}
  {\bibfnamefont {Bhavnesh}\ \bibnamefont {Jangid}}, \bibinfo {author}
  {\bibfnamefont {Matthew~R.}\ \bibnamefont {Hennefarth}}, \bibinfo {author}
  {\bibfnamefont {Valay}\ \bibnamefont {Agarawal}}, \bibinfo {author}
  {\bibfnamefont {Leon}\ \bibnamefont {Otis}}, \bibinfo {author} {\bibfnamefont
  {Soumi}\ \bibnamefont {Haldar}}, \bibinfo {author} {\bibfnamefont
  {Matthew~R.}\ \bibnamefont {Hermes}}, \ and\ \bibinfo {author} {\bibfnamefont
  {Laura}\ \bibnamefont {Gagliardi}},\ }\bibfield  {title} {\enquote {\bibinfo
  {title} {Multireference embedding and fragmentation methods for classical and
  quantum computers: From model systems to realistic applications},}\ }\href
  {\doibase 10.1021/acs.chemrev.5c00486} {\bibfield  {journal} {\bibinfo
  {journal} {Chemical Reviews}\ }\textbf {\bibinfo {volume} {126}},\ \bibinfo
  {pages} {184--203} (\bibinfo {year} {2026})}\BibitemShut {NoStop}%
\bibitem [{\citenamefont {Tran}\ \emph {et~al.}(2019)\citenamefont {Tran},
  \citenamefont {Van~Voorhis},\ and\ \citenamefont {Thom}}]{tran2019using}%
  \BibitemOpen
  \bibfield  {author} {\bibinfo {author} {\bibfnamefont {Henry~K.}\
  \bibnamefont {Tran}}, \bibinfo {author} {\bibfnamefont {Troy}\ \bibnamefont
  {Van~Voorhis}}, \ and\ \bibinfo {author} {\bibfnamefont {Alex J.~W.}\
  \bibnamefont {Thom}},\ }\bibfield  {title} {\enquote {\bibinfo {title}
  {{Using SCF metadynamics to extend density matrix embedding theory to excited
  states}},}\ }\href {\doibase 10.1063/1.5096177} {\bibfield  {journal}
  {\bibinfo  {journal} {J. Chem. Phys.}\ }\textbf {\bibinfo {volume} {151}},\
  \bibinfo {pages} {034112} (\bibinfo {year} {2019})}\BibitemShut {NoStop}%
\bibitem [{\citenamefont {Ye}\ \emph {et~al.}(2021)\citenamefont {Ye},
  \citenamefont {Tran},\ and\ \citenamefont {Van~Voorhis}}]{ye2021accurate}%
  \BibitemOpen
  \bibfield  {author} {\bibinfo {author} {\bibfnamefont {Hong-Zhou}\
  \bibnamefont {Ye}}, \bibinfo {author} {\bibfnamefont {Henry~K.}\ \bibnamefont
  {Tran}}, \ and\ \bibinfo {author} {\bibfnamefont {Troy}\ \bibnamefont
  {Van~Voorhis}},\ }\bibfield  {title} {\enquote {\bibinfo {title} {Accurate
  electronic excitation energies in full-valence active space via bootstrap
  embedding},}\ }\href {\doibase 10.1021/acs.jctc.0c01221} {\bibfield
  {journal} {\bibinfo  {journal} {J. Chem. Theory Comput.}\ }\textbf {\bibinfo
  {volume} {17}},\ \bibinfo {pages} {3335--3347} (\bibinfo {year}
  {2021})}\BibitemShut {NoStop}%
\bibitem [{\citenamefont {Mitra}\ \emph {et~al.}(2021)\citenamefont {Mitra},
  \citenamefont {Pham}, \citenamefont {Pandharkar}, \citenamefont {Hermes},\
  and\ \citenamefont {Gagliardi}}]{mitra2021excited}%
  \BibitemOpen
  \bibfield  {author} {\bibinfo {author} {\bibfnamefont {Abhishek}\
  \bibnamefont {Mitra}}, \bibinfo {author} {\bibfnamefont {Hung~Q.}\
  \bibnamefont {Pham}}, \bibinfo {author} {\bibfnamefont {Riddhish}\
  \bibnamefont {Pandharkar}}, \bibinfo {author} {\bibfnamefont {Matthew~R.}\
  \bibnamefont {Hermes}}, \ and\ \bibinfo {author} {\bibfnamefont {Laura}\
  \bibnamefont {Gagliardi}},\ }\bibfield  {title} {\enquote {\bibinfo {title}
  {Excited states of crystalline point defects with multireference density
  matrix embedding theory},}\ }\href {\doibase 10.1021/acs.jpclett.1c03229}
  {\bibfield  {journal} {\bibinfo  {journal} {J. Phys. Chem. Lett.}\ }\textbf
  {\bibinfo {volume} {12}},\ \bibinfo {pages} {11688--11694} (\bibinfo {year}
  {2021})}\BibitemShut {NoStop}%
\bibitem [{\citenamefont {Chen}\ \emph {et~al.}(2014)\citenamefont {Chen},
  \citenamefont {Booth}, \citenamefont {Sharma}, \citenamefont {Knizia},\ and\
  \citenamefont {Chan}}]{chen2014intermediate}%
  \BibitemOpen
  \bibfield  {author} {\bibinfo {author} {\bibfnamefont {Qiaoni}\ \bibnamefont
  {Chen}}, \bibinfo {author} {\bibfnamefont {George~H}\ \bibnamefont {Booth}},
  \bibinfo {author} {\bibfnamefont {Sandeep}\ \bibnamefont {Sharma}}, \bibinfo
  {author} {\bibfnamefont {Gerald}\ \bibnamefont {Knizia}}, \ and\ \bibinfo
  {author} {\bibfnamefont {Garnet Kin-Lic}\ \bibnamefont {Chan}},\ }\bibfield
  {title} {\enquote {\bibinfo {title} {Intermediate and spin-liquid phase of
  the half-filled honeycomb hubbard model},}\ }\href
  {https://doi.org/10.1103/PhysRevB.89.165134} {\bibfield  {journal} {\bibinfo
  {journal} {Phys. Rev. B}\ }\textbf {\bibinfo {volume} {89}},\ \bibinfo
  {pages} {165134} (\bibinfo {year} {2014})}\BibitemShut {NoStop}%
\bibitem [{\citenamefont {Booth}\ and\ \citenamefont
  {Chan}(2015)}]{booth2015spectral}%
  \BibitemOpen
  \bibfield  {author} {\bibinfo {author} {\bibfnamefont {George~H}\
  \bibnamefont {Booth}}\ and\ \bibinfo {author} {\bibfnamefont {Garnet
  Kin-Lic}\ \bibnamefont {Chan}},\ }\bibfield  {title} {\enquote {\bibinfo
  {title} {Spectral functions of strongly correlated extended systems via an
  exact quantum embedding},}\ }\href
  {https://doi.org/10.1103/PhysRevB.91.155107} {\bibfield  {journal} {\bibinfo
  {journal} {Phys. Rev. B}\ }\textbf {\bibinfo {volume} {91}},\ \bibinfo
  {pages} {155107} (\bibinfo {year} {2015})}\BibitemShut {NoStop}%
\bibitem [{\citenamefont {Verma}\ \emph {et~al.}(2023)\citenamefont {Verma},
  \citenamefont {Mitra}, \citenamefont {Jin}, \citenamefont {Haldar},
  \citenamefont {Vorwerk}, \citenamefont {Hermes}, \citenamefont {Galli},\ and\
  \citenamefont {Gagliardi}}]{Verma2023_Optical}%
  \BibitemOpen
  \bibfield  {author} {\bibinfo {author} {\bibfnamefont {Shreya}\ \bibnamefont
  {Verma}}, \bibinfo {author} {\bibfnamefont {Abhishek}\ \bibnamefont {Mitra}},
  \bibinfo {author} {\bibfnamefont {Yu}~\bibnamefont {Jin}}, \bibinfo {author}
  {\bibfnamefont {Soumi}\ \bibnamefont {Haldar}}, \bibinfo {author}
  {\bibfnamefont {Christian}\ \bibnamefont {Vorwerk}}, \bibinfo {author}
  {\bibfnamefont {Matthew~R.}\ \bibnamefont {Hermes}}, \bibinfo {author}
  {\bibfnamefont {Giulia}\ \bibnamefont {Galli}}, \ and\ \bibinfo {author}
  {\bibfnamefont {Laura}\ \bibnamefont {Gagliardi}},\ }\bibfield  {title}
  {\enquote {\bibinfo {title} {Optical properties of neutral f centers in bulk
  mgo with density matrix embedding},}\ }\href {\doibase
  10.1021/acs.jpclett.3c01875} {\bibfield  {journal} {\bibinfo  {journal} {The
  Journal of Physical Chemistry Letters}\ }\textbf {\bibinfo {volume} {14}},\
  \bibinfo {pages} {7703--7710} (\bibinfo {year} {2023})}\BibitemShut {NoStop}%
\bibitem [{\citenamefont {Lau}\ \emph {et~al.}(2024)\citenamefont {Lau},
  \citenamefont {Busemeyer},\ and\ \citenamefont
  {Berkelbach}}]{Lau2024_Optical}%
  \BibitemOpen
  \bibfield  {author} {\bibinfo {author} {\bibfnamefont {Bryan T.~G.}\
  \bibnamefont {Lau}}, \bibinfo {author} {\bibfnamefont {Brian}\ \bibnamefont
  {Busemeyer}}, \ and\ \bibinfo {author} {\bibfnamefont {Timothy~C.}\
  \bibnamefont {Berkelbach}},\ }\bibfield  {title} {\enquote {\bibinfo {title}
  {Optical properties of defects in solids via quantum embedding with good
  active space orbitals},}\ }\href {\doibase 10.1021/acs.jpcc.3c08185}
  {\bibfield  {journal} {\bibinfo  {journal} {The Journal of Physical Chemistry
  C}\ }\textbf {\bibinfo {volume} {128}},\ \bibinfo {pages} {2959--2966}
  (\bibinfo {year} {2024})}\BibitemShut {NoStop}%
\bibitem [{\citenamefont {Ai}\ \emph {et~al.}(2022)\citenamefont {Ai},
  \citenamefont {Sun},\ and\ \citenamefont {Jiang}}]{Ai2022_Efficient}%
  \BibitemOpen
  \bibfield  {author} {\bibinfo {author} {\bibfnamefont {Yuhang}\ \bibnamefont
  {Ai}}, \bibinfo {author} {\bibfnamefont {Qiming}\ \bibnamefont {Sun}}, \ and\
  \bibinfo {author} {\bibfnamefont {Hong}\ \bibnamefont {Jiang}},\ }\bibfield
  {title} {\enquote {\bibinfo {title} {Efficient multiconfigurational quantum
  chemistry approach to single-ion magnets based on density matrix embedding
  theory},}\ }\href {\doibase 10.1021/acs.jpclett.2c02890} {\bibfield
  {journal} {\bibinfo  {journal} {The Journal of Physical Chemistry Letters}\
  }\textbf {\bibinfo {volume} {13}},\ \bibinfo {pages} {10627--10634} (\bibinfo
  {year} {2022})}\BibitemShut {NoStop}%
\bibitem [{\citenamefont {Li}\ \emph {et~al.}(2025{\natexlab{a}})\citenamefont
  {Li}, \citenamefont {Li}, \citenamefont {Zhai},\ and\ \citenamefont
  {Chan}}]{Li2026_Towards-excitations}%
  \BibitemOpen
  \bibfield  {author} {\bibinfo {author} {\bibfnamefont {Shuoxue}\ \bibnamefont
  {Li}}, \bibinfo {author} {\bibfnamefont {Chenghan}\ \bibnamefont {Li}},
  \bibinfo {author} {\bibfnamefont {Huanchen}\ \bibnamefont {Zhai}}, \ and\
  \bibinfo {author} {\bibfnamefont {Garnet Kin-Lic}\ \bibnamefont {Chan}},\
  }\bibfield  {title} {\enquote {\bibinfo {title} {Towards excitations and
  dynamical quantities in correlated lattices with density matrix embedding
  theory},}\ }\href {\doibase 10.1103/vyvq-chlt} {\bibfield  {journal}
  {\bibinfo  {journal} {Phys. Rev. B}\ }\textbf {\bibinfo {volume} {112}},\
  \bibinfo {pages} {125154} (\bibinfo {year} {2025}{\natexlab{a}})}\BibitemShut
  {NoStop}%
\bibitem [{\citenamefont {Cernatic}\ \emph
  {et~al.}(2024{\natexlab{b}})\citenamefont {Cernatic}, \citenamefont
  {Fromager},\ and\ \citenamefont {Yalouz}}]{cernatic2024fragment}%
  \BibitemOpen
  \bibfield  {author} {\bibinfo {author} {\bibfnamefont {Filip}\ \bibnamefont
  {Cernatic}}, \bibinfo {author} {\bibfnamefont {Emmanuel}\ \bibnamefont
  {Fromager}}, \ and\ \bibinfo {author} {\bibfnamefont {Saad}\ \bibnamefont
  {Yalouz}},\ }\bibfield  {title} {\enquote {\bibinfo {title} {Fragment quantum
  embedding using the householder transformation: A multi-state extension based
  on ensembles},}\ }\href {\doibase 10.1063/5.0229787} {\bibfield  {journal}
  {\bibinfo  {journal} {The Journal of Chemical Physics}\ }\textbf {\bibinfo
  {volume} {161}},\ \bibinfo {pages} {124107} (\bibinfo {year}
  {2024}{\natexlab{b}})}\BibitemShut {NoStop}%
\bibitem [{\citenamefont {Bulik}\ \emph
  {et~al.}(2014{\natexlab{a}})\citenamefont {Bulik}, \citenamefont {Scuseria},\
  and\ \citenamefont {Dukelsky}}]{bulik2014density}%
  \BibitemOpen
  \bibfield  {author} {\bibinfo {author} {\bibfnamefont {Ireneusz~W}\
  \bibnamefont {Bulik}}, \bibinfo {author} {\bibfnamefont {Gustavo~E}\
  \bibnamefont {Scuseria}}, \ and\ \bibinfo {author} {\bibfnamefont {Jorge}\
  \bibnamefont {Dukelsky}},\ }\bibfield  {title} {\enquote {\bibinfo {title}
  {Density matrix embedding from broken symmetry lattice mean fields},}\ }\href
  {https://doi.org/10.1103/PhysRevB.89.035140} {\bibfield  {journal} {\bibinfo
  {journal} {Phys. Rev. B}\ }\textbf {\bibinfo {volume} {89}},\ \bibinfo
  {pages} {035140} (\bibinfo {year} {2014}{\natexlab{a}})}\BibitemShut
  {NoStop}%
\bibitem [{\citenamefont {Bulik}\ \emph
  {et~al.}(2014{\natexlab{b}})\citenamefont {Bulik}, \citenamefont {Chen},\
  and\ \citenamefont {Scuseria}}]{bulik2014electron}%
  \BibitemOpen
  \bibfield  {author} {\bibinfo {author} {\bibfnamefont {Ireneusz~W}\
  \bibnamefont {Bulik}}, \bibinfo {author} {\bibfnamefont {Weibing}\
  \bibnamefont {Chen}}, \ and\ \bibinfo {author} {\bibfnamefont {Gustavo~E}\
  \bibnamefont {Scuseria}},\ }\bibfield  {title} {\enquote {\bibinfo {title}
  {Electron correlation in solids via density embedding theory},}\ }\href
  {https://doi.org/10.1063/1.4891861} {\bibfield  {journal} {\bibinfo
  {journal} {J. Chem. Phys.}\ }\textbf {\bibinfo {volume} {141}},\ \bibinfo
  {pages} {054113} (\bibinfo {year} {2014}{\natexlab{b}})}\BibitemShut
  {NoStop}%
\bibitem [{\citenamefont {Fulde}\ and\ \citenamefont
  {Stoll}(2017)}]{fulde2017dealing}%
  \BibitemOpen
  \bibfield  {author} {\bibinfo {author} {\bibfnamefont {Peter}\ \bibnamefont
  {Fulde}}\ and\ \bibinfo {author} {\bibfnamefont {Hermann}\ \bibnamefont
  {Stoll}},\ }\bibfield  {title} {\enquote {\bibinfo {title} {Dealing with the
  exponential wall in electronic structure calculations},}\ }\href
  {https://doi.org/10.1063/1.4983207} {\bibfield  {journal} {\bibinfo
  {journal} {J. Chem. Phys.}\ }\textbf {\bibinfo {volume} {146}},\ \bibinfo
  {pages} {194107} (\bibinfo {year} {2017})}\BibitemShut {NoStop}%
\bibitem [{\citenamefont {Plat}\ and\ \citenamefont
  {Hotta}(2020)}]{plat2020entanglement}%
  \BibitemOpen
  \bibfield  {author} {\bibinfo {author} {\bibfnamefont {Xavier}\ \bibnamefont
  {Plat}}\ and\ \bibinfo {author} {\bibfnamefont {Chisa}\ \bibnamefont
  {Hotta}},\ }\bibfield  {title} {\enquote {\bibinfo {title} {Entanglement
  spectrum as a marker for phase transitions in the density embedding theory
  for interacting spinless fermionic models},}\ }\href
  {https://doi.org/10.1103/PhysRevB.102.140410} {\bibfield  {journal} {\bibinfo
   {journal} {Phys. Rev. B}\ }\textbf {\bibinfo {volume} {102}},\ \bibinfo
  {pages} {140410} (\bibinfo {year} {2020})}\BibitemShut {NoStop}%
\bibitem [{\citenamefont {Mordovina}\ \emph {et~al.}(2019)\citenamefont
  {Mordovina}, \citenamefont {Reinhard}, \citenamefont {Theophilou},
  \citenamefont {Appel},\ and\ \citenamefont {Rubio}}]{mordovina2019self}%
  \BibitemOpen
  \bibfield  {author} {\bibinfo {author} {\bibfnamefont {Uliana}\ \bibnamefont
  {Mordovina}}, \bibinfo {author} {\bibfnamefont {Teresa~E.}\ \bibnamefont
  {Reinhard}}, \bibinfo {author} {\bibfnamefont {Iris}\ \bibnamefont
  {Theophilou}}, \bibinfo {author} {\bibfnamefont {Heiko}\ \bibnamefont
  {Appel}}, \ and\ \bibinfo {author} {\bibfnamefont {Angel}\ \bibnamefont
  {Rubio}},\ }\bibfield  {title} {\enquote {\bibinfo {title} {Self-consistent
  density-functional embedding: A novel approach for density-functional
  approximations},}\ }\href {\doibase 10.1021/acs.jctc.9b00063} {\bibfield
  {journal} {\bibinfo  {journal} {J. Chem. Theory Comput.}\ }\textbf {\bibinfo
  {volume} {15}},\ \bibinfo {pages} {5209--5220} (\bibinfo {year}
  {2019})}\BibitemShut {NoStop}%
\bibitem [{\citenamefont {Sekaran}\ \emph {et~al.}(2022)\citenamefont
  {Sekaran}, \citenamefont {Saubanère},\ and\ \citenamefont
  {Fromager}}]{sekaran2022local}%
  \BibitemOpen
  \bibfield  {author} {\bibinfo {author} {\bibfnamefont {Sajanthan}\
  \bibnamefont {Sekaran}}, \bibinfo {author} {\bibfnamefont {Matthieu}\
  \bibnamefont {Saubanère}}, \ and\ \bibinfo {author} {\bibfnamefont
  {Emmanuel}\ \bibnamefont {Fromager}},\ }\bibfield  {title} {\enquote
  {\bibinfo {title} {Local potential functional embedding theory: A
  self-consistent flavor of density functional theory for lattices without
  density functionals},}\ }\href {\doibase 10.3390/computation10030045}
  {\bibfield  {journal} {\bibinfo  {journal} {Computation}\ }\textbf {\bibinfo
  {volume} {10}},\ \bibinfo {pages} {45} (\bibinfo {year} {2022})}\BibitemShut
  {NoStop}%
\bibitem [{\citenamefont {Capelle}\ and\ \citenamefont {{Campo
  Jr.}}(2013)}]{DFT_ModelHamiltonians}%
  \BibitemOpen
  \bibfield  {author} {\bibinfo {author} {\bibfnamefont {Klaus}\ \bibnamefont
  {Capelle}}\ and\ \bibinfo {author} {\bibfnamefont {Vivaldo~L.}\ \bibnamefont
  {{Campo Jr.}}},\ }\bibfield  {title} {\enquote {\bibinfo {title} {Density
  functionals and model hamiltonians: Pillars of many-particle physics},}\
  }\href {https://doi.org/10.1016/j.physrep.2013.03.002} {\bibfield  {journal}
  {\bibinfo  {journal} {Phys. Rep.}\ }\textbf {\bibinfo {volume} {528}},\
  \bibinfo {pages} {91} (\bibinfo {year} {2013})}\BibitemShut {NoStop}%
\bibitem [{\citenamefont {Sekaran}\ \emph {et~al.}(2023)\citenamefont
  {Sekaran}, \citenamefont {Bindech},\ and\ \citenamefont
  {Fromager}}]{sekaran2023unified}%
  \BibitemOpen
  \bibfield  {author} {\bibinfo {author} {\bibfnamefont {Sajanthan}\
  \bibnamefont {Sekaran}}, \bibinfo {author} {\bibfnamefont {Oussama}\
  \bibnamefont {Bindech}}, \ and\ \bibinfo {author} {\bibfnamefont {Emmanuel}\
  \bibnamefont {Fromager}},\ }\bibfield  {title} {\enquote {\bibinfo {title}
  {{A unified density matrix functional construction of quantum baths in
  density matrix embedding theory beyond the mean-field approximation}},}\
  }\href {\doibase 10.1063/5.0157746} {\bibfield  {journal} {\bibinfo
  {journal} {J. Chem. Phys.}\ }\textbf {\bibinfo {volume} {159}},\ \bibinfo
  {pages} {034107} (\bibinfo {year} {2023})}\BibitemShut {NoStop}%
\bibitem [{\citenamefont {Sekaran}\ \emph {et~al.}(2021)\citenamefont
  {Sekaran}, \citenamefont {Tsuchiizu}, \citenamefont {Sauban\`ere},\ and\
  \citenamefont {Fromager}}]{sekaran2021householder}%
  \BibitemOpen
  \bibfield  {author} {\bibinfo {author} {\bibfnamefont {Sajanthan}\
  \bibnamefont {Sekaran}}, \bibinfo {author} {\bibfnamefont {Masahisa}\
  \bibnamefont {Tsuchiizu}}, \bibinfo {author} {\bibfnamefont {Matthieu}\
  \bibnamefont {Sauban\`ere}}, \ and\ \bibinfo {author} {\bibfnamefont
  {Emmanuel}\ \bibnamefont {Fromager}},\ }\bibfield  {title} {\enquote
  {\bibinfo {title} {Householder-transformed density matrix functional
  embedding theory},}\ }\href {\doibase 10.1103/PhysRevB.104.035121} {\bibfield
   {journal} {\bibinfo  {journal} {Phys. Rev. B}\ }\textbf {\bibinfo {volume}
  {104}},\ \bibinfo {pages} {035121} (\bibinfo {year} {2021})}\BibitemShut
  {NoStop}%
\bibitem [{\citenamefont {Geerlings}\ \emph {et~al.}(2003)\citenamefont
  {Geerlings}, \citenamefont {De~Proft},\ and\ \citenamefont
  {Langenaeker}}]{CDFTrev2003}%
  \BibitemOpen
  \bibfield  {author} {\bibinfo {author} {\bibfnamefont {P.}~\bibnamefont
  {Geerlings}}, \bibinfo {author} {\bibfnamefont {F.}~\bibnamefont {De~Proft}},
  \ and\ \bibinfo {author} {\bibfnamefont {W.}~\bibnamefont {Langenaeker}},\
  }\bibfield  {title} {\enquote {\bibinfo {title} {Conceptual density
  functional theory},}\ }\href {\doibase 10.1021/cr990029p} {\bibfield
  {journal} {\bibinfo  {journal} {Chemical Reviews}\ }\textbf {\bibinfo
  {volume} {103}},\ \bibinfo {pages} {1793--1874} (\bibinfo {year}
  {2003})}\BibitemShut {NoStop}%
\bibitem [{\citenamefont {Filatov}\ \emph {et~al.}(2024)\citenamefont
  {Filatov}, \citenamefont {Mironov},\ and\ \citenamefont
  {Kraka}}]{Filatov24_Unraveling}%
  \BibitemOpen
  \bibfield  {author} {\bibinfo {author} {\bibfnamefont {Michael}\ \bibnamefont
  {Filatov}}, \bibinfo {author} {\bibfnamefont {Vladimir}\ \bibnamefont
  {Mironov}}, \ and\ \bibinfo {author} {\bibfnamefont {Elfi}\ \bibnamefont
  {Kraka}},\ }\bibfield  {title} {\enquote {\bibinfo {title} {Unraveling the
  effect of aromaticity for the dynamics of excited states of single benzene
  fluorophores},}\ }\href {\doibase https://doi.org/10.1002/jcc.27304}
  {\bibfield  {journal} {\bibinfo  {journal} {Journal of Computational
  Chemistry}\ }\textbf {\bibinfo {volume} {45}},\ \bibinfo {pages} {1033--1045}
  (\bibinfo {year} {2024})}\BibitemShut {NoStop}%
\bibitem [{\citenamefont {K\"ummel}\ and\ \citenamefont
  {Perdew}(2003)}]{Perdew03OEP}%
  \BibitemOpen
  \bibfield  {author} {\bibinfo {author} {\bibfnamefont {Stephan}\ \bibnamefont
  {K\"ummel}}\ and\ \bibinfo {author} {\bibfnamefont {John~P.}\ \bibnamefont
  {Perdew}},\ }\bibfield  {title} {\enquote {\bibinfo {title} {Simple iterative
  construction of the optimized effective potential for orbital functionals,
  including exact exchange},}\ }\href {\doibase 10.1103/PhysRevLett.90.043004}
  {\bibfield  {journal} {\bibinfo  {journal} {Phys. Rev. Lett.}\ }\textbf
  {\bibinfo {volume} {90}},\ \bibinfo {pages} {043004} (\bibinfo {year}
  {2003})}\BibitemShut {NoStop}%
\bibitem [{\citenamefont {Yang}\ \emph {et~al.}(2004)\citenamefont {Yang},
  \citenamefont {Ayers},\ and\ \citenamefont {Wu}}]{Weitao04PFT}%
  \BibitemOpen
  \bibfield  {author} {\bibinfo {author} {\bibfnamefont {Weitao}\ \bibnamefont
  {Yang}}, \bibinfo {author} {\bibfnamefont {Paul~W.}\ \bibnamefont {Ayers}}, \
  and\ \bibinfo {author} {\bibfnamefont {Qin}\ \bibnamefont {Wu}},\ }\bibfield
  {title} {\enquote {\bibinfo {title} {Potential functionals: Dual to density
  functionals and solution to the $v$-representability problem},}\ }\href
  {\doibase 10.1103/PhysRevLett.92.146404} {\bibfield  {journal} {\bibinfo
  {journal} {Phys. Rev. Lett.}\ }\textbf {\bibinfo {volume} {92}},\ \bibinfo
  {pages} {146404} (\bibinfo {year} {2004})}\BibitemShut {NoStop}%
\bibitem [{\citenamefont {Heaton-Burgess}\ \emph {et~al.}(2007)\citenamefont
  {Heaton-Burgess}, \citenamefont {Bulat},\ and\ \citenamefont
  {Yang}}]{heaton2007optimized}%
  \BibitemOpen
  \bibfield  {author} {\bibinfo {author} {\bibfnamefont {Tim}\ \bibnamefont
  {Heaton-Burgess}}, \bibinfo {author} {\bibfnamefont {Felipe~A}\ \bibnamefont
  {Bulat}}, \ and\ \bibinfo {author} {\bibfnamefont {Weitao}\ \bibnamefont
  {Yang}},\ }\bibfield  {title} {\enquote {\bibinfo {title} {Optimized
  effective potentials in finite basis sets},}\ }\href@noop {} {\bibfield
  {journal} {\bibinfo  {journal} {Phys. Rev. Lett.}\ }\textbf {\bibinfo
  {volume} {98}},\ \bibinfo {pages} {256401} (\bibinfo {year}
  {2007})}\BibitemShut {NoStop}%
\bibitem [{\citenamefont {Goshen}\ and\ \citenamefont
  {Kraisler}(2026)}]{goshen2026manyelectronsystemsfractionalelectron}%
  \BibitemOpen
  \bibfield  {author} {\bibinfo {author} {\bibfnamefont {Yuli}\ \bibnamefont
  {Goshen}}\ and\ \bibinfo {author} {\bibfnamefont {Eli}\ \bibnamefont
  {Kraisler}},\ }\href {https://arxiv.org/abs/2601.03012} {\enquote {\bibinfo
  {title} {Many-electron systems with fractional electron number and spin:
  exact properties above and below the equilibrium total spin value},}\ }
  (\bibinfo {year} {2026}),\ \Eprint {http://arxiv.org/abs/2601.03012}
  {arXiv:2601.03012 [cond-mat.mtrl-sci]} \BibitemShut {NoStop}%
\bibitem [{\citenamefont {Goshen}\ and\ \citenamefont
  {Kraisler}(2024)}]{Goshen24_Ensemble}%
  \BibitemOpen
  \bibfield  {author} {\bibinfo {author} {\bibfnamefont {Yuli}\ \bibnamefont
  {Goshen}}\ and\ \bibinfo {author} {\bibfnamefont {Eli}\ \bibnamefont
  {Kraisler}},\ }\bibfield  {title} {\enquote {\bibinfo {title} {Ensemble
  ground state of a many-electron system with fractional electron number and
  spin: Piecewise-linearity and flat-plane condition generalized},}\ }\href
  {\doibase 10.1021/acs.jpclett.3c03509} {\bibfield  {journal} {\bibinfo
  {journal} {The Journal of Physical Chemistry Letters}\ }\textbf {\bibinfo
  {volume} {15}},\ \bibinfo {pages} {2337--2343} (\bibinfo {year}
  {2024})}\BibitemShut {NoStop}%
\bibitem [{\citenamefont {Hayman}\ \emph {et~al.}(2025)\citenamefont {Hayman},
  \citenamefont {Levy}, \citenamefont {Goshen}, \citenamefont {Fraenkel},
  \citenamefont {Kraisler},\ and\ \citenamefont
  {Stein}}]{Hayman25_spin-migration}%
  \BibitemOpen
  \bibfield  {author} {\bibinfo {author} {\bibfnamefont {Alon}\ \bibnamefont
  {Hayman}}, \bibinfo {author} {\bibfnamefont {Nevo}\ \bibnamefont {Levy}},
  \bibinfo {author} {\bibfnamefont {Yuli}\ \bibnamefont {Goshen}}, \bibinfo
  {author} {\bibfnamefont {Malachi}\ \bibnamefont {Fraenkel}}, \bibinfo
  {author} {\bibfnamefont {Eli}\ \bibnamefont {Kraisler}}, \ and\ \bibinfo
  {author} {\bibfnamefont {Tamar}\ \bibnamefont {Stein}},\ }\bibfield  {title}
  {\enquote {\bibinfo {title} {Spin migration in density functional theory:
  Energy, potential, and density perspectives},}\ }\href {\doibase
  10.1063/5.0241200} {\bibfield  {journal} {\bibinfo  {journal} {The Journal of
  Chemical Physics}\ }\textbf {\bibinfo {volume} {162}},\ \bibinfo {pages}
  {114301} (\bibinfo {year} {2025})}\BibitemShut {NoStop}%
\bibitem [{\citenamefont {Burgess}\ \emph {et~al.}(2024)\citenamefont
  {Burgess}, \citenamefont {Linscott},\ and\ \citenamefont
  {O'Regan}}]{Burgess24_Tilted-Plane}%
  \BibitemOpen
  \bibfield  {author} {\bibinfo {author} {\bibfnamefont {Andrew~C.}\
  \bibnamefont {Burgess}}, \bibinfo {author} {\bibfnamefont {Edward}\
  \bibnamefont {Linscott}}, \ and\ \bibinfo {author} {\bibfnamefont {David~D.}\
  \bibnamefont {O'Regan}},\ }\bibfield  {title} {\enquote {\bibinfo {title}
  {Tilted-plane structure of the energy of finite quantum systems},}\ }\href
  {\doibase 10.1103/PhysRevLett.133.026404} {\bibfield  {journal} {\bibinfo
  {journal} {Phys. Rev. Lett.}\ }\textbf {\bibinfo {volume} {133}},\ \bibinfo
  {pages} {026404} (\bibinfo {year} {2024})}\BibitemShut {NoStop}%
\bibitem [{\citenamefont {Mori-S\'anchez}\ \emph {et~al.}(2009)\citenamefont
  {Mori-S\'anchez}, \citenamefont {Cohen},\ and\ \citenamefont
  {Yang}}]{Mori-Sanchez09_Discontinuous}%
  \BibitemOpen
  \bibfield  {author} {\bibinfo {author} {\bibfnamefont {Paula}\ \bibnamefont
  {Mori-S\'anchez}}, \bibinfo {author} {\bibfnamefont {Aron~J.}\ \bibnamefont
  {Cohen}}, \ and\ \bibinfo {author} {\bibfnamefont {Weitao}\ \bibnamefont
  {Yang}},\ }\bibfield  {title} {\enquote {\bibinfo {title} {Discontinuous
  nature of the exchange-correlation functional in strongly correlated
  systems},}\ }\href {\doibase 10.1103/PhysRevLett.102.066403} {\bibfield
  {journal} {\bibinfo  {journal} {Phys. Rev. Lett.}\ }\textbf {\bibinfo
  {volume} {102}},\ \bibinfo {pages} {066403} (\bibinfo {year}
  {2009})}\BibitemShut {NoStop}%
\bibitem [{\citenamefont {Su}\ \emph {et~al.}(2018)\citenamefont {Su},
  \citenamefont {Li},\ and\ \citenamefont {Yang}}]{Neil-Qiang-Su18_Describing}%
  \BibitemOpen
  \bibfield  {author} {\bibinfo {author} {\bibfnamefont {Neil~Qiang}\
  \bibnamefont {Su}}, \bibinfo {author} {\bibfnamefont {Chen}\ \bibnamefont
  {Li}}, \ and\ \bibinfo {author} {\bibfnamefont {Weitao}\ \bibnamefont
  {Yang}},\ }\bibfield  {title} {\enquote {\bibinfo {title} {Describing strong
  correlation with fractional-spin correction in density functional theory},}\
  }\href {\doibase 10.1073/pnas.1807095115} {\bibfield  {journal} {\bibinfo
  {journal} {Proceedings of the National Academy of Sciences}\ }\textbf
  {\bibinfo {volume} {115}},\ \bibinfo {pages} {9678--9683} (\bibinfo {year}
  {2018})}\BibitemShut {NoStop}%
\bibitem [{\citenamefont {Smith}\ \emph {et~al.}(2016)\citenamefont {Smith},
  \citenamefont {Pribram-Jones},\ and\ \citenamefont {Burke}}]{smith2016exact}%
  \BibitemOpen
  \bibfield  {author} {\bibinfo {author} {\bibfnamefont {J.~C.}\ \bibnamefont
  {Smith}}, \bibinfo {author} {\bibfnamefont {A.}~\bibnamefont
  {Pribram-Jones}}, \ and\ \bibinfo {author} {\bibfnamefont {K.}~\bibnamefont
  {Burke}},\ }\bibfield  {title} {\enquote {\bibinfo {title} {Exact thermal
  density functional theory for a model system: Correlation components and
  accuracy of the zero-temperature exchange-correlation approximation},}\
  }\href {\doibase 10.1103/PhysRevB.93.245131} {\bibfield  {journal} {\bibinfo
  {journal} {Phys. Rev. B}\ }\textbf {\bibinfo {volume} {93}},\ \bibinfo
  {pages} {245131} (\bibinfo {year} {2016})}\BibitemShut {NoStop}%
\bibitem [{\citenamefont {Mermin}(1965)}]{Mermin65_Thermal}%
  \BibitemOpen
  \bibfield  {author} {\bibinfo {author} {\bibfnamefont {N.~David}\
  \bibnamefont {Mermin}},\ }\bibfield  {title} {\enquote {\bibinfo {title}
  {Thermal properties of the inhomogeneous electron gas},}\ }\href {\doibase
  10.1103/PhysRev.137.A1441} {\bibfield  {journal} {\bibinfo  {journal} {Phys.
  Rev.}\ }\textbf {\bibinfo {volume} {137}},\ \bibinfo {pages} {A1441--A1443}
  (\bibinfo {year} {1965})}\BibitemShut {NoStop}%
\bibitem [{\citenamefont {Hilleke}\ \emph {et~al.}(2025)\citenamefont
  {Hilleke}, \citenamefont {Karasiev}, \citenamefont {Trickey}, \citenamefont
  {Goshadze},\ and\ \citenamefont {Hu}}]{Hilleke25_Fully}%
  \BibitemOpen
  \bibfield  {author} {\bibinfo {author} {\bibfnamefont {Katerina~P.}\
  \bibnamefont {Hilleke}}, \bibinfo {author} {\bibfnamefont {Valentin~V.}\
  \bibnamefont {Karasiev}}, \bibinfo {author} {\bibfnamefont {S.~B.}\
  \bibnamefont {Trickey}}, \bibinfo {author} {\bibfnamefont {R.~M.~N.}\
  \bibnamefont {Goshadze}}, \ and\ \bibinfo {author} {\bibfnamefont {S.~X.}\
  \bibnamefont {Hu}},\ }\bibfield  {title} {\enquote {\bibinfo {title} {Fully
  thermal meta-gga exchange correlation free-energy density functional},}\
  }\href {\doibase 10.1103/PhysRevMaterials.9.L050801} {\bibfield  {journal}
  {\bibinfo  {journal} {Phys. Rev. Mater.}\ }\textbf {\bibinfo {volume} {9}},\
  \bibinfo {pages} {L050801} (\bibinfo {year} {2025})}\BibitemShut {NoStop}%
\bibitem [{\citenamefont {Pittalis}\ \emph {et~al.}(2011)\citenamefont
  {Pittalis}, \citenamefont {Proetto}, \citenamefont {Floris}, \citenamefont
  {Sanna}, \citenamefont {Bersier}, \citenamefont {Burke},\ and\ \citenamefont
  {Gross}}]{PRL11_Pittalis_exact_conds_thermalDFT}%
  \BibitemOpen
  \bibfield  {author} {\bibinfo {author} {\bibfnamefont {S.}~\bibnamefont
  {Pittalis}}, \bibinfo {author} {\bibfnamefont {C.~R.}\ \bibnamefont
  {Proetto}}, \bibinfo {author} {\bibfnamefont {A.}~\bibnamefont {Floris}},
  \bibinfo {author} {\bibfnamefont {A.}~\bibnamefont {Sanna}}, \bibinfo
  {author} {\bibfnamefont {C.}~\bibnamefont {Bersier}}, \bibinfo {author}
  {\bibfnamefont {K.}~\bibnamefont {Burke}}, \ and\ \bibinfo {author}
  {\bibfnamefont {E.~K.~U.}\ \bibnamefont {Gross}},\ }\bibfield  {title}
  {\enquote {\bibinfo {title} {Exact conditions in finite-temperature
  density-functional theory},}\ }\href {\doibase
  10.1103/PhysRevLett.107.163001} {\bibfield  {journal} {\bibinfo  {journal}
  {Phys. Rev. Lett.}\ }\textbf {\bibinfo {volume} {107}},\ \bibinfo {pages}
  {163001} (\bibinfo {year} {2011})}\BibitemShut {NoStop}%
\bibitem [{\citenamefont {Harsha}\ \emph {et~al.}(2019)\citenamefont {Harsha},
  \citenamefont {Henderson},\ and\ \citenamefont
  {Scuseria}}]{Harsha19_Thermofield}%
  \BibitemOpen
  \bibfield  {author} {\bibinfo {author} {\bibfnamefont {Gaurav}\ \bibnamefont
  {Harsha}}, \bibinfo {author} {\bibfnamefont {Thomas~M.}\ \bibnamefont
  {Henderson}}, \ and\ \bibinfo {author} {\bibfnamefont {Gustavo~E.}\
  \bibnamefont {Scuseria}},\ }\bibfield  {title} {\enquote {\bibinfo {title}
  {Thermofield theory for finite-temperature quantum chemistry},}\ }\href
  {\doibase 10.1063/1.5089560} {\bibfield  {journal} {\bibinfo  {journal} {The
  Journal of Chemical Physics}\ }\textbf {\bibinfo {volume} {150}},\ \bibinfo
  {pages} {154109} (\bibinfo {year} {2019})}\BibitemShut {NoStop}%
\bibitem [{\citenamefont {Harsha}\ \emph {et~al.}(2022)\citenamefont {Harsha},
  \citenamefont {Xu}, \citenamefont {Henderson},\ and\ \citenamefont
  {Scuseria}}]{Harsha22_Thermal}%
  \BibitemOpen
  \bibfield  {author} {\bibinfo {author} {\bibfnamefont {Gaurav}\ \bibnamefont
  {Harsha}}, \bibinfo {author} {\bibfnamefont {Yi}~\bibnamefont {Xu}}, \bibinfo
  {author} {\bibfnamefont {Thomas~M.}\ \bibnamefont {Henderson}}, \ and\
  \bibinfo {author} {\bibfnamefont {Gustavo~E.}\ \bibnamefont {Scuseria}},\
  }\bibfield  {title} {\enquote {\bibinfo {title} {Thermal coupled cluster
  theory for su(2) systems},}\ }\href {\doibase 10.1103/PhysRevB.105.045125}
  {\bibfield  {journal} {\bibinfo  {journal} {Phys. Rev. B}\ }\textbf {\bibinfo
  {volume} {105}},\ \bibinfo {pages} {045125} (\bibinfo {year}
  {2022})}\BibitemShut {NoStop}%
\bibitem [{\citenamefont {Benavides-Riveros}\ \emph {et~al.}(2022)\citenamefont
  {Benavides-Riveros}, \citenamefont {Chen}, \citenamefont {Schilling},
  \citenamefont {Mantilla},\ and\ \citenamefont
  {Pittalis}}]{Benavides-Riveros22_Excitations}%
  \BibitemOpen
  \bibfield  {author} {\bibinfo {author} {\bibfnamefont {Carlos~L.}\
  \bibnamefont {Benavides-Riveros}}, \bibinfo {author} {\bibfnamefont {Lipeng}\
  \bibnamefont {Chen}}, \bibinfo {author} {\bibfnamefont {Christian}\
  \bibnamefont {Schilling}}, \bibinfo {author} {\bibfnamefont {Sebasti\'an}\
  \bibnamefont {Mantilla}}, \ and\ \bibinfo {author} {\bibfnamefont {Stefano}\
  \bibnamefont {Pittalis}},\ }\bibfield  {title} {\enquote {\bibinfo {title}
  {Excitations of quantum many-body systems via purified ensembles: A
  unitary-coupled-cluster-based approach},}\ }\href {\doibase
  10.1103/PhysRevLett.129.066401} {\bibfield  {journal} {\bibinfo  {journal}
  {Phys. Rev. Lett.}\ }\textbf {\bibinfo {volume} {129}},\ \bibinfo {pages}
  {066401} (\bibinfo {year} {2022})}\BibitemShut {NoStop}%
\bibitem [{\citenamefont {Sun}\ \emph {et~al.}(2020)\citenamefont {Sun},
  \citenamefont {Ray}, \citenamefont {Cui}, \citenamefont {Stoudenmire},
  \citenamefont {Ferrero},\ and\ \citenamefont
  {Chan}}]{Sun20_Finite-temperature}%
  \BibitemOpen
  \bibfield  {author} {\bibinfo {author} {\bibfnamefont {Chong}\ \bibnamefont
  {Sun}}, \bibinfo {author} {\bibfnamefont {Ushnish}\ \bibnamefont {Ray}},
  \bibinfo {author} {\bibfnamefont {Zhi-Hao}\ \bibnamefont {Cui}}, \bibinfo
  {author} {\bibfnamefont {Miles}\ \bibnamefont {Stoudenmire}}, \bibinfo
  {author} {\bibfnamefont {Michel}\ \bibnamefont {Ferrero}}, \ and\ \bibinfo
  {author} {\bibfnamefont {Garnet Kin-Lic}\ \bibnamefont {Chan}},\ }\bibfield
  {title} {\enquote {\bibinfo {title} {Finite-temperature density matrix
  embedding theory},}\ }\href {\doibase 10.1103/PhysRevB.101.075131} {\bibfield
   {journal} {\bibinfo  {journal} {Phys. Rev. B}\ }\textbf {\bibinfo {volume}
  {101}},\ \bibinfo {pages} {075131} (\bibinfo {year} {2020})}\BibitemShut
  {NoStop}%
\bibitem [{\citenamefont {Fromager}\ and\ \citenamefont
  {Lasorne}(2024)}]{Fromager_2024_Density}%
  \BibitemOpen
  \bibfield  {author} {\bibinfo {author} {\bibfnamefont {Emmanuel}\
  \bibnamefont {Fromager}}\ and\ \bibinfo {author} {\bibfnamefont {Benjamin}\
  \bibnamefont {Lasorne}},\ }\bibfield  {title} {\enquote {\bibinfo {title}
  {Density functional theory beyond the born–oppenheimer approximation: exact
  mapping onto an electronically non-interacting kohn–sham molecule},}\
  }\href {\doibase 10.1088/2516-1075/ad45d5} {\bibfield  {journal} {\bibinfo
  {journal} {Electronic Structure}\ }\textbf {\bibinfo {volume} {6}},\ \bibinfo
  {pages} {025002} (\bibinfo {year} {2024})}\BibitemShut {NoStop}%
\bibitem [{\citenamefont {Dupuy}\ \emph {et~al.}(2026)\citenamefont {Dupuy},
  \citenamefont {Lasorne},\ and\ \citenamefont
  {Fromager}}]{dupuy2026exactlyfactorizedmolecularkohnsham}%
  \BibitemOpen
  \bibfield  {author} {\bibinfo {author} {\bibfnamefont {Lucien}\ \bibnamefont
  {Dupuy}}, \bibinfo {author} {\bibfnamefont {Benjamin}\ \bibnamefont
  {Lasorne}}, \ and\ \bibinfo {author} {\bibfnamefont {Emmanuel}\ \bibnamefont
  {Fromager}},\ }\href {https://arxiv.org/abs/2601.03972} {\enquote {\bibinfo
  {title} {Exactly factorized molecular kohn-sham density functional theory},}\
  } (\bibinfo {year} {2026}),\ \Eprint {http://arxiv.org/abs/2601.03972}
  {arXiv:2601.03972 [physics.chem-ph]} \BibitemShut {NoStop}%
\bibitem [{\citenamefont {Requist}\ and\ \citenamefont
  {Gross}(2016)}]{Requist16_Exact}%
  \BibitemOpen
  \bibfield  {author} {\bibinfo {author} {\bibfnamefont {Ryan}\ \bibnamefont
  {Requist}}\ and\ \bibinfo {author} {\bibfnamefont {E.~K.~U.}\ \bibnamefont
  {Gross}},\ }\bibfield  {title} {\enquote {\bibinfo {title} {Exact
  factorization-based density functional theory of electrons and nuclei},}\
  }\href {\doibase 10.1103/PhysRevLett.117.193001} {\bibfield  {journal}
  {\bibinfo  {journal} {Phys. Rev. Lett.}\ }\textbf {\bibinfo {volume} {117}},\
  \bibinfo {pages} {193001} (\bibinfo {year} {2016})}\BibitemShut {NoStop}%
\bibitem [{\citenamefont {Li}\ \emph {et~al.}(2025{\natexlab{b}})\citenamefont
  {Li}, \citenamefont {Requist},\ and\ \citenamefont
  {Gross}}]{li2025bornoppenheimertimedependentdensityfunctional}%
  \BibitemOpen
  \bibfield  {author} {\bibinfo {author} {\bibfnamefont {Chen}\ \bibnamefont
  {Li}}, \bibinfo {author} {\bibfnamefont {Ryan}\ \bibnamefont {Requist}}, \
  and\ \bibinfo {author} {\bibfnamefont {E.~K.~U.}\ \bibnamefont {Gross}},\
  }\href {https://arxiv.org/abs/2511.09899} {\enquote {\bibinfo {title} {Beyond
  born-oppenheimer time-dependent density functional theory},}\ } (\bibinfo
  {year} {2025}{\natexlab{b}}),\ \Eprint {http://arxiv.org/abs/2511.09899}
  {arXiv:2511.09899 [physics.chem-ph]} \BibitemShut {NoStop}%
\end{thebibliography}%


\end{document}